\documentclass[fleqn,usenatbib]{mnras}

\usepackage{newtxtext,newtxmath}

\usepackage[T1]{fontenc}
\usepackage{ae,aecompl}

\usepackage{graphicx}	
\usepackage{amsmath}	
\usepackage{amssymb}	
\usepackage{colortbl}

\title[GSP Nebular SNe\,Ia]{Nebular-Phase Spectra of Type Ia Supernovae from the Las Cumbres Observatory Global Supernova Project}

\author[M.~L.~Graham et al.]{M.~L.~Graham$^{1}$\thanks{E-mail: mlg3k@uw.edu},
T.~D.~Kennedy$^{1}$,
S.~Kumar$^{2}$,
R.~C.~Amaro$^{3}$,
D.~J.~Sand$^{3}$,
\newauthor
S.~W.~Jha$^{4}$,
L.~Galbany$^{5}$,
J.~Vinko$^{6,7,8,9}$,
J.~C.~Wheeler$^{9}$,
E.~Y.~Hsiao$^{2}$,
\newauthor
K.~A.~Bostroem$^{1,10}$,
J.~Burke$^{11,12}$,
D.~Hiramatsu$^{11,12,13,14}$,
G.~Hosseinzadeh$^{3}$,
\newauthor
C.~McCully$^{11}$,
D.~A.~Howell$^{11,12}$,
T.~Diamond$^{15}$,
P.~Hoeflich$^{2}$,
X.~Wang$^{16,17}$,
W.~Li$^{18}$
\\
$^{1}$DiRAC Institute, Department of Astronomy, University of Washington, Box 351580, U.W., Seattle, WA 98195, USA \\
$^{2}$Department of Physics, Florida State University, 77 Chieftan Way, Tallahassee, FL 32306, USA \\
$^{3}$Steward Observatory, University of Arizona, 933 North Cherry Avenue, Tucson, AZ 85721-0065, USA \\
$^{4}$Department of Physics and Astronomy, Rutgers, the State University of New Jersey, 136 Frelinghuysen Road, Piscataway, NJ 08854, USA \\
$^{5}$Institute of Space Sciences (ICE, CSIC), Campus UAB, Carrer de Can Magrans, s/n, E-08193 Barcelona, Spain.\\
$^{6}$Konkoly Observatory, ELKH Research Center for Astronomy and Earth Sciences, Konkoly Thege M. ut 15-17, Budapest, 1121 Hungary \\
$^{7}$Department of Optics and Quantum Electronics, University of Szeged, Dom ter 9, Szeged, 6720 Hungary \\
$^{8}$Institute of Physics, ELTE E\"otv\"os Lor\'and University, P\'azmany P\'eter s\'et\'any 1/A, Budapest, 1117, Hungary \\
$^{9}$Department of Astronomy, University of Texas at Austin, 2515 Speedway, Stop C1400 Austin, Texas 78712-1205, USA \\
$^{10}$Department of Physics and Astronomy, University of California, 1 Shields Avenue, Davis, CA 95616-5270, USA \\
$^{11}$Las Cumbres Observatory, 6740 Cortona Drive, Suite 102, Goleta, CA 93117-5575, USA \\
$^{12}$Department of Physics, University of California, Santa Barbara, CA 93106-9530, USA \\
$^{13}$Center for Astrophysics, Harvard \& Smithsonian, 60 Garden Street, Cambridge, MA 02138, USA \\
$^{14}$The NSF AI Institute for Artificial Intelligence and Fundamental Interactions \\
$^{15}$Private astronomer, tiaradiamond@gmail.com \\
$^{16}$ Physics Department \& Tsinghua Center for Astrophysics, Tsinghua University, Beijing, 100084, China \\
$^{17}$ Beijing Planetarium, Beijing Academy of Science and Technology, Beijing 100044, China \\
$^{18}$ The School of Physics and Astronomy, Tel Aviv University, Tel Aviv 69978, Israel.
}

\date{Accepted XXX. Received YYY; in original form ZZZ}

\pubyear{2021}

\hypersetup{draft}

\begin{document}
\label{firstpage}
\pagerange{\pageref{firstpage}--\pageref{lastpage}}
\maketitle

\begin{abstract}

The observed diversity in Type Ia supernovae (SNe\,Ia) -- the thermonuclear explosions of carbon-oxygen white dwarf stars used as cosmological standard candles -- is currently met with a variety of explosion models and progenitor scenarios.
To help improve our understanding of whether and how often different models contribute to the occurrence of SNe\,Ia and their assorted properties, we present a comprehensive analysis of seven nearby SNe\,Ia.
We obtained one to two epochs of optical spectra with Gemini Observatory during the nebular phase ($>$200 days past peak) for each of these events, all of which had time-series of photometry and spectroscopy at early times (the first $\sim$8 weeks after explosion).
We use the combination of early- and late-time observations to assess the predictions of various models for the explosion (e.g., double-detonation, off-center detonation, stellar collisions), progenitor star (e.g., ejecta mass, metallicity), and binary companion (e.g., another white dwarf or a non-degenerate star).
Overall, we find general consistency in our observations with spherically-symmetric models for SN\,Ia explosions, and with scenarios in which the binary companion is another degenerate star.
We also present an in-depth analysis of SN\,2017fzw, a member of the sub-group of SNe\,Ia which appear to be transitional between the subluminous ``91bg-like" events and normal SNe\,Ia, and for which nebular-phase spectra are rare.

\end{abstract}


\section{Introduction}\label{sec:int}

Type Ia supernovae (SNe\,Ia) are well-established to be the thermonuclear explosions of carbon-oxygen white dwarf stars \citep[e.g.,][]{2000ARA&A..38..191H,2014ARA&A..52..107M}.
The detonation synthesizes radioactive $^{56}$Ni, which decays to $^{56}$Co and then $^{56}$Fe, powering the light curve.
The mass of synthesized $^{56}$Ni is known to be the primary physical cause of the light-curve width-luminosity correlation which allows SNe\,Ia to be used as cosmological standard candles \citep[e.g.,][]{1977SvA....21..675P,1993ApJ...413L.105P}.
During the photospheric phase in the months after explosion, while the ejecta material is still optically thick, SN\,Ia spectra are dominated by high-velocity absorption features of \ion{Si}{II}, \ion{S}{II}, \ion{Mg}{II}, and \ion{Ca}{II}. 
However, at $>$200 days after peak brightness, the ejecta material becomes optically thin and the optical spectra are dominated by forbidden emission lines from the nucleosynthetic products: iron, cobalt, and nickel.
The velocity, width, and flux in these lines can reveal the amount and spatial distribution of these elements within the nebula.
Correlations between late- and early-time observations are particularly useful to understand some of the open questions regarding the progenitor white dwarf, the explosion mechanism, and the binary companion star type.

\textit{Explosion Mechanism --}
Some of the open questions regarding SN\,Ia explosions include whether the detonation is preceded by a deflagration phase of slow burning in the core (the ``delayed detonation" model) and how long that phase might be \citep[e.g.,][]{1991A&A...245..114K,1994ApJ...427..315A,2013MNRAS.429.1156S}; 
whether and how a surface detonation of accrued material can initiate a core detonation (the ``double detonation" model; \citealt{1998MNRAS.301..405W,2007A&A...476.1133F}); 
whether and how often a detonation can be instigated by a violent merger with a binary companion star \citep[e.g.,][]{2011A&A...528A.117P,2016MNRAS.459.4428K};
and whether and how the white dwarf's mass affects the explosion and the observed SN\,Ia properties, especially when it is less than the Chandrasekhar mass of 1.4 $\rm M_{\odot}$ \citep[e.g.,][]{2017MNRAS.470..157B,2019ApJ...873...84P}.
For example, the fact that nickel is seen in late-time spectra indicates that the density of the progenitor white dwarf was high enough to burn to stable nickel (as the radioactive $^{56}$Ni has decayed by the nebular phase), and this density requirement has traditionally excluded white dwarfs that are substantially below the Chandrasekhar mass \citep{2018MNRAS.474.3187W}.
However, recent simulations by \citet{2021ApJ...909L..18S} have shown for the first time that white dwarf explosions for a wide range of masses, including sub-Chandrasekhar mass, can reproduce the observed characteristics of normal SNe\,Ia.
Furthermore, many explosion models predict different spatial/velocity distributions for the nucleosynthetic material, and/or different mass ratios for stable and radioactive materials, which can be investigated with nebular-phase spectra. 

\textit{Explosion Asymmetry --}
\citet{2010Natur.466...82M} present a model for asymmetric explosions in which SNe\,Ia with a detonation that is offset ``away" from the observer results in both a  red-shifted nebular feature and a  quickly-declining velocity for the \ion{Si}{II}~$\lambda$6355~\AA\ absorption feature in the two weeks after peak brightness (a ``high velocity gradient", HVG; \citealt{2005ApJ...623.1011B}), and vice versa for offsets ``towards" the observer (a blue-shifted nebular feature and a low velocity gradient, LVG).
So far, SN\,Ia observations show that most HVG SNe\,Ia exhibit red-shifted nebular features, but it remains unclear whether explosion asymmetry is the unique explanation for this trend.
Another source of asymmetry in SN\,Ia explosions could be head-on collisions between white dwarf stars (WD-WD collisions), potentially driven into highly elliptical orbits by a tertiary star \citep[e.g.,][]{2009ApJ...705L.128R,2009MNRAS.399L.156R,2013ApJ...778L..37K}.
If such a collision was aligned with the observer's line-of-sight, the velocities of the WD-WD pair could cause double-peaked nebular-phase emission features \citep[e.g.,][]{2015MNRAS.454L..61D}.
However, \citet{2013Sci...340..170W} argued that the observed diversity in SNe\,Ia might be due to progenitor environment (i.e., progenitor metallicity), instead of explosion asymmetry.

\textit{Progenitor Metallicity --}
\citet{2003ApJ...590L..83T} describe how the additional neutrons available in higher-metallicity white dwarf stars might lead to a higher ratio of stable-to-radioactive nucleosynthetic products.
At early times, higher-metallicity progenitors might exhibit a depressed near-ultraviolet (NUV) flux due to line blanketing \citep{2000ApJ...530..966L}.
In the nebular phase, higher metallicity could manifest as a higher Ni/Fe ratio, as all the radioactive $^{56}$Ni would have decayed, leaving stable nickel as the source of forbidden emission line [\ion{Ni}{II}]~$\lambda$7378~\AA.

\textit{Non-degenerate Companions --}
Whether and how often the binary companion star is another white dwarf (the double-degenerate scenario), or a main sequence or red giant star (the single-degenerate scenario), is not yet well constrained. 
The presence of a main sequence or red giant companion star could be revealed by a ``blue bump" in the very early-time light curve (within a few days of explosion; \citealt{2010ApJ...708.1025K}), and/or by a narrow H$\alpha$ emission feature from hydrogen swept off of a non-degenerate companion and embedded in the expanding nebula \citep[e.g.,][]{2005A&A...443..649M,
2007ApJ...670.1275L}.
So far, no SN\,Ia has exhibited \emph{both} potential signatures of a non-degenerate companion star.

In this work, we present nebular-phase optical and infrared spectroscopy from Gemini Observatory for seven nearby SNe\,Ia with early-time (first $\lesssim$2 months after explosion) optical photometry and spectroscopy from the Las Cumbres Observatory \citep{2013PASP..125.1031B} and, in most cases, from other facilities as well.
We describe the targeted sample of SNe\,Ia and present relevant early-time data in Section~\ref{sec:sne}, and present and analyze our nebular-phase observations -- and measure parameters for the forbidden emission lines of nickel, cobalt, and iron -- in Section~\ref{sec:nebobs}.
In Section~\ref{sec:ana} we analyze our sample of nebular-phase spectra as a whole, and compare with previously published samples, in order to assess general physical models for SN\,Ia explosions and progenitor scenarios. 
In Section~\ref{sec:disc} we provide unique in-depth physical analyses based on the early- and late-time data for each individual SN\,Ia.
A summary of our conclusions is provided in Section~\ref{sec:con}.
In this work we assume a flat cosmology of $H_0=70$~$\rm km\ s^{-1}\ Mpc^{-1}$, $\Omega_M=0.3$, and $\Omega_{\Lambda}=0.7$, and quote all light curve phases in days from peak brightness unless otherwise specified.

\section{The Sample of Nearby SNe\,Ia}\label{sec:sne}

\begin{figure*}
\includegraphics[width=4.cm]{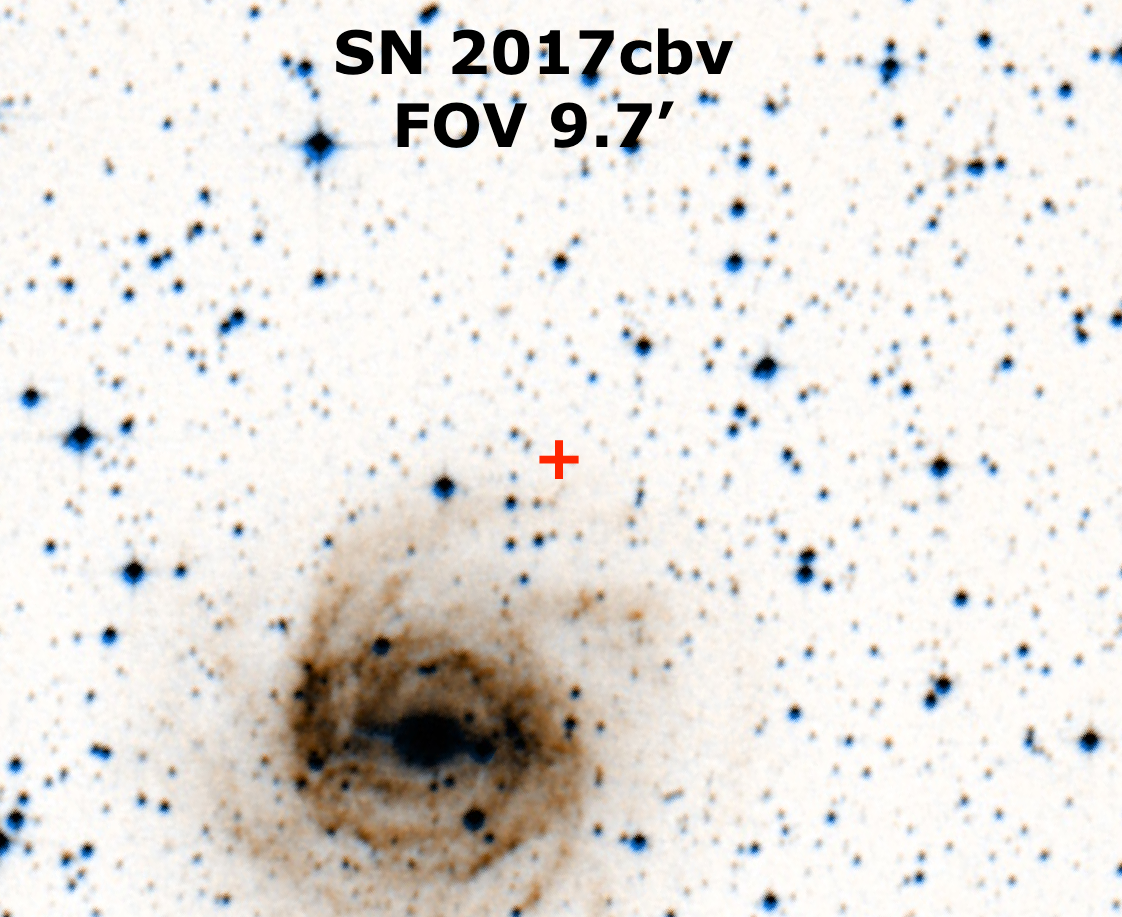}
\includegraphics[width=4.cm]{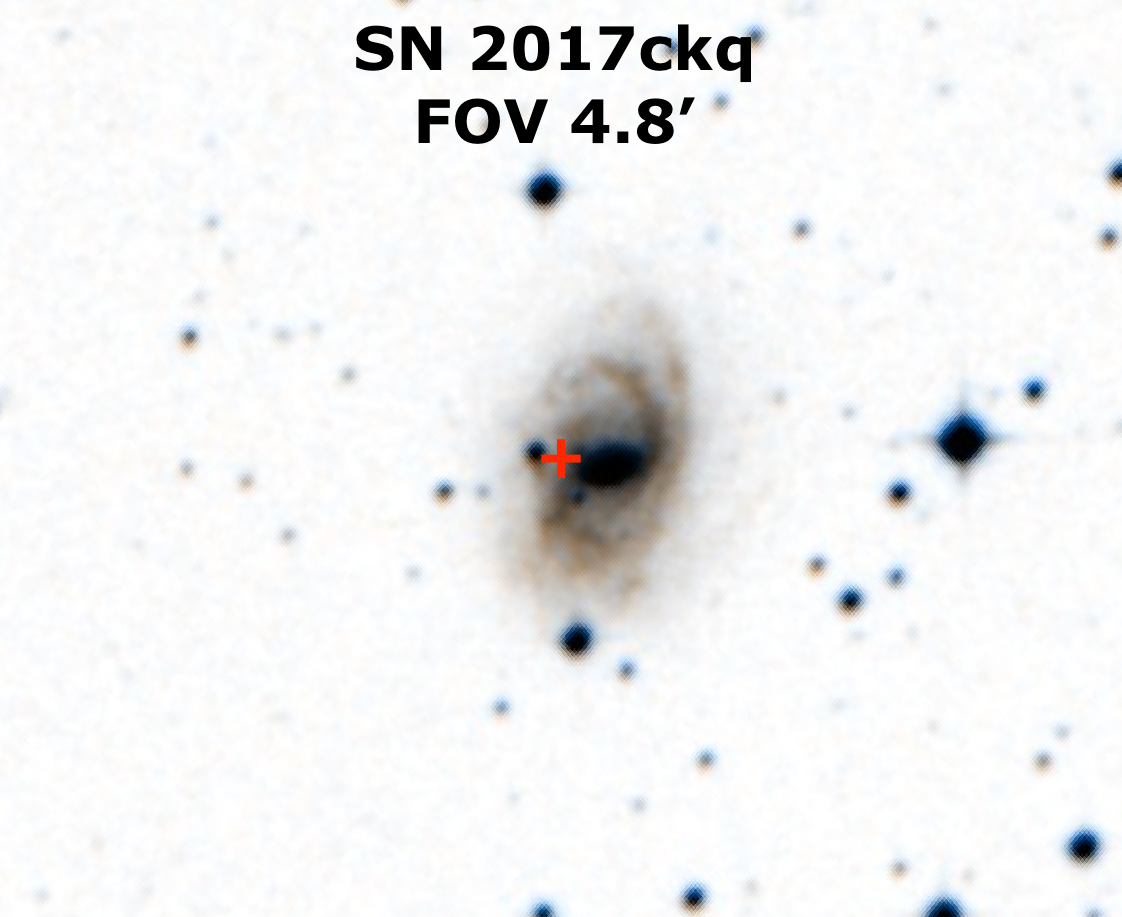}
\includegraphics[width=4.cm]{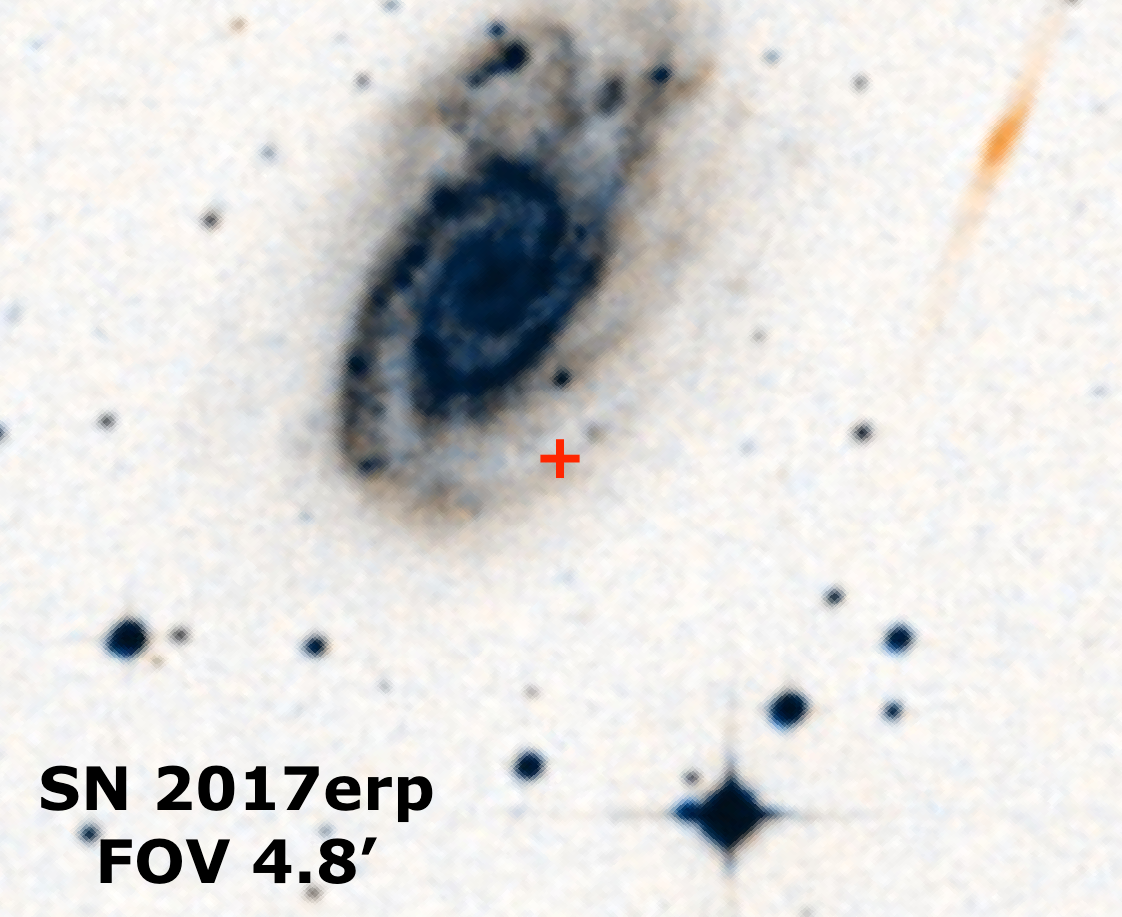}
\includegraphics[width=4.cm]{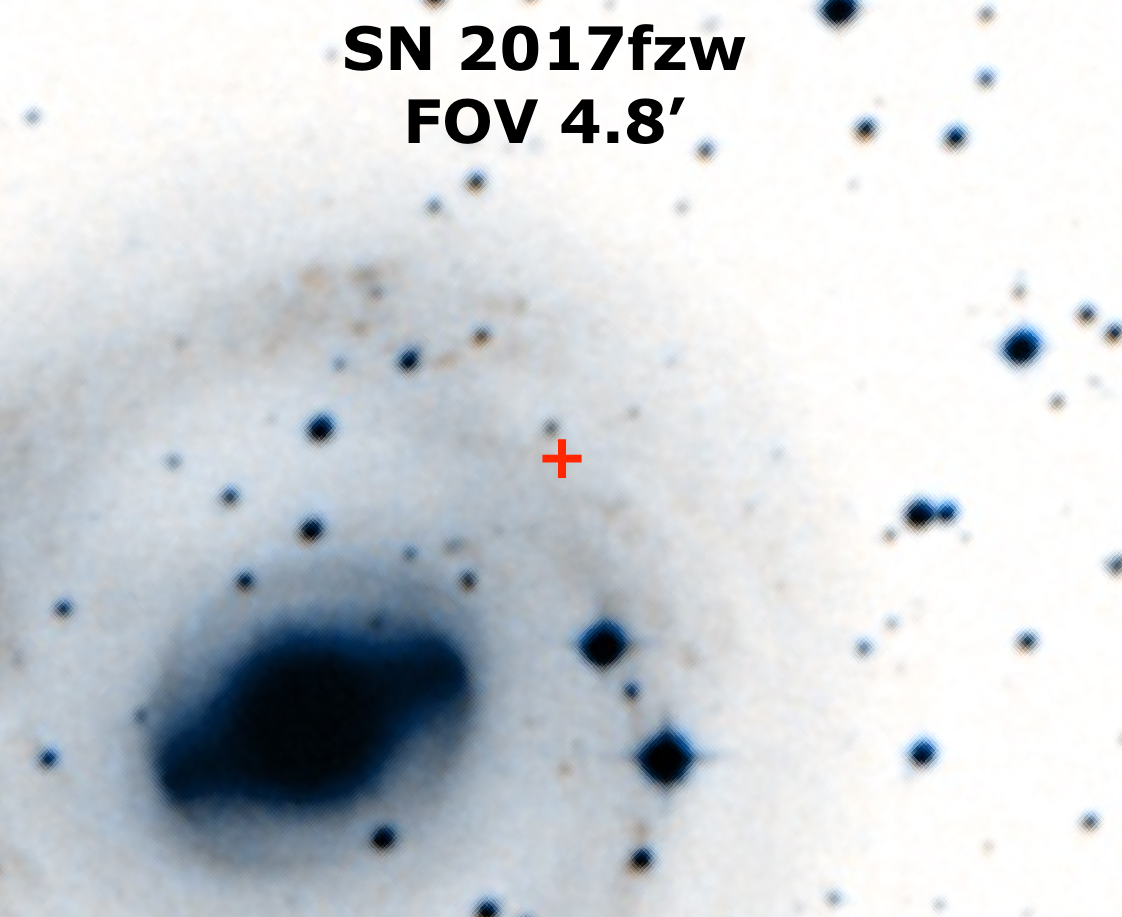}
\includegraphics[width=4.cm]{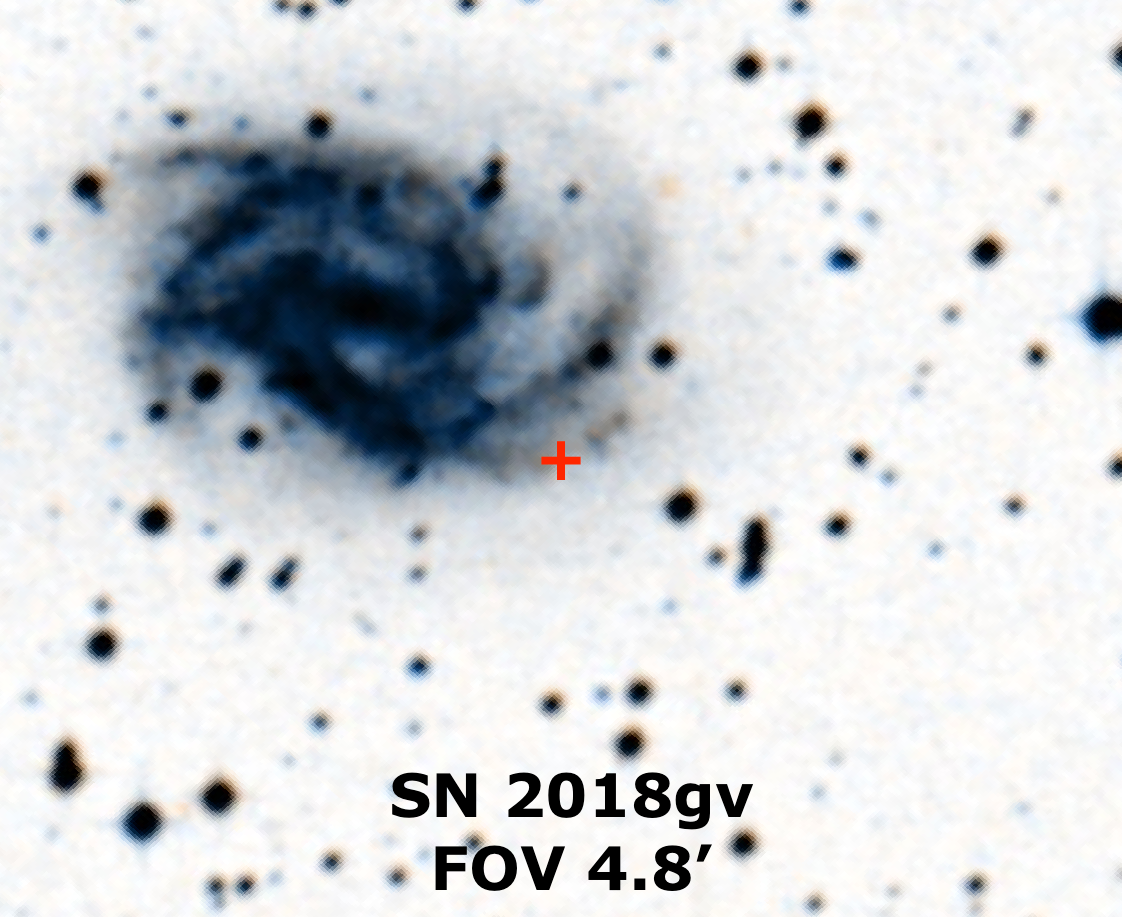}
\includegraphics[width=4.cm]{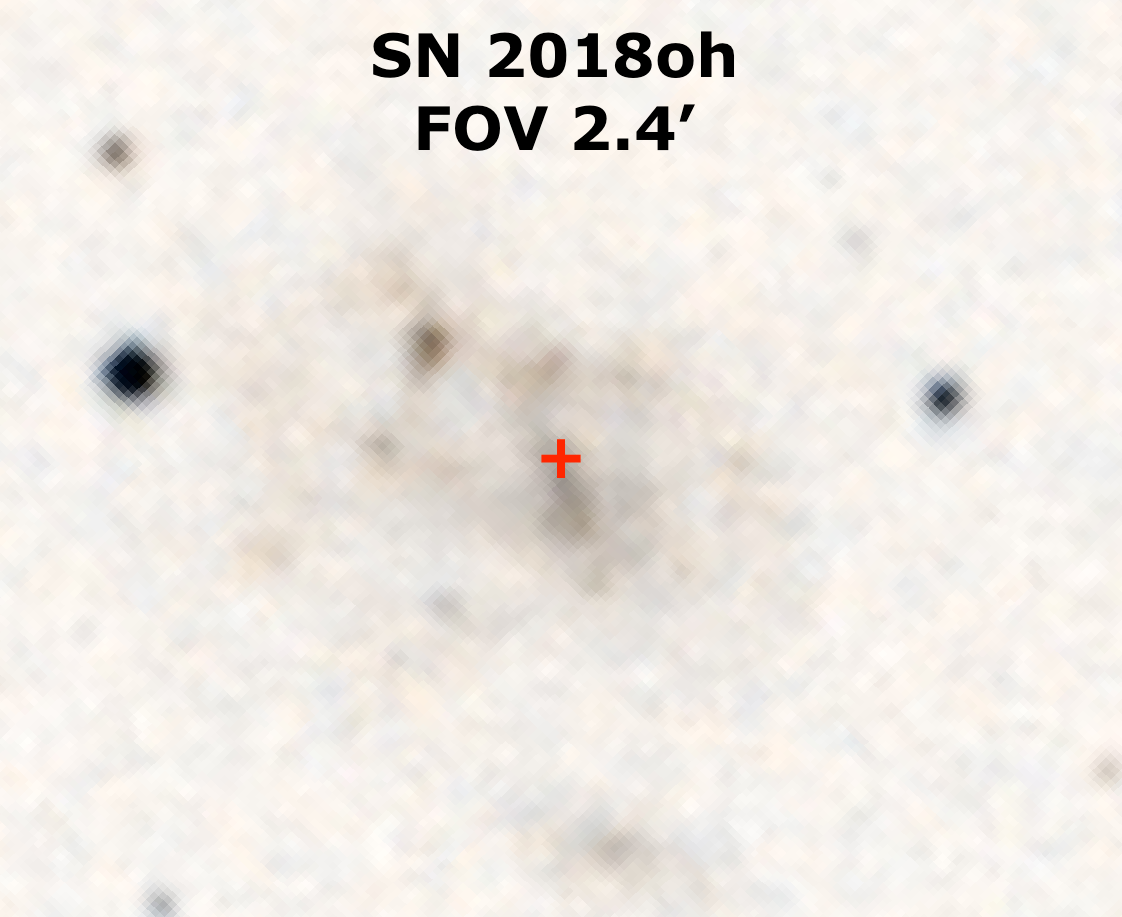}
\includegraphics[width=4.cm]{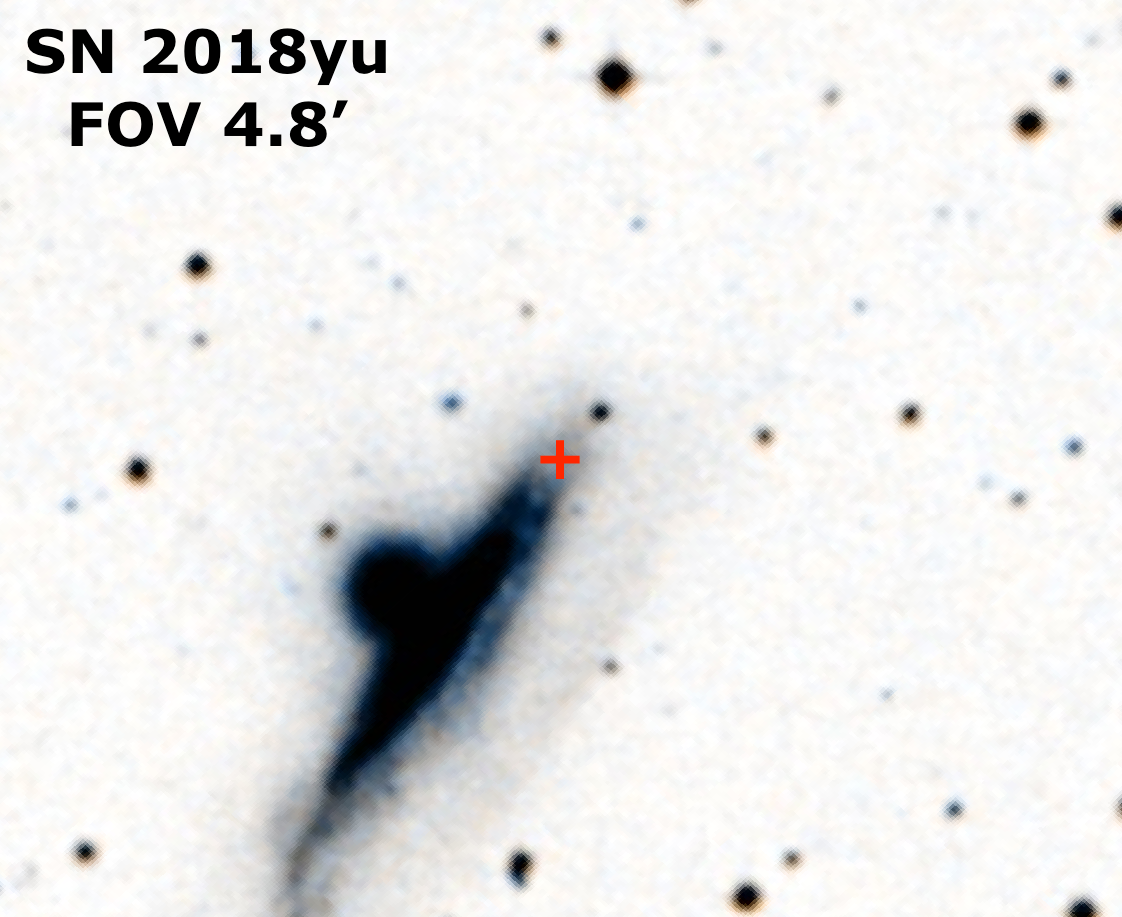}
\caption{Image stamps from the Digital Sky Survey (inverse color), centered on the location of each of our SN\,Ia (red cross), to illustrate the location of each with respect to its host galaxy. All images are north-up east-left, and have fields of view (FOV) as indicated.}
\label{fig:stamps}
\end{figure*}

The Las Cumbres Observatory's \citep{2013PASP..125.1031B} Global Supernova Project (GSP) obtains optical photometry and spectroscopy of nearby SNe with a regular cadence.
The GSP engages in this monitoring to enable studies of both large samples of SNe and of rare events, such as SNe that are very nearby or peculiar in some way.
From the GSP-monitored sample we selected SNe\,Ia for nebular-phase spectroscopy at Gemini Observatory that were well offset from their host galaxy (Figure~\ref{fig:stamps}), were nearby enough to be $B\lesssim22$ mag at $>$200 days after peak brightness, had a well-sampled early-time light curve through $\sim$30 days post-peak (so that the phase and brightness at maximum can be measured), and had at least 2 spectral epochs in the $\sim$2 weeks after peak brightness (so that the photospheric velocity and its gradient can be measured).
After applying those criteria, we prioritized SNe\,Ia which were otherwise peculiar, such as those that showed deviations in their early-time light curve or belonged to sub-classes such as SN\,1991bg-like \citep{1992AJ....104.1543F}.

In Table~\ref{tab:sne} we list the names and coordinates of the SNe\,Ia included in this analysis, along with their redshift and distance derived from their host galaxy properties, and the line-of-sight Galactic extinction for their coordinates.
In Table~\ref{tab:early} we list the early-time qualities of our sample's light curves and spectra, based either on data from the Las Cumbres Observatory or other published works as cited in Table~\ref{tab:early} and as described in Sections~\ref{ssec:sne_earlyphot} and \ref{ssec:sne_earlyspec}.
These early-time qualities are incorporated into the analysis and discussion of these individual SNe\,Ia in Section~\ref{sec:disc}. 

\begin{table*}
\centering
\begin{tabular}{|lccccc|}
\hline
SN Name & Coordinates & Redshift & Distance & \multicolumn{2}{c|}{Galactic Extinction$^{[1]}$} \\
 & RA,Dec (J2000) & ($z$) & (Mpc) & $A_B$ (mag) & $E(B-V)$ mag \\
\hline
2017cbv & 14:32:34.38, -44:08:03.1 & 0.00399$^{\rm [2]}$                        & $16.9\pm3.1^{\rm [3]}$         & 0.5956 & 0.1452\\
2017ckq & 10:44:25.39, -32:12:32.8 & $0.0100\pm0.0001^{\rm [4]}$         & 42.9$^{\rm [5]}$                     & 0.2657 & 0.0648  \\
2017erp & 15:09:14.81, -11:20:03.2  & $0.006174\pm0.000003^{\rm [6]}$ &  $29.4 \pm 1.3^{\rm [7]}$      & 0.3800 & 0.0928  \\
2017fzw & 06:21:34.77, -27:12:53.5 & 0.0054$^{\rm [2]}$                           & 19.5$^{\rm [3]}$                    & 0.1512 & 0.0369   \\
2018gv  & 08:05:34.61, -11:26:16.3  & 0.0053$^{\rm [2]}$                           & 16.8$^{\rm [5]}$                    & 0.2040 & 0.0497   \\
2018oh  & 09:06:39.59, +19:20:17.5 & 0.010981$^{\rm [9]}$                       & $47.9 \pm 1.9$ $^{\rm [7]}$ & 0.1552 & 0.0368   \\
2018yu  & 05:22:32.36, -11:29:13.8  & 0.00811$^{\rm [2]}$                         & 37.1$^{\rm [8]}$                    & 0.5363 & 0.1306   \\
\hline
\end{tabular}
\caption{A list of the SNe\,Ia included in this study.
[1] Extinction values are from \protect\citet{2011ApJ...737..103S}, and assume Galactic $R_V=3.1$. 
[2] \protect{\citet{1991rc3..book.....D}}. 
[3] \protect{\citet{Tully1988}}. 
[4] \protect{\citet{2000ApJ...529..786M}}. 
[5] \protect{\citet{2013AJ....146...86T}}. 
[6] \protect{\citet{2005A&A...430..373T}}. 
[7] \protect{\citet{2020ApJ...892..121K}}.
[8] \protect{\citet{2007A&A...465...71T}}. 
[9] \protect{\citet{1990ApJS...72..245S}}. 
\label{tab:sne}}
\end{table*}

\begin{table*}
\centering
\begin{tabular}{|lc|ccc|ccc|} 
\hline
 & & \multicolumn{3}{c|}{Early-Time $B$-Band Light Curve} & \multicolumn{3}{c|}{Early-Time Photospheric \ion{Si}{II}} \\
SN\,Ia & Sub- & Peak Date & Peak Brightness & Decline Rate                 & Velocity$^{[1]}$   & Gradient$^{[2]}$           & Class$^{[3]}$ \\
Name   & Type & (in $B$; UT)      & ($B$-band mag)  & $\Delta m_{\rm 15}(B)$ (mag) & ($\rm km\ s^{-1}$) & ($\rm km\ s^{-1}\ d^{-1}$) &   \\
\hline
2017cbv & 99aa         & 2017-03-29.1$^{\rm [4]}$   & -19.25$^{\rm [5]}$                & 1.06$^{\rm [4]}$                    & $-9300\pm60$     & $19\pm8$   & LVG \\
2017ckq &                  & 2017-04-08.1                     & $-18.9\pm0.3$                      & $0.96\pm0.20$                     & $-10500\pm200$ & $71\pm23$  & $\sim$HVG \\
2017erp &                   & 2017-06-30.9$^{\rm [6]}$  & $-19.50\pm0.15^{\rm [6,7]}$ & $1.129 \pm 0.011^{\rm [7]}$ & $-10500\pm100$ & $78\pm13$  & $\sim$HVG \\
2017fzw & 91bg/trans & 2017-08-22.9$^{\rm [8]}$  & $-18.81\pm0.18^{\rm [8]}$    & $1.60\pm0.02^{\rm [8]}$      & $-13800\pm400$ & $300\pm50$ & FAINT \\
2018gv  &                   & 2018-01-31$^{\rm [9]}$      & $-19.1^{\rm [9]}$                   & $0.96^{\rm [9]}$                   & $-11100\pm90$   & $29\pm4$   & LVG \\
2018oh  &                   & 2018-02-13.7$^{\rm [10]}$ & $-19.47\pm0.10^{\rm [10]}$  & $0.96\pm0.03^{\rm [10]}$    & $-10100\pm10$  & $69\pm4^{\rm [10]}$  & LVG \\
2018yu  &                   & 2018-03-17.9                      & $-19.1\pm0.5$                      & $1.05\pm0.05$                    & $-9500\pm120$  & $45\pm21$  & LVG \\
\hline
\end{tabular}
\caption{The characteristic quantities from the early-time data for our targeted SNe\,Ia. All values derived in this work unless otherwise cited.
[1] \ion{Si}{II} velocity at peak brightness.
[2] Rate of decline of \ion{Si}{II} during the first two weeks after peak brightness.
[3] Low (L) or high (H) velocity gradient (VG) group (division at $70$~$\rm km\ s^{-1}\ day^{-1}$) or the FAINT group \protect{\citep{2005ApJ...623.1011B}}.
[4] \protect{\citet{2017ApJ...845L..11H}}.
[5] \protect{\citet{2018ApJ...863...24S}}.
[6] \protect{\citet{2019ApJ...877..152B}}.
[7] \protect{\citet{2020ApJ...892..121K}}.
[8] Galbany et al., (in prep.) .
[9] \protect{\citet{2020ApJ...902...46Y}}.
[10] \protect{\citet{2019ApJ...870...12L}}.
\label{tab:early} }
\end{table*}

\subsection{Near-Peak Optical Photometry}\label{ssec:sne_earlyphot}

\begin{table}
\centering
\begin{tabular}{|lcc|} 
\hline
MJ Date & $B$ Mag & $V$ Mag \\
\hline
\multicolumn{3}{|l|}{\it SN\,2017ckq:} \\
57844.1 & $14.58\pm0.25$ & $14.51\pm0.27$  \\
57848.7 & $14.27\pm0.15$ & $14.28\pm0.20$  \\
57851.9 & $14.29\pm0.17$ & $14.24\pm0.21$  \\
57853.6 & $14.26\pm0.09$ & $13.97\pm0.09$  \\
57857.6 & $14.55\pm0.09$ & $14.34\pm0.11$  \\
57861.8 & $14.70\pm0.30$ & $14.44\pm0.19$  \\
57866.2 & $15.26\pm0.26$ & $14.76\pm0.25$  \\
57870.2 & $15.71\pm0.38$ & $14.96\pm0.26$  \\
\hline
\multicolumn{3}{|l|}{\it SN\,2018yu:} \\
58186.0 & $14.66\pm0.09$ & $14.54\pm0.08$ \\
58189.0 & $14.25\pm0.05$ & $14.12\pm0.06$ \\
58192.0 & $14.06\pm0.07$ & $13.91\pm0.06$ \\
58193.1 & $14.02\pm0.05$ & $14.03\pm0.04$ \\
58195.0 & $14.02\pm0.07$ & $13.87\pm0.06$ \\
58196.0 & $14.03\pm0.05$ & $13.87\pm0.06$ \\
58197.0 & $14.06\pm0.05$ & $13.88\pm0.06$ \\
58201.4 & $14.26\pm0.08$ & $14.06\pm0.06$ \\
58205.0 & $14.51\pm0.05$ & $14.22\pm0.06$ \\
58211.1 & $15.20\pm0.07$ & $14.56\pm0.06$ \\
\hline
\end{tabular}
\caption{The near-peak $B$- and $V$-band photometry from Las Cumbres Observatory used to measure the light curve properties of SNe\,Ia 2017ckq and 2018yu. For all of our other targets, the light curve parameters that we use are from previously published work. \label{tab:lcs} }
\end{table}

The existence of near-peak optical photometry from the Las Cumbres Observatory was one of the selection criteria for the SNe that we targeted for nebular-phase spectroscopy with Gemini Observatory.
However, since most of our targets were interesting nearby events, by the time of this nebular-phase analysis five of our seven targets already had publications based on analyses of early-time photometry.
These publications included the derived light curve parameters of peak date, peak brightness, and decline rate in the $B$ filter which are needed for contextual analysis of our nebular-phase spectra.
Instead of recalculating these properties based on Las Cumbres photometry, for these five SNe\,Ia we have adopted the published values and cited the appropriate literature in Table~\ref{tab:early}. 
For the two SNe\,Ia which did not have published light curve parameters, SNe\,2017ckq and 2018yu, we provide their near-peak $B$- and $V$-band photometry from the Las Cumbres Observatory in Table~\ref{tab:lcs}, and discuss their light curves below.

\textit{SN\,2017ckq --} Photometric monitoring with the Las Cumbres Observatory shows that SN\,2017ckq peaked in $B$-band brightness on 2017-04-08.1 UT at $14.3 \pm 0.1$ mag and exhibited a decline rate of $\Delta m_{\rm 15}(B) \approx 0.96 \pm 0.2$ mag.
These measurements are based on a third-order polynomial fit to the host-subtracted SN photometry listed in Table~\ref{tab:lcs}.
The underlying host galaxy surface brightness at the location of SN\,2017ckq is quite faint, $\mu_B \approx 22.6$ $\rm mag\ arcsec^{-2}$ (despite the impression given by the DSS image in Figure~\ref{fig:stamps}).
Photometry was measured using Source Extractor \citep{1996A&AS..117..393B} and calibrated using the APASS star catalog \citep{2016yCat.2336....0H}.
Based on the $V$-band observations near $B$-band peak brightness, the color $B-V$ is consistent with $0$, suggesting a low amount of host galaxy extinction.
Given the host galaxy distance, the peak absolute magnitude of SN\,2017ckq is $M_B\approx-18.9\pm0.3$ mag.

\textit{SN\,2018yu --} Photometric monitoring with the Las Cumbres Observatory shows that SN\,2018yu peaked in $B$-band brightness on 2018-03-17.9 UT at $14.02\pm0.05$ mag and exhibited a decline rate of $\Delta m_{\rm 15}(B) \approx 1.05\pm0.05$ mag.
These measurements are based on a third-order polynomial fit to the host-subtracted SN photometry listed in Table~\ref{tab:lcs}.
The underlying host galaxy surface brightness at the location of SN\,2018yu is very faint, $\mu_B \approx 23.6$ $\rm mag\ arcsec^{-2}$.
As with SN\,2017ckq, photometry was measured using Source Extractor \citep{1996A&AS..117..393B} and calibrated using the APASS star catalog \citep{2016yCat.2336....0H}.
At peak $B$-band brightness, the SN's color is $B-V\approx0.14$ mag, suggesting $\leq0.2$ mag of extinction due to the host galaxy.
We correspondingly adjust the peak apparent brightness to be $B\approx 13.8\pm0.1$ mag.
However, we also note that early-time optical spectra obtained by the Las Cumbres Observatory exhibit a host-galaxy component of \ion{Na}{I}D at $z=0.01$ with an equivalent width of $\approx 0.3$~\AA, and that this corresponds to an approximate host-galaxy $E(B-V)\approx0.03$ based on the empirical relation derived in \citet{2012MNRAS.426.1465P}, which suggests less host-galaxy extinction ($<0.1$ mag, assuming $R_V=3.1$).
The peak absolute magnitude of SN\,2018yu is $M_B\approx-19.1\pm0.5$ mag, with most of that uncertainty due to the host galaxy's distance error.

\subsection{Photospheric Velocity Classifications}\label{ssec:sne_earlyspec}

\begin{table}
\centering
\begin{tabular}{|ll|ll|}
\hline
SN Name & Date & SN Name & Date  \\
\hline
2017cbv      & 2017-03-31   & SN2017fzw & 2017-08-13 \\
                    & 2017-04-03   &                    & 2017-09-02 \\
                    & 2017-04-07   &                     & 2017-09-07 \\
                    & 2017-04-07  &                     & 2017-09-14 \\
                    & 2017-04-13  & SN2018gv  & 2018-02-04 \\
SN2017ckq & 2017-04-09  &                     & 2018-02-08 \\
                    & 2017-04-13   &                     & 2018-02-13 \\
                    & 2017-04-17  & SN2018oh  & 2018-02-14 \\
                    & 2017-04-21  &                     & 2018-02-20 \\
SN2017erp & 2017-06-30   & SN2018yu  & 2018-03-16 \\
                    & 2017-07-06  &                     & 2018-03-19 \\
                    & 2017-07-12  &                     & 2018-03-20 \\
                    & 2017-07-15  &                     & 2018-03-24 \\
                    &                      &                     & 2018-03-28 \\
\hline
\end{tabular}
\caption{The early-time spectra from FLOYDS at Las Cumbres that were used to determine the \ion{Si}{II} velocity gradients.
\label{tab:early_spec_list}}
\end{table}

We use early-time optical spectroscopy from the FLOYDS robotic spectrograph on the Las Cumbres Observatory 2 meter telescopes, as listed in Table~\ref{tab:early_spec_list}, to measure each SN's photospheric velocity and velocity gradient, and to classify each SN as belonging to the low- or high-velocity gradient sub-classes \citep[LVG/HVG][]{2005ApJ...623.1011B}.
To measure the photospheric velocity, we smooth the FLOYDS optical spectra with a Savitsky-Golay filter of width $\sim20$--$50$~\AA, use the minimum of the \ion{Si}{II}~$\lambda$~6355~\AA\ line as representative of the velocity of the photosphere, and bootstrap (shuffle the flux uncertainties between pixels, simulate a new observed flux values, and remeasure the line velocity) to estimate the uncertainty in the velocity measurement.
We measure the photospheric velocity in this way for all spectra obtained by the Las Cumbres Observatory between phases of $0$ to $15$ days after peak brightness, and then do a linear regression to interpolate the velocity at peak ($v_{\rm SiII}$) and to estimate velocity gradient ($\dot{v}_{\rm SiII}$), which are quoted in columns 6 and 7 of Table~\ref{tab:early}.
Several modifications made to this technique for a few of our SNe\,Ia, as discussed in the paragraphs below.
In column 8 of Table~\ref{tab:early} we assign each SN\,Ia to either the low- and high-velocity gradient groups (L/HVG) based on whether their $\dot{v}_{\rm SiII}$ is below or above $70$~$\rm km\ s^{-1}\ day^{-1}$, as defined by \cite{2005ApJ...623.1011B}.
For SNe\,Ia which have an uncertainty that overlaps the boundary we represent this uncertainty by assigning the type as $\sim$L/HVG.

\textit{SN\,2017erp --} Our measured velocity gradient for SN\,2017erp of $\dot{v}_{\rm SiII} = 78\pm13$~$\rm km\ s^{-1}\ day^{-1}$ was obtained from FLOYDS spectra at $-0.8$, $5$, and $11$, and $14$ days (Table~\ref{tab:early_spec_list}), combined with two spectra from Lick Observatory which were obtained at phases of $0.1$ and $4$ days \citep{2020MNRAS.492.4325S}\footnote{Publicly available in the Weizmann Interactive Supernova Data Repository (WISeREP) at \url{www.wiserep.org}.}.
This velocity gradient is greater than the $70$~$\rm km\ s^{-1}\ day^{-1}$ minimum for classifying a SN\,Ia as ``high velocity gradient" (HVG), but since the error bar does overlap with that cutoff, we list SN\,2017erp as ``$\sim$HVG" in Table~\ref{tab:early}.
Our measured photospheric velocity for SN\,2017erp of $v_{\rm SiII}=-10500\pm100$ $\rm km\ s^{-1}$ is an average of the velocities from the $-0.8$ and $0.1$ day spectra.

\textit{SN\,2017fzw --} Our estimate for the photospheric velocity is a special case because 91bg-like and transitional events evolve quickly, but our spectral sampling is relatively sparse -- the four near-peak spectral epochs are at phases $-12$, $8$, $13$, and $20$ days -- and there are no supplemental publicly-available near-peak optical spectra for this object.
We fit a line between the first two epochs in order to estimate the velocity at peak brightness, and between the second two epochs to estimate the velocity gradient, and bootstrap our uncertainty estimates based on line fits with two, three, and all four epochs.
Since the low- and high-velocity gradient groups apply to normal SNe\,Ia, we do not assign an L/HVG subclass for SN\,2017fzw in Table~\ref{tab:early}, and instead label it as FAINT to be consistent with \citet{2005ApJ...623.1011B}.

\textit{SN\,2018oh --} The peak velocity is taken directly from a FLOYDS spectrum with a phase of $0$ days.
There is only one other FLOYDS spectrum within the first two weeks after peak brightness, at $6$ days, and both spectra are relatively noisy.
We find that the application of a Monte Carlo analysis to estimate the slope between these two points finds a velocity gradient of $75\pm65$~$\rm km\ s^{-1}\ day^{-1}$.
This is an insufficient confidence to declare SN\,2018oh a member of the HVG subgroup, especially since its velocity at peak brightness is low.
Instead, we use the work of \cite{2019ApJ...870...12L}, who present a dense time-series of optical spectroscopy which exhibit a photospheric \ion{Si}{II} velocity at peak brightness of $10300$~$\rm km\ s^{-1}$, and a velocity gradient over the first 10 days after peak brightness of $69\pm4$~$\rm km\ s^{-1}\ day^{-1}$, which is on the high side but puts SN\,2018oh in the LVG class.

\section{Nebular-Phase Observations}\label{sec:nebobs}

We obtained spectroscopic observations of the seven SNe\,Ia in our sample at $>200$ days after peak brightness primarily via a targeted follow-up program at Gemini Observatory, but also used late-time observations from other facilities. 
This section describes the acquisition, reduction, and calibration of these spectra, and measurements of nebular-phase emission line parameters such as velocity, full-width at half-max (FWHM), and integrated flux.

\subsection{Optical Spectra from Gemini Observatory}\label{ssec:nebobs_gmos}

The observation dates, instrument configurations, and exposure times for the longslit optical spectroscopy of the SNe\,Ia in our sample that we obtained with Gemini Observatory's Gemini Multi-Object Spectrograph (GMOS) in longslit mode \citep{2004PASP..116..425H} are listed in Table~\ref{tab:specobs} in Appendix~\ref{app:tables}.
To reduce and calibrate our data, internal spectroscopic flats and CuAr arc lamps were obtained during the night, before or after the object exposures, bias frames during the day, and standard stars were observed in the same configuration within a few nights.
Data were reduced with IRAF using custom scripts based on the Gemini data reduction cookbook \citep{GeminiCookboox}.
One-dimensional spectra were extracted from each reduced, sky-subtracted, two-dimensional spectrum, and adjacent pixels were used to fit for and remove as much contaminating flux from the host galaxy as possible (but some residuals remain; e.g., from emission lines with spatial variation at the SN's location).

\begin{figure*}
\includegraphics[width=8cm]{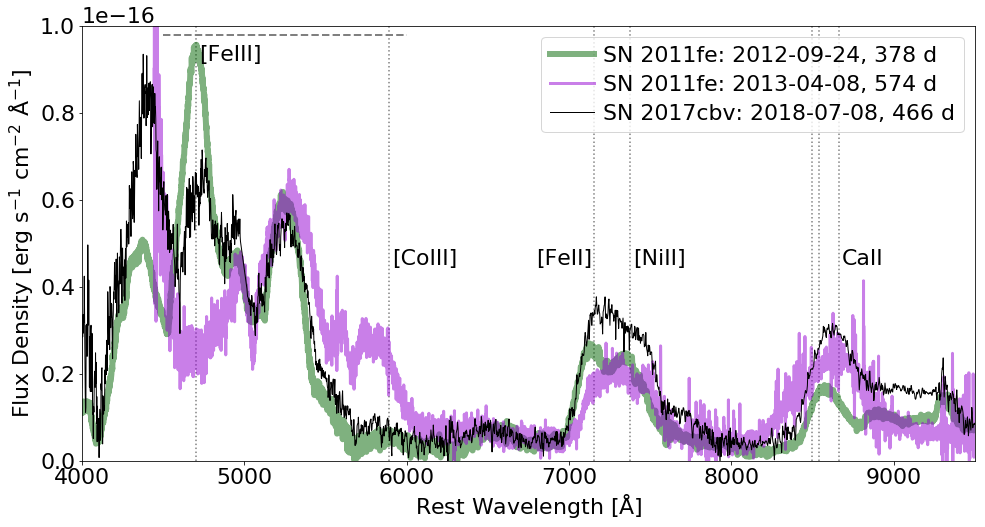}
\includegraphics[width=8cm]{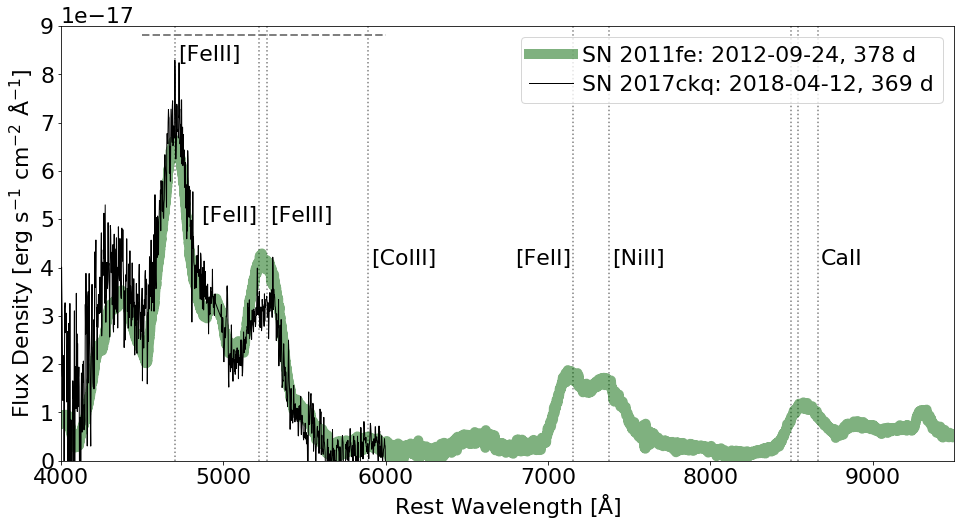}
\includegraphics[width=8cm]{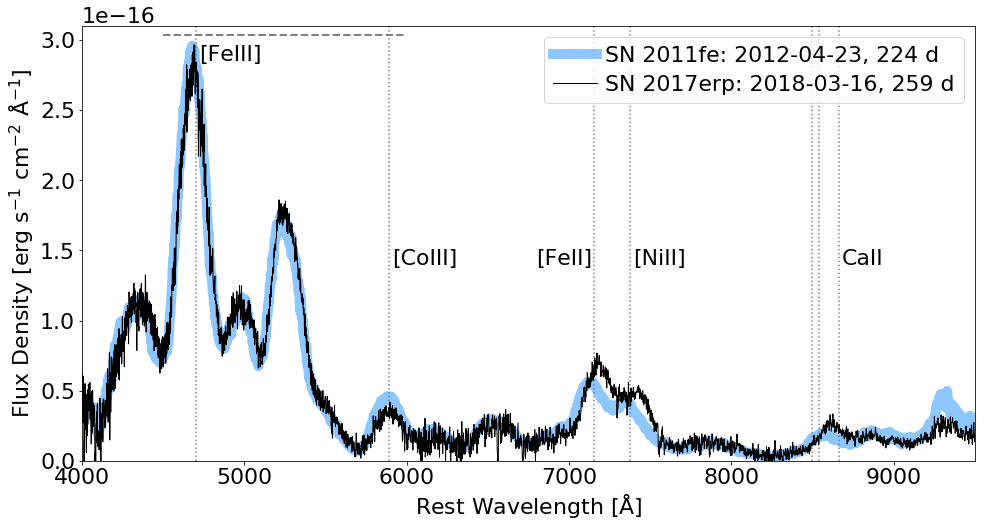}
\includegraphics[width=8cm]{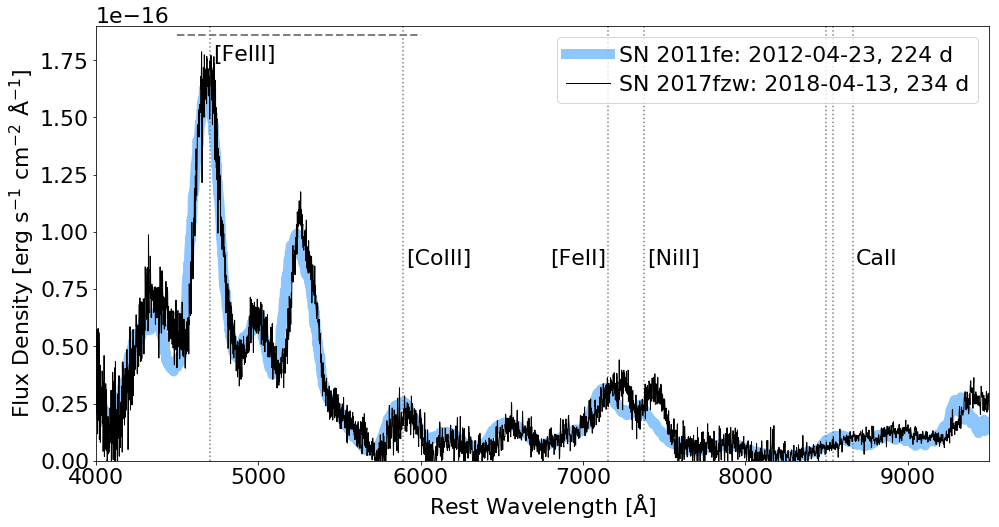}
\includegraphics[width=8cm]{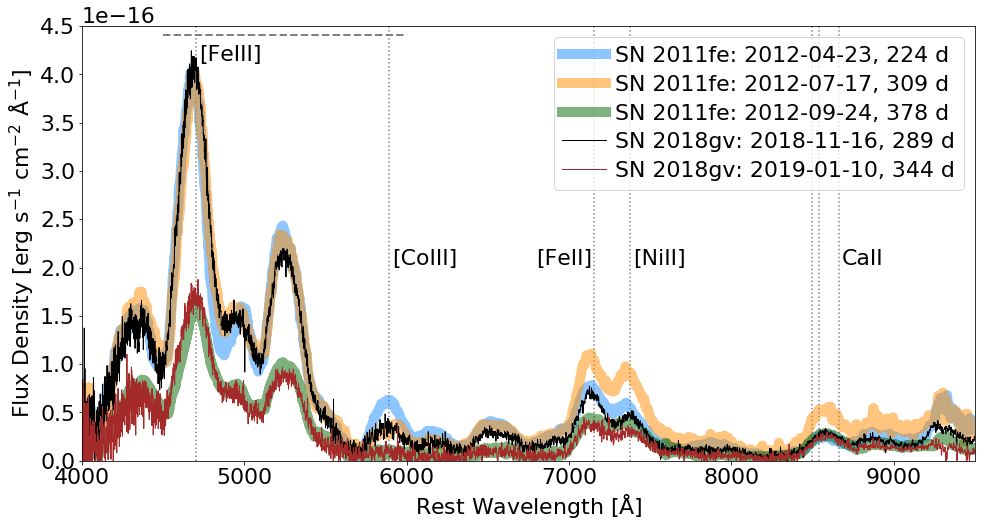}
\includegraphics[width=8cm]{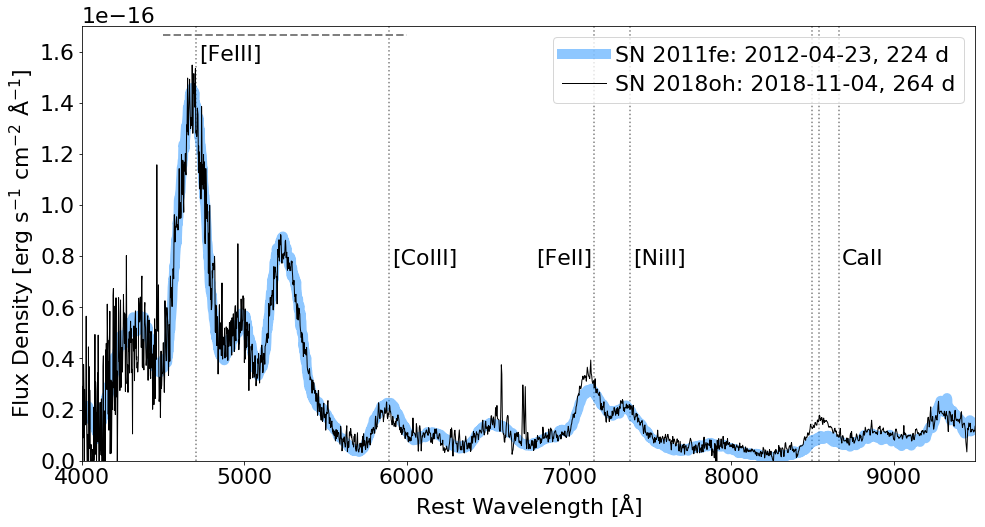}
\includegraphics[width=8cm]{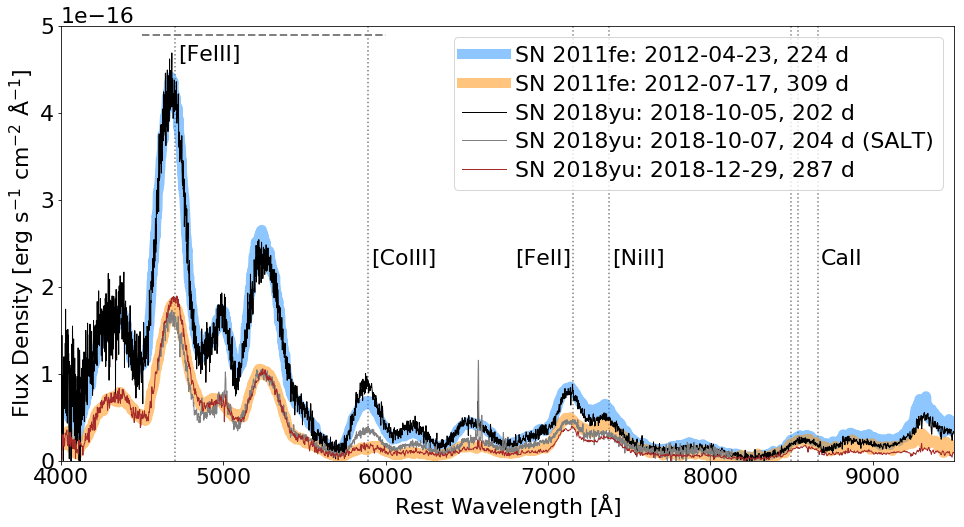}
\caption{Comparisons of the nebular-phase optical spectra in our Gemini sample (thin black lines; brown for second epochs; grey for SALT) with spectra of SN\,2011fe at a similar phase (thick semi-transparent colored lines). The 224 (blue) and 309 (orange) day spectra of SN\,2011fe are from \protect\cite{2015MNRAS.446.2073G}; the 378 (green) day from \protect\cite{2020MNRAS.492.4325S}; and the 574 (violet) day from \protect\cite{2015MNRAS.454.1948G}. Spectra of SN\,2011fe have been scaled to match the integrated flux of our spectra from $\sim$4500 to $\sim$6000~\AA\ (dashed horizontal line). The dominant species for some emission features of interest are labeled with dotted vertical lines.}
\label{fig:nebspec_wSN11fe}
\end{figure*}

These extracted 1D spectra were corrected for atmospheric extinction and the instrumental sensitivity function, sigma-clipped to reject artifact pixels from, e.g., cosmic rays, and then median-combined.
To create a single joined spectrum for each observational epoch, the R400 spectra were flux-matched to the B600 spectra and two were joined in the middle of the overlap region (${\sim}5500$~\AA).
The spectra were then flux-calibrated to the late-time photometry estimates listed in Table~\ref{tab:photobs}.
These estimates were interpolated or extrapolated from imaging obtained within days to weeks of our nebular spectra, but very precise calibration is not necessary for our analysis and mainly used for display in Figure~\ref{fig:nebspec_wSN11fe}.
After this approximate flux calibration, spectra were corrected for line-of-sight dust extinction using the {\tt astropy}-affiliated {\tt dust\_extinction} package\footnote{\url{https://github.com/karllark/dust_extinction}} and the $E(B-V)$ parameters listed in Table~\ref{tab:sne}, and corrected to rest-frame wavelengths using the redshifts listed in Table~\ref{tab:sne}.
Plots of all of our Gemini spectra are shown in Figure~\ref{fig:nebspec_wSN11fe}, with scaled spectra of the fiducial SN\,Ia 2011fe at similar phases for comparison.

\begin{table}
\centering
\begin{tabular}{|lccl|}
\hline
SN Name & Phase & Photometry & Reference \\
		 & [days] & [mag] & \\
\hline
2017cbv & 466 & V$\sim$20.8 & [2] \\
2017ckq & 369 & G$\sim$21.7 & Gaia17bhb [1] \\
2017erp & 259 & V$\sim$19.6 & [2] \\
2017fzw & 234 & G$\sim$20.1 & Gaia17cbe [1] \\
2018gv  & 289 & G$\sim$19.6 & Gaia18bat [1] \\
2018gv  & 344 & G$\sim$20.5 & Gaia18bat [1] \\
2018oh  & 264 & G$\sim$20.6 & Gaia18awj [1] \\
2018yu  & 202 & V$\sim$19.3 & [2] \\
2018yu  & 287 & V$\sim$20.3 & [2] \\
\hline
\end{tabular}
\caption{Late-time photometric estimates for the SNe\,Ia in our sample. Epochs that were near in time to our spectroscopic observations were interpolated (or extrapolated) to estimate photometric measurements. [1] Gaia Alerts, \protect\cite{2012gfss.conf...21W}. [2] Las Cumbres Observatory Global Supernova Project. \label{tab:photobs}}
\end{table}

\subsection{Optical Spectra from the South African Large Telescope}\label{ssec:nebobs_salt}

One nebular-phase spectrum of SN\,2018yu was obtained with the South African Large Telescope (SALT) Robert Stobie Spectrograph (RSS) using the 1.5\arcsec\ longslit and the PG0900 grating (resolution $\sim5.6$~\AA).
The grating's central wavelength was shifted between six exposures to cover the full optical range $4600$ to $8200$~\AA, using a filter to block second order contamination for red settings.
The six exposures had a total exposure time of 2386 seconds.
This spectrum was corrected for line-of-sight dust extinction (as described above), de-redshifted, and is co-plotted in Figure~\ref{fig:nebspec_wSN11fe}.
Without any flux scaling, this SALT spectrum for SN\,2018yu at $204$ days past peak appears to match the Gemini spectrum at $202$ days quite well.

\subsection{Near-Infrared Spectroscopy from Gemini Observatory}\label{ssec:nebobs_nir}

A nebular phase near-infrared (NIR) spectrum of SN\,2018yu at 282 days past maximum was obtained using the long slit mode of Flamingos-2 on Gemini South \citep{2008SPIE.7014E..0VE}.
The spectrum was observed using the JH filter and grism with a 3 pixel slit centered at 1.39 $\mu$m.
Using the standard ABBA nod-along-the-slit convention, we obtained a total of 104 exposures of 120 seconds each for SN\,2018yu over the course of two consecutive nights.
Standard star observations of a nearby A0V star were obtained immediately before and after the SN observations, as well as a flat and arc for wavelength calibrations.
NIR spectra were reduced using Gemini IRAF and the Flamingos-2 Data Reduction Cookbook\footnote{\url{https://gemini-iraf-flamingos-2-cookbook.readthedocs.io}}.
The spectra were also corrected for the atmosphere's telluric absorption using {\sc xtellcor} and the method presented in \citet{2003PASP..115..389V}.
The final NIR spectrum of SN\,2018yu, presented in Section \ref{ssec:disc_18yu}, is a combination of observations obtained over two consecutive nights that was combined after each night's observations were reduced, and telluric corrected separately using that night's standard star observations and associated calibrations.

\subsection{Nebular Emission Line Parameters}\label{ssec:nebobs_linepars}

We use two methods, ``direct measure" and ``Gaussian fit", to determine the velocity, FWHM, and integrated flux of the forbidden emission lines of [\ion{Fe}{III}]$\lambda$4701\AA, [\ion{Co}{III}]$\lambda$5891\AA, [\ion{Fe}{II}]$\lambda$7155\AA, and [\ion{Ni}{II}]$\lambda$7378\AA.
The latter two are the strongest [\ion{Fe}{II}] and [\ion{Ni}{II}] in the region, but they are blended with multiple weaker features of iron and nickel.
For the bulk of our analysis we use two-component fits to the iron and nickel feature in order to compare with previous work \citep[e.g.][]{2013MNRAS.430.1030S,2015MNRAS.454.3816C,2017MNRAS.472.3437G}.
In Section~\ref{ssec:ana_NiFe} we present a special analysis based on a multi-component fit to the iron and nickel feature. 

Both the ``direct measure" and ``Gaussian fit" methods begin with estimating the continuum flux by linearly interpolating between the local minimums on either side of the line and subtracting the continuum (sometimes called the pseudo-continuum), and smooth the continuum-subtracted flux with a Savitsky-Golay filter of window size $\sim$50~\AA\ using the {\tt scipy} package's {\tt signal.savgol\_filter} function.
This continuum-subtracted smoothed flux is used for both types of measurements.
In this work we re-measure the line parameters for the nebular-phase spectra from \citet{2017MNRAS.472.3437G} using the same codes as applied to the spectra presented here, for consistency; we found that any differences in the results are small.

{\bf Direct Measure: } We measure the full width of the feature at half the maximum smoothed flux (FWHM), using the pixel of peak flux as the maximum.
We use the midpoint of the FWHM as the line's central wavelength, and use this central wavelength to calculate the line's velocity with respect to the expected rest-frame emission line wavelength (quoted above).
We do not use the pixel of maximum flux to define the line's central wavelength because this is more susceptible to systematic error introduced random fluctuations and line asymmetry, even though we use the smoothed flux.
We numerically integrate the smoothed flux to derive the integrated line flux.

To estimate the uncertainty for the FWHM and velocity, we use a bootstrap method: shuffle the flux given the errors in order to generate a new continuum-subtracted flux array, then apply the same process of smoothing and measuring the line parameters.
For the flux error we use the difference between the original and the smoothed fluxes; we then randomly reassign the errors to each pixel \textit{with replacement} and add them to the smoothed flux in order to synthesize a new flux array.
This process of error-shuffling, flux synthesis, smoothing, and line parameter measurement is repeated $1000$ times, and the standard deviation in the FWHM and velocity measurements is taken as the error.
However, the error in the integrated flux is a quantity we measure directly (without bootstrapping), by simply integrating the absolute flux errors over the line region.
The directly measured line parameters and their errors are listed in Table~\ref{tab:linepars_direct} in Appendix~\ref{app:tables}

{\bf Gaussian Fit: } The {\tt scipy.optimize.curve\_fit} function is used to fit either a single- ([\ion{Fe}{III}] and [\ion{Co}{III}]) or double-component ([\ion{Fe}{II}]+[\ion{Ni}{II}]) Gaussian function to the continuum-subtracted smoothed flux of the feature.
Flux errors are estimated for each pixel as the difference between the original and the smoothed flux, and passed to {\tt curve\_fit}. 
We convert the Gaussian's standard deviation to the FWHM ($2.35\sigma$), use its peak wavelength to calculate the line's velocity, and numerically integrate the function to obtain the integrated line flux.

As in the direct method, to estimate an uncertainty on these three line parameters we use a bootstrap method of shuffling the flux given the errors and re-fitting $1000$ times, and using the standard deviation in the measurements is taken as the error in each line parameter's measurement.
The Gaussian-fit line parameters and their errors are listed in Table~\ref{tab:linepars_gaussian} in Appendix~\ref{app:tables}.
See also Section~\ref{ssec:ana_NiFe} for multi-component Gaussian fits for this feature, which incorporates additional iron and nickel lines in this region that have low relative intensities. 

\begin{figure}
\includegraphics[width=8cm]{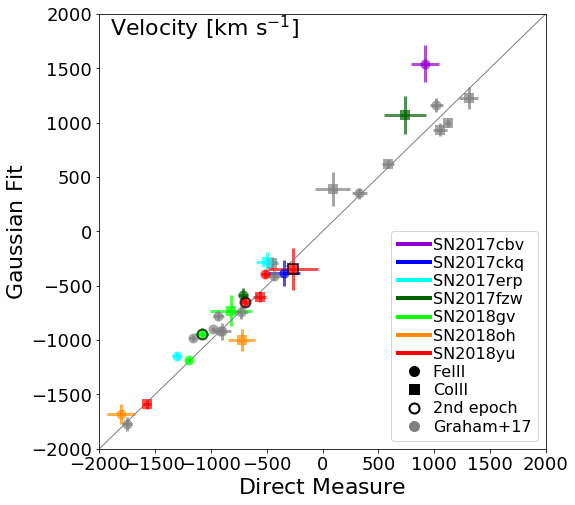}
\includegraphics[width=8cm]{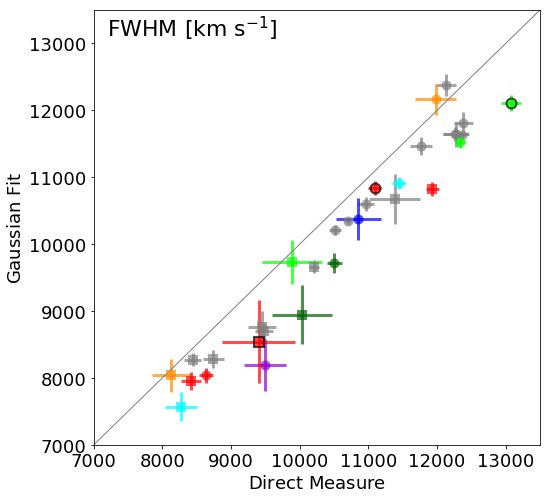}
\includegraphics[width=8cm]{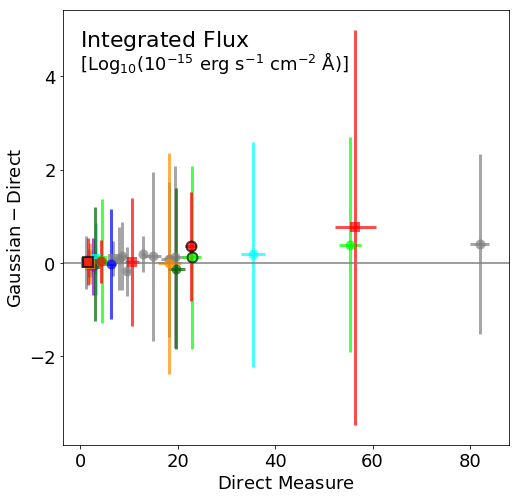}
\caption{Comparisons of the directly measured and Gaussian fit parameters for the [\ion{Fe}{III}] and [\ion{Co}{III}] lines in the nebular spectra of our seven SNe\,Ia. }
\label{fig:line_meas}
\end{figure}

\medskip
In Figure~\ref{fig:line_meas} we compare the ``direct measure" and ``Gaussian fit" results for the [\ion{Fe}{III}] and [\ion{Co}{III}] lines (circles and squares, respectively).
For SN\,2018gv and 2018yu, data points from the second epochs ($344$ and $287$ days, respectively) are denoted with a black outline.
The top and middle panels of Figure~\ref{fig:line_meas} show that, compared to the direct measures, the Gaussian fit results tends to produce larger and smaller values for the velocity and FWHM, respectively, due to the fact that the nebular-phase emission lines are not perfectly symmetric Gaussian features.
It is worth noting that for velocity, the direct measure and the Gaussian fit always agree on its {\it sign} (i.e., both are always blue- or red-shifted with respect to the rest frame), and that for the line FWHM the average difference between the direct measure and Gaussian fit results is relatively small, $-660\pm370$~$\rm km\ s^{-1}$.

In the bottom panel of Figure~\ref{fig:line_meas} we plot the difference between the Gaussian fit and the direct measure of the integrated flux as a function of the directly measured flux.
We do this instead of plotting the Gaussian fit versus the direct measure because the range of the integrated flux covers multiple orders of magnitude and the error bars are too small to be visible.
Most of the error in the {\it y-}axis error bar comes from the directly-measured error in integrated flux.
We can see that the Gaussian fit integrated flux tends to be larger than the direct measure (especially for higher-flux lines), but in most cases the two agree.
Overall, the general agreement between the direct measured and Gaussian fit results supports the use of Gaussian fits for the [\ion{Fe}{II}]+[\ion{Ni}{II}] feature, for which direct measure is not an option due to line blending.

\section{Sample Analysis}\label{sec:ana}

We present the derived properties of our sample of nebular-phase SNe\,Ia in context with other analyses in order to address some of the fundamental open questions about SNe\,Ia explosions.
In the following sections we discuss how the nebular spectra can reveal the evolving physical state of the nebular material (\S~\ref{ssec:ana_nebevol}); asymmetries in the explosion (\S~\ref{ssec:ana_asym}); a WD-WD collision model (\S~\ref{ssec:ana_coiii}); the nature of the explosion mechanism (\S~\ref{ssec:ana_NiFe}).
Additionally, for our sample of SNe\,Ia nebular spectra, Sand et al. (in prep.) shows that the upper limits on narrow H$\alpha$ emission are up to three orders of magnitude lower than what is expected from non-degenerate companion progenitor systems \citep{2005A&A...443..649M, 2018ApJ...852L...6B, 2020A&A...638A..80D}, which is consistent with other statistical analyses of nebular-phase SN\,Ia spectra that are revealing how rare a phenomenon this may be \citep[e.g.][]{2016MNRAS.457.3254M,2017MNRAS.472.3437G,2019ApJ...877L...4S,2020MNRAS.493.1044T}.

\subsection{Evolution of the Nebula}\label{ssec:ana_nebevol}

The complex of iron features from $\sim$4400 to $\sim$5500~\AA\ can be used to explore the time-changing physical state of the nebular material.
For example, \citet{2016MNRAS.462..649B} focused on this iron complex in late-time spectra for over two dozen SNe\,Ia (although not all with spectra $>$200 days, the phases explored in this work). 
They used their extensive data set and synthetic model spectra to study the red-ward evolution of the $\lambda$4700\AA\ emission line (primarily [\ion{Fe}{III}]), which had been noted by many past works but was not well understood.
\citet{2016MNRAS.462..649B} found that in addition to changing contributions from forbidden lines (e.g., decaying $^{56}$Co, evolving contributions of [\ion{Fe}{II}] and [\ion{Fe}{III}]), the opacity from permitted iron absorption lines played a role in this evolution.
Here we add our spectra to the ongoing study of this complex of iron features. 

\begin{figure}
\includegraphics[width=8cm]{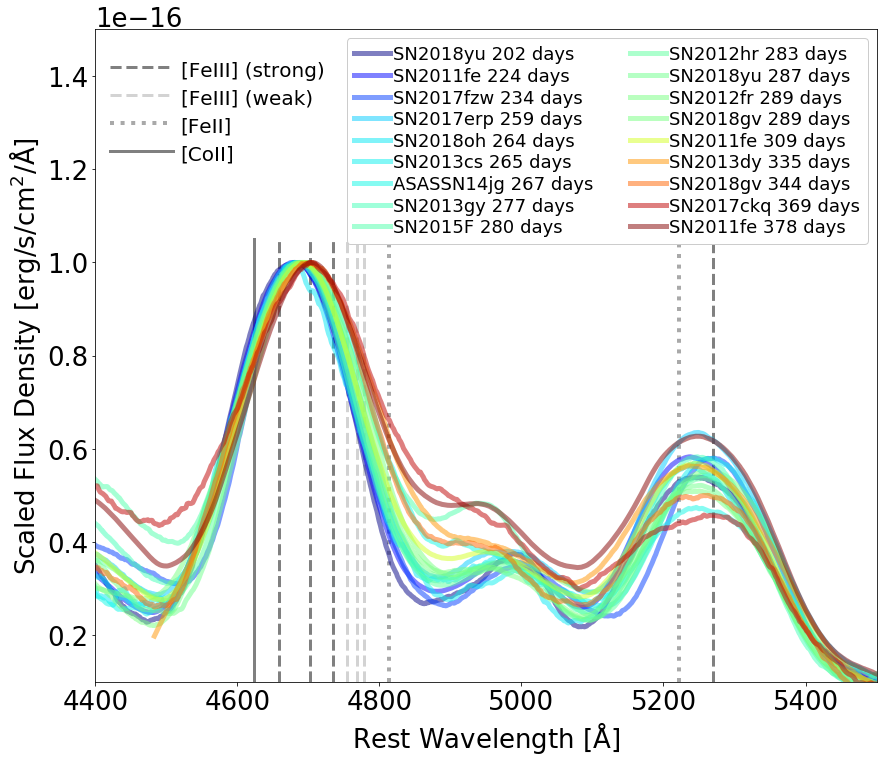}
\includegraphics[width=8cm]{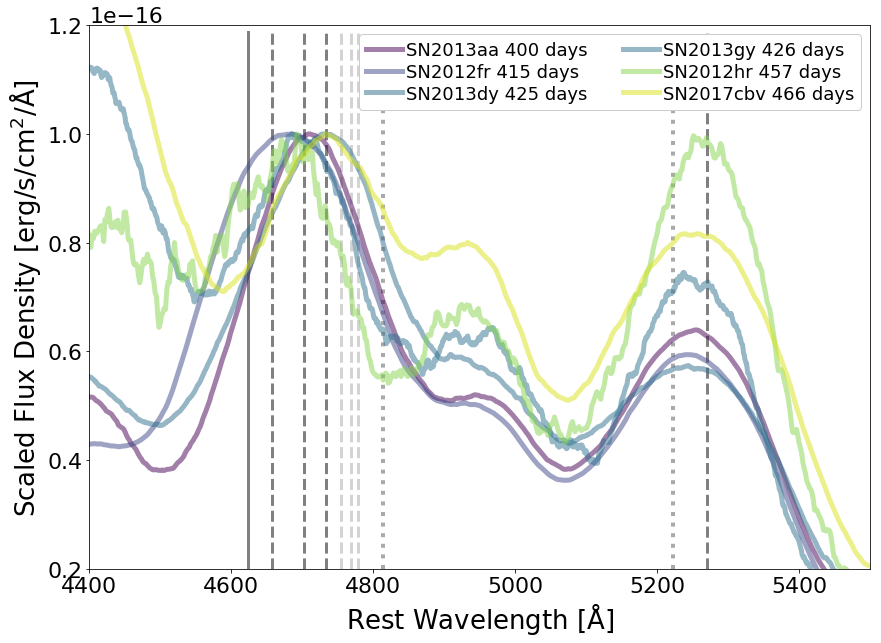}
\includegraphics[width=8cm]{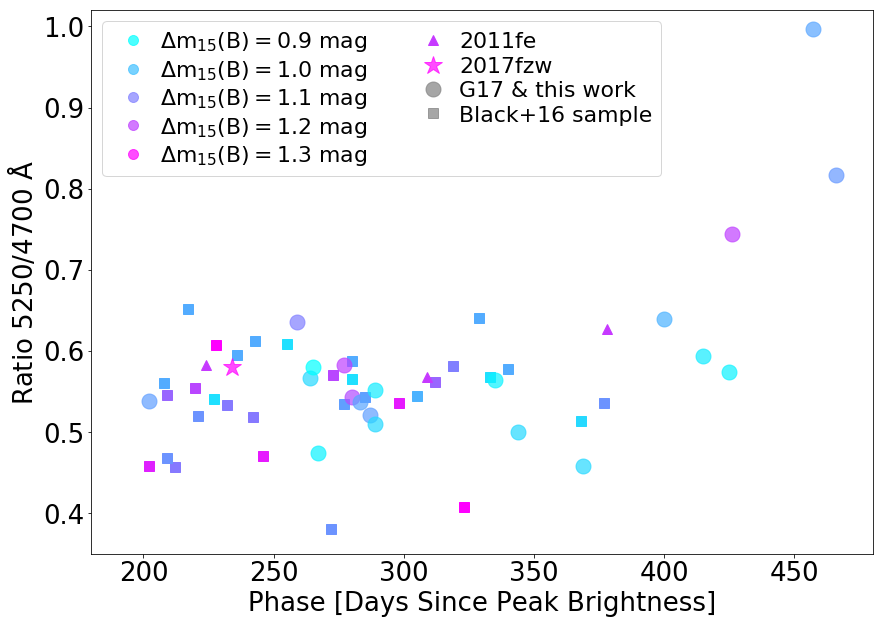}
\caption{\textit{Top:} Scaled nebular spectra from \citet{2017MNRAS.472.3437G} and this work (with spectra of SN\,2011fe for comparison), shown in a color-map based on the phase as described in the lenged, in the region of the strongest [\ion{Fe}{III}] and [\ion{Fe}{II}] lines (dashed and dotted vertical lines). \textit{Middle:} Similar to the top panel, but for spectra at phases $>$400 days past peak. \textit{Bottom:} The ratio of the peak flux of the 5250 to the 4700~\AA\ feature as a function of phase, with points colored by the SNe\,Ia light-curve decline rate, $\rm \Delta m_{15}(B)$, as described in the legend.}
\label{fig:nebevolspec}
\end{figure}

In the top two panels of Figure~\ref{fig:nebevolspec} we compare two dozen spectra of SNe\,Ia at a variety of nebular phases in order to demonstrate the time evolution of the $\lambda$4700\AA\ emission line.
These spectra have been smoothed with a Savitsky-Golay filter of 100~\AA, have had a pseudo-continuum at $\lambda \approx 5600$~\AA\ subtracted, and have been flux scaled such that the $\lambda$4700\AA\ line peak flux is $10^{-16}$ $\rm erg\ s^{-1}\ cm^{-2}\ \AA$. 
The spectra are colored by the phase of their observation (as in the panels' legends), so that the red-ward evolution of the $\lambda$4700\AA\ line stands out clearly.
Vertical dashed, dotted, and solid lines in Figure~\ref{fig:nebevolspec} show the locations of [\ion{Fe}{III}], [\ion{Fe}{II}], and [\ion{Co}{II}], respectively.

\begin{figure}
\includegraphics[width=8cm]{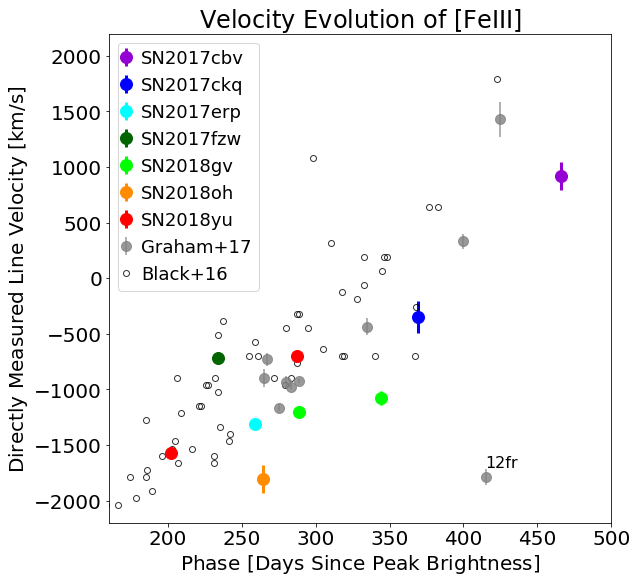}
\includegraphics[width=8cm]{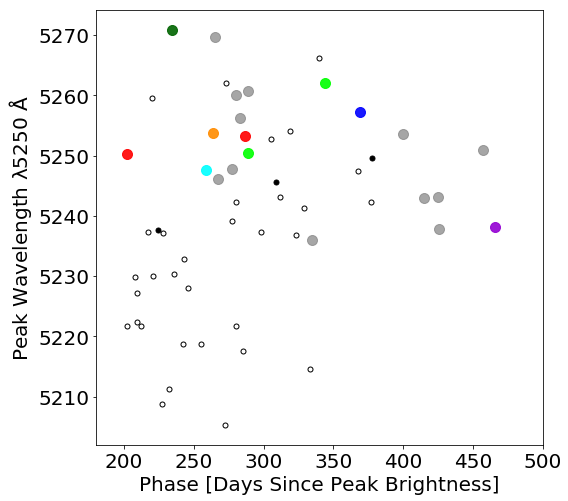}
\caption{\textit{Top:} The directly measured velocity of the [\ion{Fe}{III}]~$\lambda$4700\AA\ emission line as a function of phase in days since peak brightness for our Gemini nebular-phase spectra (colored symbols). Grey filled symbols represent data from \citet{2017MNRAS.472.3437G} and small open circles data from  \citet{2016MNRAS.462..649B}, for comparison.
\textit{Bottom}: The directly measured peak wavelength of the $\lambda$5250 \AA\ iron feature as a function of phase in days since peak brightness for the same data set as above.}
\label{fig:nebevol}
\end{figure}

In the top panel of Figure~\ref{fig:nebevol} we show the evolution of the nebular-phase $\lambda$4700\AA\ line velocity as a function of phase for the same spectra in Figure~\ref{fig:nebevolspec}, with the sample of \citet{2016MNRAS.462..649B} also plotted for comparison.
Thanks to the sensitivity of the 8m Gemini telescopes, this work is stretching the sample of nebular spectra with measurable $\lambda$4700~\AA\ lines past $450$ days.
Despite adding seven new SNe\,Ia to this plot, SN\,2012fr remains an unique outlier in this trend, as presented in  \citet{2017MNRAS.472.3437G}.

For the $\lambda$4700\AA\ complex, the decline of emission from the radioactive decay of [\ion{Co}{II}] and, potentially, an increasing amount of [\ion{Fe}{II}] in the cooling nebula might at first be suspected as the culprit for the red-ward evolution of the peak (i.e., because they contribute to the blue and red sides of the feature, respectively).
While they both likely do contribute, an increasing dominance of [\ion{Fe}{II}] would also be expected to cause a \emph{blue}-ward shift in the peak of the $\lambda$5250\AA\ line (i.e., towards the dotted line of [\ion{Fe}{II}] at $\lambda$5200~\AA).
In fact, a tentative trend in the \emph{opposite} direction is seen in our plot of the $\lambda$5250\AA\ line's peak wavelength as a function of phase in the bottom panel of Figure~\ref{fig:nebevol}. 
This red-ward shift was also shown for the nebular iron features at $\lambda{\sim}$5000~\AA\ by \citet{2016MNRAS.462..649B}.
The fact that we also see it for the $\lambda$5250 feature is consistent with the measurements of \citet{2016MNRAS.462..649B}, and with their conclusions that other factors, primarily opacity from the absorption by permitted lines, are the cause of the red-ward evolution exhibited by the nebular-phase iron emission features.

Aside from red-ward shifts in these iron lines' velocities (and broadening of the FWHM, as noted in \citealt{2016MNRAS.462..649B}), we were curious as to whether there was a link between the relative fluxes of the $\lambda$4700 and $\lambda$5250~\AA\ lines, and the properties of SN\,Ia early-time light curves.
For example, if the $\lambda$5250~\AA\ line was increasingly dominated by [\ion{Fe}{II}] as the ionization state of the nebula evolved, perhaps the flux ratio of the $\lambda$5250 to the $\lambda$4700~\AA\ line would be larger at a given phase for SNe\,Ia that synthesized less $^{56}$Ni.
In other words, perhaps the the transition to [\ion{Fe}{II}] occurs earlier for less energetic, faster-cooling SN\,Ia nebulae.

To investigate this, in the bottom panel of Figure~\ref{fig:nebevolspec} we plot the flux ratio of the $\lambda$5250 to $\lambda$4700~\AA\ lines as a function of phase, and color the symbols by the decline-rate parameter $\rm \Delta m_{15}(B)$, which is correlated with the peak $B$-band magnitude which is known to be a proxy for $^{56}$Ni mass \citep[e.g.,][]{1982ApJ...253..785A,1993ApJ...413L.105P}. 
In this color scheme, SNe\,Ia that decline slowly are cyan (i.e., synthesized more $^{56}$Ni), and SNe\,Ia that decline more rapidly are magenta (i.e., synthesized less $^{56}$Ni).
In addition to the nebular-phase spectra presented in this work and in \citet{2017MNRAS.472.3437G}, this plot includes the spectra analyzed by \citet{2016MNRAS.462..649B} as smaller square points\footnote{
A list of the SNe\,Ia from the \citet{2016MNRAS.462..649B} sample and references to their spectra, which were obtained from the WISeREP database \citep{2012PASP..124..668Y}:
1990N   \citep{1998AJ....115.1096G};
1994ae  \citep{2012AJ....143..126B};
1995D   \citep{2012AJ....143..126B};
1996X   \citep{2001MNRAS.321..254S};
1998aq  \citep{2012AJ....143..126B};
1998bu  \citep{2012AJ....143..126B,2012MNRAS.425.1789S,2001ApJ...549L.215C};
2002dj  \citep{2008MNRAS.388..971P};
2003du  \citep{2007A&A...469..645S};
2003hv  \citep{2009A&A...505..265L};
2004eo  \citep{2007MNRAS.377.1531P};
2005cf  \citep{2009ApJ...697..380W};
2007le  \citep{2012MNRAS.425.1789S};
2008Q   \citep{2012MNRAS.425.1789S}; and
2011by  \citep{2013MNRAS.430.1030S}.
}.

The most obvious trend in the bottom panel of Figure~\ref{fig:nebevolspec} is the increasing $\lambda$5250/$\lambda$4700 flux ratio at phases later than 350 days.
This could be due to the transition from the emission being dominated by [\ion{Fe}{III}] to [\ion{Fe}{II}] as the ionization state of the nebula evolves and the evolving temperature of the nebula (the [\ion{Fe}{III}] line strengths are temperature dependent). 
For example, the modeling work of \citet{2015ApJ...814L...2F} shows that the nebula's temperature starts a dramatic drop at about $\sim$350 days (their Figure 1) as the dominant cooling mechanism switches from optical to NIR emission lines, known as the SN\,Ia Infrared Catastrophe (IRC).
The models of \citet{2015ApJ...814L...2F} predict a plateau in the NIR light curves of SNe\,Ia, which was recently confirmed by observations presented by \citet{2020NatAs...4..188G}; see also the evidence presented for the IRC based on psuedo-bolometric light-curve modeling for SN\,Ia 2011fe by \citet{2017MNRAS.468.3798D}.
Additionally, we note that this increase in the $\lambda$5250/$\lambda$4700 flux ratio at $\sim$350 days is similar in its timing to the increase in NIR to optical flux ratio shown in \citet[][their Figure 9]{2018MNRAS.477.3567M}.
With NIR spectra for SN\,2013aa at 360 and 425 days, \citet[][their Figure 12]{2018MNRAS.477.3567M} also show that this increase in the NIR/optical flux ratio is due to the [\ion{Fe}{II}] emission lines at $\lambda$12000 and 16000~\AA\ remaining constant while the optical emission declines.
As also suggested by \citet{2018MNRAS.477.3567M}, these NIR observations support the hypothesis that the increasing $\lambda$5250/$\lambda$4700 flux ratio is due to the cooling nebula's evolution from doubly- to singly-ionized iron.

In making the bottom panel of Figure~\ref{fig:nebevolspec} we were looking for a correlation wherein faster-declining SNe\,Ia (magenta points) have a higher $\lambda$5250/$\lambda$4700 flux ratio at earlier phases compared to slower-declining SNe\,Ia (cyan points).
This trend would manifest in this plot as a magenta-to-cyan gradient from high-to-low flux ratio, and the gradient might not be strictly horizontal but might appear on an angle from upper-left to lower-right, due to a correlation between $\lambda$5250/$\lambda$4700 flux ratio and phase.
Such a gradient is only very tentatively seen, and perhaps only emerges at $>$350 days, but there is also a bias here in that faster-declining (magenta) SNe\,Ia are not as frequently observable so late into the nebular phase.
Thus we conclude that the data at hand cannot confirm or reject our hypothesis of a correlation between $^{56}$Ni mass and the $\lambda$5250/$\lambda$4700 flux ratio as an indicator of the ionization state or temperature of the nebula.
A larger number of $>$350 day spectra in future samples might help to clarify this.

\subsection{Explosion Asymmetry}\label{ssec:ana_asym}

\citet{2010ApJ...708.1703M} was the first to interpret the correlation between the photospheric velocity gradient\footnote{Recall from Section~\ref{sec:sne} that the photospheric velocity gradient is measured from the rate of decrease of the velocity of the photospheric \ion{Si}{II} $\lambda$6355~\AA\ absorption feature during the two weeks after light curve peak brightness.} and the velocity of the nebular-phase [\ion{Fe}{II}]+[\ion{Ni}{II}]~$\lambda$7200~\AA\ feature as a signature of explosion asymmetry.
Their proposed physical model is an off-center explosion which, when aligned \emph{away} the observer along their line-of-sight, results in \emph{red}-shifted [\ion{Fe}{II}] and [\ion{Ni}{II}] emission lines because the bulk of the nucleosynthetic material is on the \emph{far} side of the nebula.
This scenario causes the SN\,Ia ejecta's outer layers of the \textit{near} side to be of lower density compared to the far side, and since the photosphere can recede more rapidly into this lower density material, a larger photospheric velocity gradient is observed at early times.

\begin{figure}
\includegraphics[width=8cm]{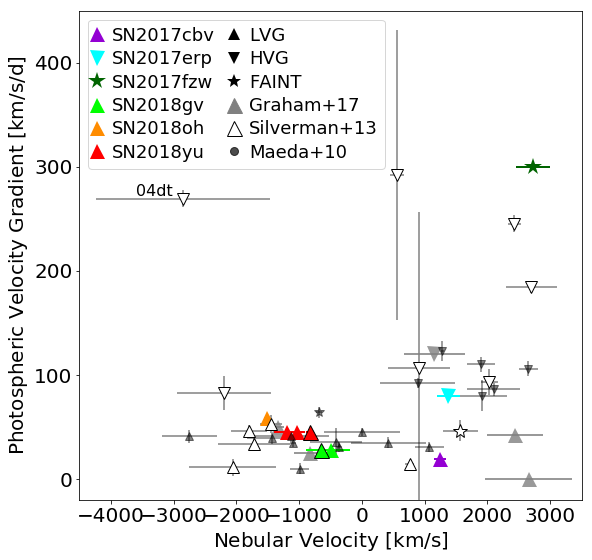}
\includegraphics[width=8.2cm]{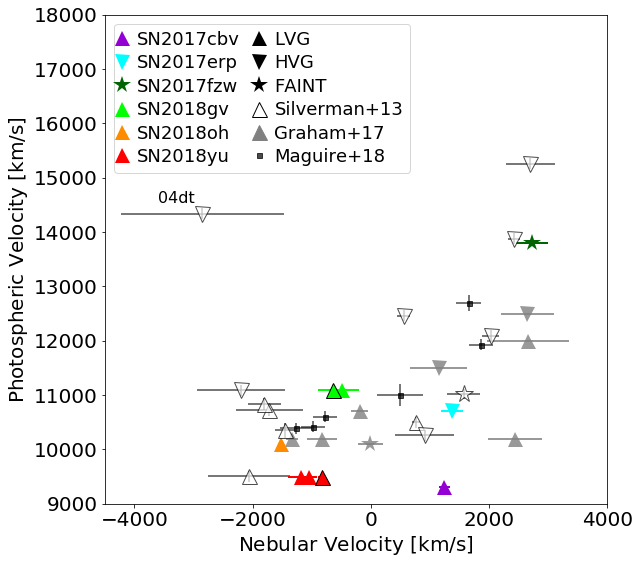}
\caption{The photospheric velocity gradient of \ion{Si}{II} $\lambda$6355\AA\ during the two weeks after peak brightness (\textit{top}), and the photospheric velocity of \ion{Si}{II}~$\lambda$6355\AA\ near light-curve peak brightness (\textit{bottom}), versus the nebular line velocity as measured using Gaussian fits to the [\ion{Fe}{II}] and [\ion{Ni}{II}] blended feature in our nebular-phase spectra from Gemini Observatory (colored symbols; a black outline indicates data from a second epoch).}
\label{fig:asym}
\end{figure}

In the top panel of Figure~\ref{fig:asym} we plot the photospheric velocity gradient \textit{vs}. the nebular velocity from the original data of \citet{2010ApJ...708.1703M} as small symbols; from similar measurements presented in \citet{2013MNRAS.430.1030S} and  \citet{2017MNRAS.472.3437G} as larger symbols; and from our sample of SNe\,Ia as colored symbols.
It is only possible to measure the velocity gradient for SNe\,Ia with multiple photospheric-phase spectra, but it has been shown that the velocity gradient is correlated with the photospheric velocity (as measured from the \ion{Si}{II} $\lambda$6355~\AA\ absorption feature in a single spectroscopic observation near peak brightness; \citealt{2005ApJ...623.1011B,2009ApJ...699L.139W,2013Sci...340..170W,2013MNRAS.430.1030S}).
In the bottom panel we plot the photospheric velocity versus the nebular velocity, including SNe\,Ia for which only a single spectrum was obtained.
Measurements from \citet{2018MNRAS.477.3567M} are shown as small symbols for comparison.
We note that \citet{2018MNRAS.477.3567M} measures the nebular velocity from the [\ion{Fe}{II}]~$\lambda$7155~\AA\ feature only, whereas we (and others) use an average of the [\ion{Fe}{II}] and [\ion{Ni}{II}] line velocities, and that the difference is typically only a few hundred~$\rm km\ s^{-1}$.

It is clear from both panels of Figure \ref{fig:asym} that SNe\,Ia with large photospheric velocity gradients are almost exclusively associated with red-shifted nebular-phase emission lines.
These plots reinforce the previously observed correlation, and agree with the asymmetry models of \citet{2010ApJ...708.1703M}. 

\subsection{The WD-WD Collision Scenario}\label{ssec:ana_coiii}

\begin{figure*}
\includegraphics[width=15.5 cm]{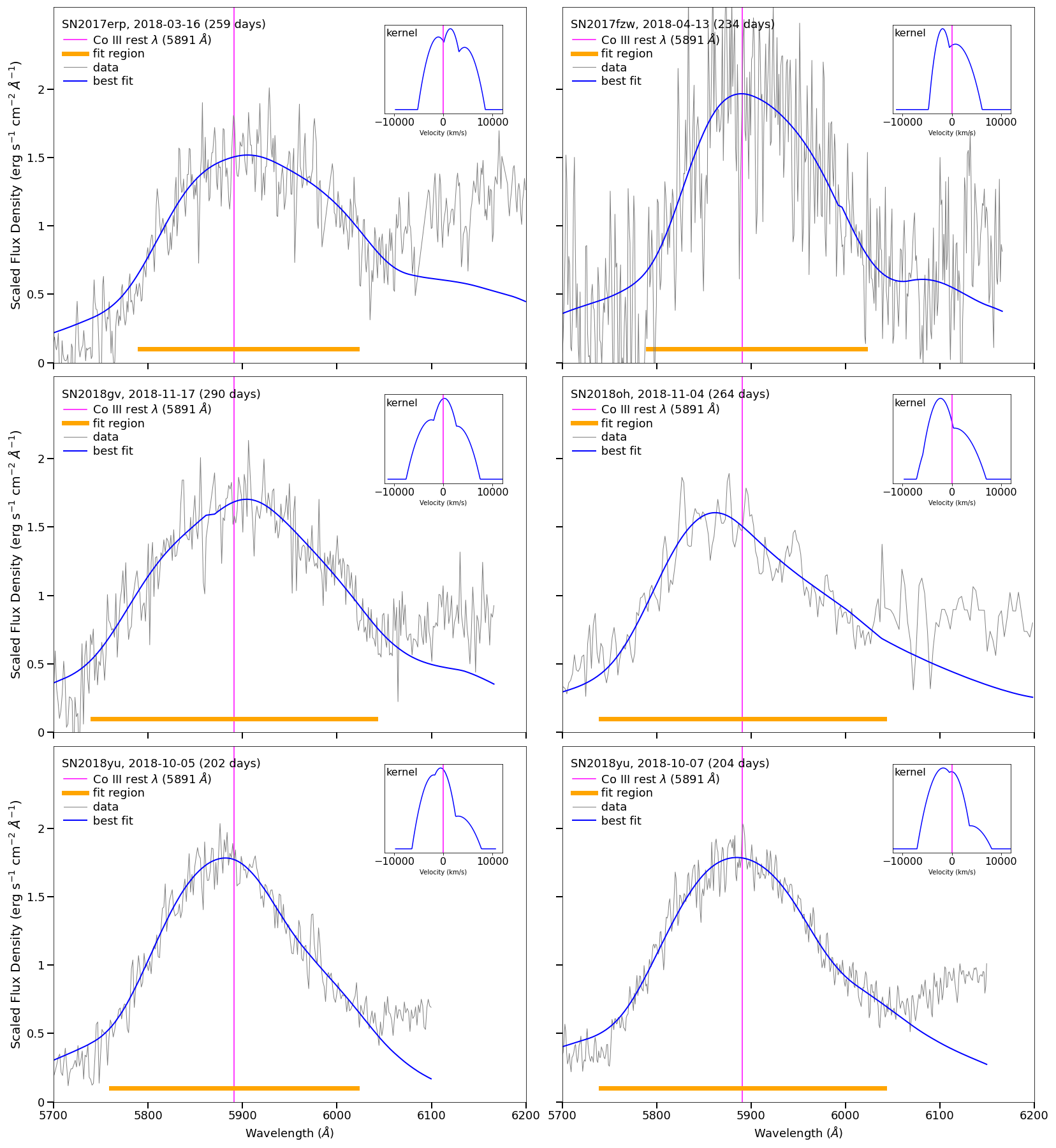}
\caption{The convolution fits and velocity kernels (inset) for the [\ion{Co}{III}] lines of each of our spectra with a discernable cobalt feature (i.e., that all the cobalt had not yet decayed). To identify a SN as a bimodal candidate, the two quadratics in the kernel must be separated by more than the sum of their standard devations. All of the kernels shown above are too blended to meet that criteria. }
\label{fig:CoKernels}
\end{figure*}

The head-on collision of two white dwarf stars, potentially as a result of orbital evolution driven by a distant tertiary in the system, has been proposed as a potential progenitor scenario for SNe\,Ia \citep[e.g., ][]{2011ApJ...741...82T,2012arXiv1211.4584K}.
One predicted observable signature of a WD-WD collision scenario is bimodal emission lines in nebular phase spectra due to the ejecta from the two progenitors moving at high speeds in different directions \citep{2013ApJ...778L..37K}.
\citet{2015MNRAS.454L..61D} identify the [\ion{Co}{III}]~$\lambda$5900~\AA\ nebular feature as the best indicator of this phenomenon, because it is both a decay product of $^{56}$Ni and does not suffer from significant contamination from other species.
Depending on the viewing angle, a WD-WD collision could cause this emission line profile to appear clearly double-peaked, broadened, or single-peaked.
Although none of the [\ion{Co}{III}]~$\lambda$5900~\AA\ emission features in our sample exhibited clear double peaks upon visual inspection (Figure~\ref{fig:CoKernels}), since the bimodality might not be obvious to the eye we apply a more stringent statistical analysis to evaluate the likelihood of double-peaked [\ion{Co}{III}] emission features in our spectra.

To test our sample for [\ion{Co}{III}]~$\lambda$5900~\AA\ lines with two components we use a convolution fitting method similar to those used by \citet{2015MNRAS.454L..61D} and \citet{2020MNRAS.492.3553V}.
We use the {\tt emcee} sampler \citep{2013PASP..125..306F} to find the two-component kernel which, when convolved with the nebular spectrum of SN\,1999by, provides a best-fit (minimum log likelihood) to the [\ion{Co}{III}]~$\lambda$5900~\AA\ feature in each of our spectra.
The best fit results (blue lines) are shown for six spectra (grey lines) for five of our SNe\,Ia\footnote{A bimodal fit was also done for the 167 day spectrum of SN\,2017fzw (Section~\ref{ssec:disc_17fzw}), but because it yielded the same result as the 234 days spectrum (top right), it was omitted from Figure~\ref{fig:CoKernels}.
SN\,2017cbv is not included in Figure~\ref{fig:CoKernels} because the [\ion{Co}{III}] feature was very weak due to the late phase, at 466 days past peak brightness.
SN\,2017ckq is not included in Figure~\ref{fig:CoKernels} because we did not obtain a red-side spectrum for that object.
} in the panels of Figure~\ref{fig:CoKernels}, where the orange bar denotes the fit region which begins and ends at the `edges' of the [\ion{Co}{III}] features, and the inset panels to show the best fit bimodal kernels.

Although we find that all features \emph{can} be fit with a bimodal kernel, this does not mean that they are likely to physically \emph{be} bimodal.
We apply the separation criteria described by \citet[][their Section 3]{2020MNRAS.492.3553V}: that the separation of the two components must be greater than the sum of their standard deviations.
Ultimately we find that none of the best-fit kernels meet the criteria to be declared a likely indication of bimodal emission, and we are therefore unable to identify any of our objects as WD-WD collisional candidates by this metric.
Our null result is consistent with the theoretical triple-system population synthesis models and evolution simulations of \citet{2018A&A...610A..22T}, who report that the rate of SNe\,Ia in triple systems is only $\sim$0.1\% the total rate of SNe\,Ia from binary systems.
However, as \citet{2020MNRAS.492.3553V} points out, bimodal nebular-phase emission lines can be the result of asymmetric detonation, which could be much more common \citep[e.g.,][]{2007ApJ...661..995G}.

\subsection{The Ni/Fe Ratio}\label{ssec:ana_NiFe}

Different explosion models for SNe\,Ia, such as the double-detonation (DDT) model \citep[e.g.,][]{2013MNRAS.429.1156S} or the sub-Chandrasekhar mass (sub-$\rm M_{Ch}$) models \citep{2010ApJ...714L..52S,2018ApJ...854...52S}, predict different nebular-phase Ni/Fe ratios with only a small dependence on phase, as $^{56}$Co continues to decay and add to the amount of $^{56}$Fe \citep[e.g.][their Figure 10]{2018MNRAS.477.3567M}.
The metallicity of the white dwarf progenitor star also has an impact on the nebular-phase Ni/Fe ratio: higher metallicity progenitors have more neutrons available to synthesize stable products \citep[e.g.][]{2003ApJ...590L..83T}.
Practically all of the nickel remaining in the nebular phase was formed stably, as $^{56}$Ni has a half-life of 6 days and the relatively smaller amount of $^{57}$Ni a half-life of just 35.6 hours (observations have shown that the mass ratio of $^{57}$Ni to $^{56}$Ni is $<$5\%; \citealt{2016ApJ...819...31G,2018A&A...620A.200F}).

\begin{figure*}
\centering
\includegraphics[width=4cm]{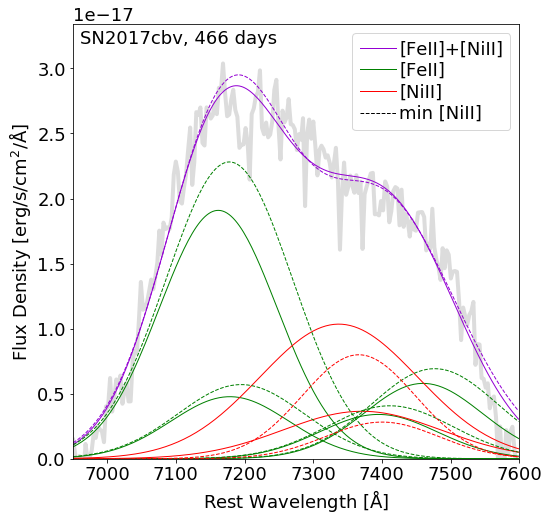}
\includegraphics[width=4cm]{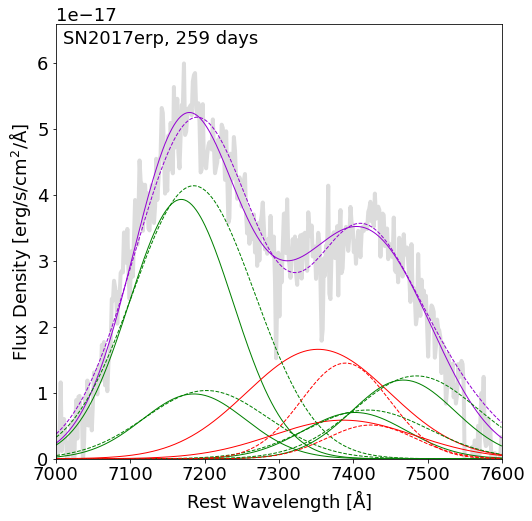}
\includegraphics[width=4cm]{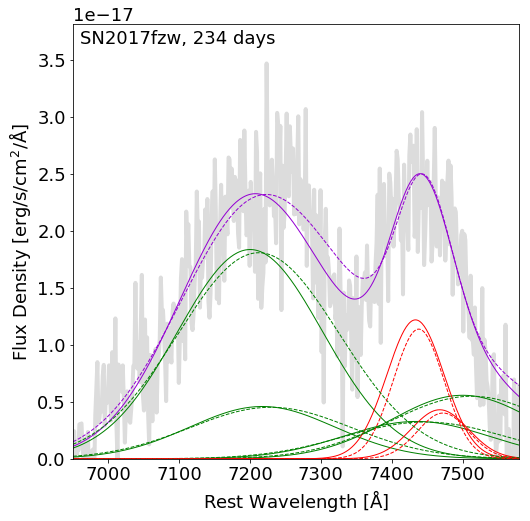}
\includegraphics[width=4cm]{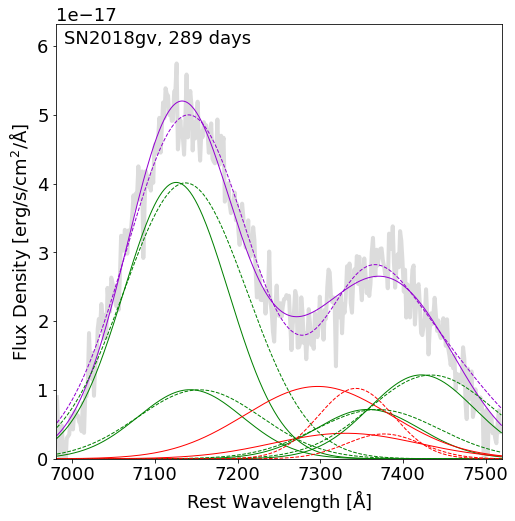}
\includegraphics[width=4cm]{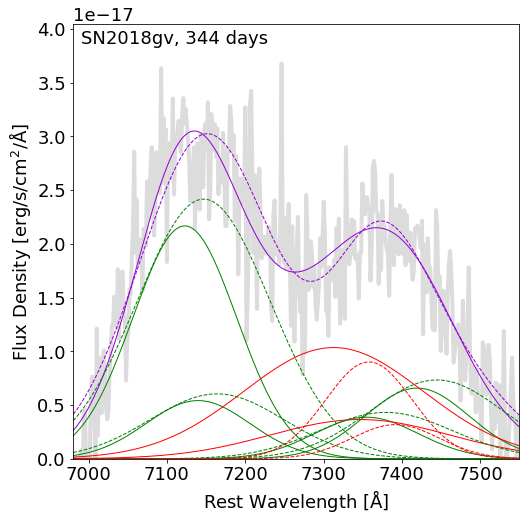}
\includegraphics[width=4cm]{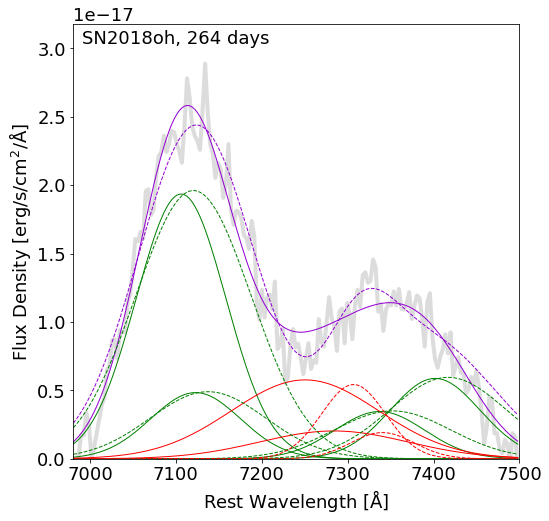}
\includegraphics[width=4cm]{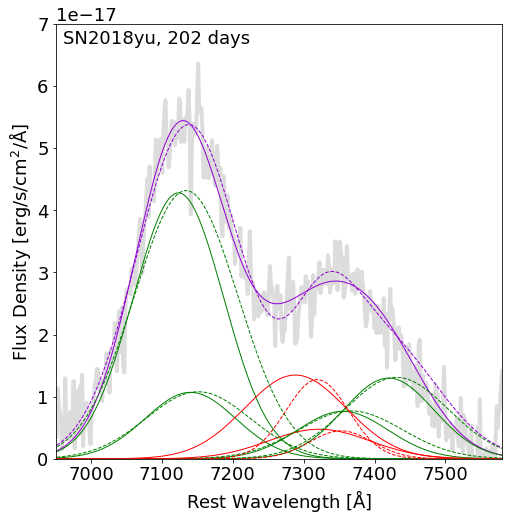}
\includegraphics[width=4cm]{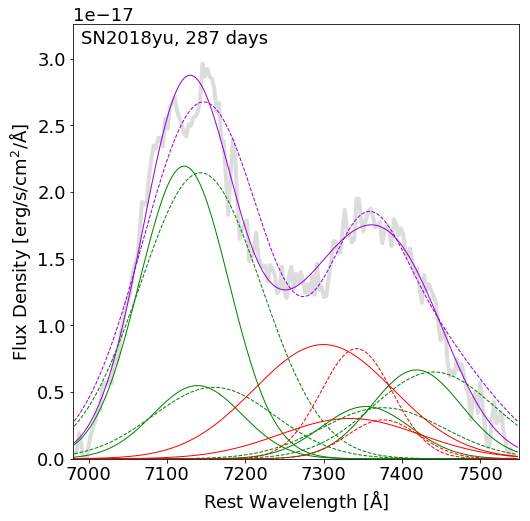}
\caption{Multi-component Gaussian fits to the blended [\ion{Fe}{II}]+[\ion{Ni}{II}] emission feature in our nebular-phase spectra.
The thick grey line is the pseudo-continuum subtracted flux density at rest wavelengths; the thick green and thin red lines show the best-fit iron and nickel features, respectively; and the purple lines show the combined best fit.
The results from a simultaneous fit of iron and nickel are shown with solid lines, and the results from a two-stage fit of first iron, and then a \emph{minimum} amount of nickel to the remaining flux, are shown with dashed lines.}
\label{fig:FeNifits}
\end{figure*}

In order to estimate the Ni/Fe ratio from our sample of nebular-phase SNe\,Ia spectra, we follow the example of \citet{2018MNRAS.477.3567M} and fit the four [\ion{Fe}{II}] and two [\ion{Ni}{II}] emission features in this region: [\ion{Fe}{II}]$\lambda$ 7155, 7172, 7388, and 7453~\AA, and [\ion{Ni}{II}]$\lambda$ 7378 and 7412~\AA.
The [\ion{Fe}{II}]$\lambda$ 7172, 7388, and 7453~\AA\ features have relative intensities of 0.24, 0.19, and 0.31 compared to the 7155~\AA\ feature, and the [\ion{Ni}{II}]$\lambda$ 7453~\AA\ feature has a relative intensity of 0.31 compared to the 7378~\AA\ feature \citep{2015MNRAS.448.2482J}.
As in Section~\ref{ssec:nebobs_linepars} we first fit and subtract a linear pseudo-continuum and then use the {\tt scipy.optimize.curve\_fit} function with six Gaussian parameters: the width, velocity, and peak of the [\ion{Ni}{II}] and [\ion{Fe}{II}] lines.
The only boundary we place is on the width of the nickel lines, which is limited to $\lesssim13000$~$\rm km\ s^{-1}$ to avoid extremely broad and shallow Gaussian features being fit (especially for the older or lower-resolution spectra).
In Figure~\ref{fig:FeNifits} we show the results of these multi-component Gaussian fits for all the spectra presented in this work\footnote{Except for the SALT spectrum of SN\,2018yu -- which we did fit, but do not show because its phase is so similar to our Gemini Observatory spectrum of SN\,2018yu -- and SN\,2017ckq, for which we did not obtain a red-side spectrum.}. 
We also perform these fits for the spectra from \citet{2017MNRAS.472.3437G}, but do not include them in Figure~\ref{fig:FeNifits}.

As mentioned above, in some cases the best fits are for extremely broad nickel lines, especially for the older and lower-resolution spectra.
Thus we adopt a two-stage fit method to estimate the \emph{minimum} amount of nickel, and obtain a \emph{lower limit} on the Ni/Fe ratio.
First, we fit iron to the blue half of the feature, and then allow nickel to make up the rest of the flux in the blended feature.
The results of these two-stage \emph{minimum-Ni} fits are shown as dashed lines in Figure~\ref{fig:FeNifits}.
It is immediately clear to the eye that in many cases fits from these two methods are generally similar (purple lines), yet can have quite different relative contributions from iron and nickel (green and red lines).

\begin{figure*}
\centering
\includegraphics[width=8cm]{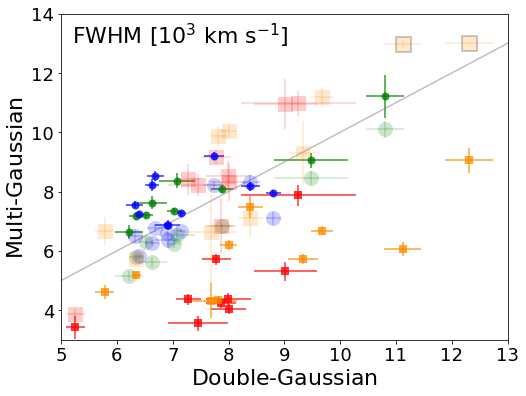}
\includegraphics[width=8cm]{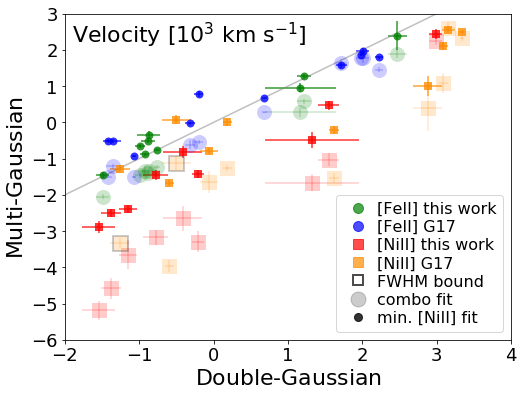}
\includegraphics[width=8cm]{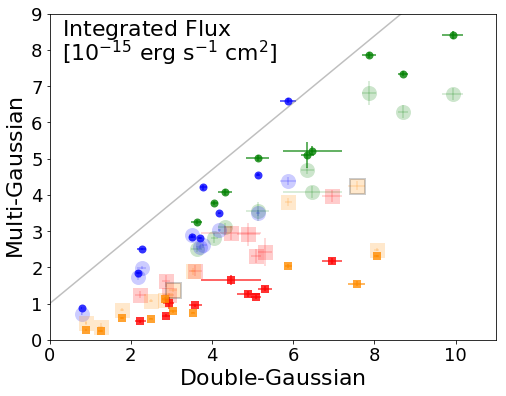}
\includegraphics[width=8cm]{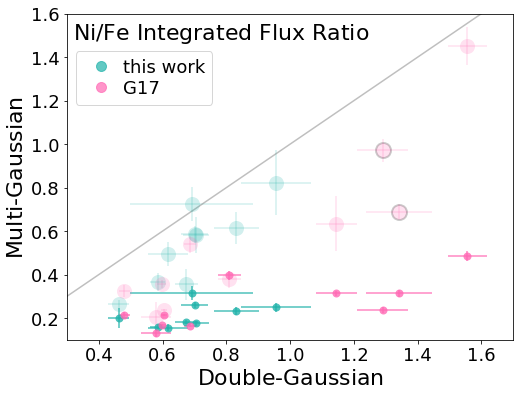}
\caption{Panels show comparisons of the emission line parameters FWHM (\textit{top left}), velocity (\textit{top right}), and integrated flux (\textit{bottom left}) measured from multi- and double-Gaussian fits to the [\ion{Fe}{II}]~$\lambda$~7155~\AA\ and [\ion{Ni}{II}]~$\lambda$~7378~\AA\ features for spectra presented in this work (green/red points) and spectra from \citet[][G17, blue/orange points]{2017MNRAS.472.3437G}.
Large transparent points represent the results of fitting iron and nickel simultaneously, whereas small opaque points represent the two-stage \emph{minimum-Ni} fit method.
Points with black boundaries indicate cases where the FWHM of the nickel feature reached the fit boundary of $13000$~$\rm km\ s^{-2}$.
The \textit{bottom right} panel compares the ratio of the integrated fluxes of the nickel and iron lines with the same stylistic conventions as the other panels.}
\label{fig:FeNicomp}
\end{figure*}

\begin{figure}
\centering
\includegraphics[width=8.5cm]{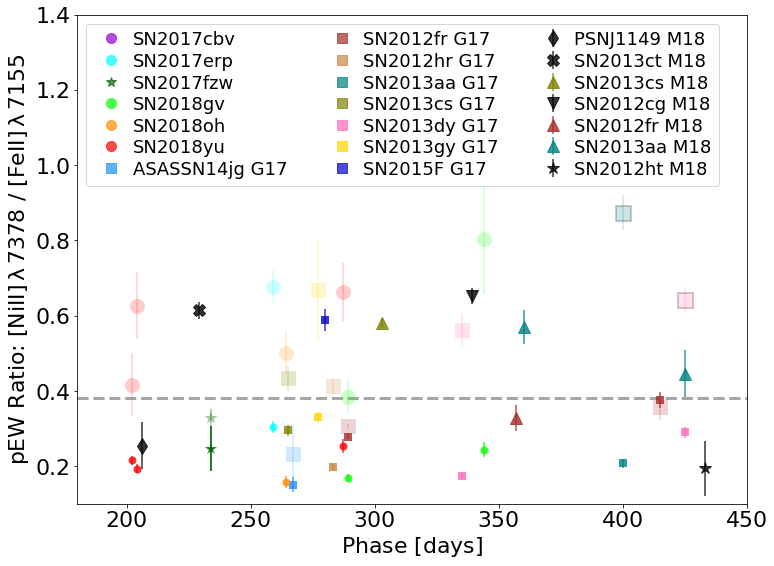}
\caption{The ratio of the pseudo-equivalent widths (pEW) of iron and nickel as a function of phase for SN\,Ia.  Data come from the current work, as well as  \citet[][G17]{2017MNRAS.472.3437G}, and  \citet[][M18, their Table 3]{2018MNRAS.477.3567M}.
Large transparent points represent the simultaneous fit results, and small opaque points the \emph{minimum-Ni} fit results.
Note that the Ni/Fe pEW ratio of SN\,2015F from the simultaneous fit method lies off the plot ($\sim1.75$ at 280 days). 
The dashed horizontal line, as in \citet[][their Figure 10]{2018MNRAS.477.3567M}, is at the solar Ni/Fe ratio and also marks the approximate division between predictions for the Ni/Fe ratio from delayed detonation (DDT) models and high-metallicity sub-$\rm M_{Ch}$ models (above the line) and lower-metallicity sub-$\rm M_{Ch}$ models (below the line).}
\label{fig:NiFeRatio}
\end{figure}

In Figure~\ref{fig:FeNicomp} we present a four-panel comparison of the results of these multiple-component Gaussian fits to the results of our double-component Gaussian fit parameters from Section~\ref{ssec:nebobs_linepars}.
The large transparent symbols show the results of the combined fit, whereas the smaller opaque symbols show the results of the \emph{minimum-Ni} fit.
The panels in Figure~\ref{fig:FeNicomp} compare the FWHM, velocity, integrated flux, and the ratio of the Ni/Fe integrated fluxes.
The top two panels demonstrate how the FWHM of nickel is broader, and the velocity of nickel is bluer, when the nickel and iron are fit simultaneously; i.e., when nickel is allowed to contribute to the blue-half of the feature (large transparent points).
The top left panel also shows that the nickel and iron have more similar velocities from the two-stage fit method (small opaque points). 
This suggests the \emph{minimum-Ni} method might be more accurate in some cases, because a significantly broader/bluer velocity for nickel compared to iron is not expected (e.g., as shown in Figure 4 of \citealt{2015MNRAS.450.2631M}).
We emphasize that the incorporation of the \emph{minimum-Ni} line-fitting method is not motivated by a desire to challenge the existence of stable Ni in the nebular SN\,Ia material, the presence of which is well-established \citep[e.g.,][]{2021arXiv210913840B}, but by our need to obtain a \emph{lower limit} on the Ni/Fe ratio (Figure~\ref{fig:NiFeRatio}) to inform our discussion of SN\,Ia explosion models, below.

The bottom-left panel of Figure~\ref{fig:FeNicomp} demonstrates how the \emph{minimum-Ni} fitting method (small opaque points) does in fact lead to minimal flux in the nickel component, by design.
Note that because the multi-Gaussian fit measures of integrated flux only include the primary lines, [\ion{Fe}{II}]~$\lambda$~7155~\AA\ and [\ion{Ni}{II}]~$\lambda$~7378~\AA,
their fluxes are systematically lower than those from the double-Gaussian fit, which includes all lines.
The bottom-right panel shows that the Ni/Fe ratio is systematically lower for a multi-Gaussian fit compared to the double-Gaussian fit.
This is primarily due to proper accounting for iron contributions in the red-half of the feature, which are attributed to nickel in the double-Gaussian fit.
We can also see that the \emph{minimum-Ni} fitting method (small opaque points) leads to significantly lower Ni/Fe ratios, also by design.

Instead of integrated flux, \citet{2018MNRAS.477.3567M} use the ratio of the pseudo-equivalent width (pEW) of the [\ion{Fe}{II}]~$\lambda$~7155~\AA\ and [\ion{Ni}{II}]~$\lambda$~7378~\AA\ features to estimate the Ni/Fe abundance ratio.
We measure the pEW for the 7378~\AA\ and 7155~\AA\ features in the SN\,Ia spectra presented in this work, and from \citet{2017MNRAS.472.3437G}.
In Figure~\ref{fig:NiFeRatio} we show the Ni/Fe pEW ratio as a function of phase, and compare with those from \citet[][their Table 3]{2018MNRAS.477.3567M}.
As in Figure~\ref{fig:FeNifits}, we use larger transparent symbols to represent the simultaneous fit of iron and nickel, and smaller opaque symbols to represent the two-stage \emph{minimum-Ni} fit. 
It is clear that the errors on the Ni/Fe pEW ratio are underestimated: there are a few cases of measurements from different spectra of the same object (e.g., SNe\,Ia 2013aa, 2013cs, 2018gv, and 2018yu) which differ by more than the error.
However, the greater problem is that the \emph{minimum} amount of nickel required to fit the blended [\ion{Fe}{II}]+[\ion{Ni}{II}] feature is significantly lower than the amount of nickel that \emph{can} be fit, leading to a lot of scatter in the $y$-axis of Figure \ref{fig:NiFeRatio}.

Specifically, the problem is that for most of the SNe\,Ia in our sample (but not all -- see Section~\ref{sssec:ana_NiFe_low}), the results of the simultaneous fit and the \emph{minimum-Ni} fit methods lie above and below the dashed line in Figure~\ref{fig:NiFeRatio}, respectively.
This dashed line is the same as that shown in \citet[][their Figure 10]{2018MNRAS.477.3567M}.
This line represents the solar Ni/Fe ratio, and also coincides with the approximate division between model predictions:
\textit{above the line} are Ni/Fe ratios from delayed detonation (DDT) models \citep{2013MNRAS.429.1156S} and super-solar metallicity ($\gtrsim$3--6$\times$) sub-$\rm M_{Ch}$ models \citep{2010ApJ...714L..52S,2013MNRAS.429.1425R}; 
\textit{below the line} are Ni/Fe ratios from lower-metallicity ($\lesssim2$ $\rm M_{\odot}$) sub-$\rm M_{Ch}$ models \citep{2018ApJ...854...52S}.

Thus, the degeneracy between iron and nickel contributions to the blended feature precludes us from drawing any general conclusions about the physical nature of the explosion mechanism based on our optical nebular phase spectra alone.
Future data sets which have higher resolution, better SNR, and/or nebular-phase NIR spectra -- in which relatively isolated [\ion{Fe}{II}] features could be used to better constrain the iron line parameters -- might offer improvements in the effort to physically constrain SNe\,Ia models.
We furthermore direct the reader to the work of \citet{2020MNRAS.491.2902F}, who apply full spectral synthesis models to optical-NIR nebular-phase spectroscopy.
They are able to draw much more robust conclusions for their entire sample, finding that the majority of Ni/Fe ratios for SNe\,Ia align with predictions of the sub-Chandrasekhar mass models (i.e., fall below the dashed line in our Figure~\ref{fig:NiFeRatio}).
That type of analysis is beyond the scope of this work.

\subsubsection{Five SNe\,Ia with truly low Ni/Fe ratios}\label{sssec:ana_NiFe_low}

In our sample of SNe\,Ia there are three with Ni/Fe ratios that are consistently below the dashed line in Figure~\ref{fig:NiFeRatio}: SN\,2017fzw, ASASSN-14jg, and SN\,2012fr.
In addition, \citet{2018MNRAS.477.3567M} find that PSNJ1149 and SN\,2012ht have low Ni/Fe ratios.
Thus it likely that these five SNe\,Ia have truly low Ni/Fe ratios, and consider them as candidates for sub-$\rm M_{Ch}$ mass and/or lower-metallicity progenitors. 
Two of these events, 2012ht and 2017fzw (both represented with 5-point star symbols in Figure \ref{fig:NiFeRatio}), are transitional SNe\,Ia which exhibit some signatures of 91bg-likes, such as faint peak magnitudes and quicker decline rates (larger $\Delta m_{15}(B)$ values).
Transitional SNe\,Ia may experience a different explosion mechanism and are discussed further in Section~\ref{ssec:disc_17fzw}.
No published or publicly available photometric data can be found for PSNJ1149, which was spectroscopically classified as a normal SN\,Ia by \citet{2015ATel.7825....1R}.

The other two of these events, SN\,2012fr and ASASSN-14jg, have similar $\Delta m_{15}(B)$ parameters of $0.93$ and $0.92$ mag, and similar peak $B$-band brightnesses of $-19.4$ and $-19.3$ mag, respectively \citep{2017MNRAS.472.3437G}.
SN\,2012fr exhibited a low velocity gradient (LVG; but did have a high \ion{Si}{II} velocity at peak, $12000$ $\rm km\ s^{-1}$, \citealt{2017MNRAS.472.3437G}), and ASASSN-14jg a low velocity, in their photospheric \ion{Si}{II} absorption feature from spectra obtained near and shortly after peak brightness.
They also both exhibit red-shifted velocities for the [\ion{Fe}{II}]+[\ion{Ni}{II}] nebular emission feature.
Perhaps the detonation of a lower-density, sub-$\rm M_{Ch}$ progenitor which spread through a greater volume of the white dwarf -- compacting the outer layers all around to exhibit an LVG for its \ion{Si}{II} -- but that was still offset `away' from the observer to exhibit red-shifted nebular lines could explain these observations.
A detonation in a larger volume of the white dwarf could also generate more $^{56}$Ni, which could cause the slightly over-luminous light curves (and lower decline rates) for 2012fr and ASASSN-14jg, and potentially also the lower Ni/Fe ratio at late times.
As previously mentioned, metallicity could also play a role here: lower (higher) metallicity white dwarfs could produce relatively less (more) stable nucleosynthetic products due to the under (over) abundance of neutrons \citep{2003ApJ...590L..83T}. 
Although both SN\,2012fr and ASASSN-14jg have spiral host galaxies and are thus candidates for having higher-metallicity progenitors, they could have originated from the spirals' metal-poor stellar populations.
Ultimately, with the data at hand it is not possible to separate the various potential factors involved, or to definitively pin the low Ni/Fe ratio on a single underlying cause.

\section{Discussion of Individual Objects}\label{sec:disc}

In Sections \ref{ssec:disc_17cbv} to \ref{ssec:disc_18yu} we explore the physical implications of the early-time data and nebular-phase spectra for each of our individual objects.

\subsection{SN\,2017cbv, a ``Blue Bump" SN\,Ia} \label{ssec:disc_17cbv}

\subsubsection{Early-time SN\,2017cbv}

\citet{2017ApJ...845L..11H} discovered SN\,2017cbv in the outskirts of spiral galaxy NGC~5643 at $z = 0.00399$ \citep{2004AJ....128...16K} by the Distance Less Than 40 Mpc (DLT40; \citealt{2018ApJ...853...62T}) survey on Mar 10.14 2017 UT with an apparent magnitude of $R=15.6$ \citep{2017ATel10158....1T}.
SN\,2017cbv was spectroscopically classified as a young Type Ia SN based on an optical spectrum from the Las Cumbres Observatory FLOYDS-S spectrograph on Mar 10.7 2017 UT \citep{2017ATel10164....1H}, with a good fit to the peculiar SN\,1999aa, \citealt{2004AJ....128..387G}.
Light curve fitting by \citet{2017ApJ...845L..11H}, who assume that SN\,2017cbv experienced a host galaxy extinction of $E(B-V)_{\rm host}\approx0$ based on its remote location and lack of \ion{Na}{I}~D absorption in higher resolution spectra \citep{2017ApJ...851L..43F}, finds that SN\,2017cbv reached a peak magnitude of $B = 11.72$ mag on Mar 29.1 2017 UT,
and exhibited a decline rate of $\Delta m_{\rm 15,B} = 1.06$ mag.
A re-analysis of the light curve by \citet{2018ApJ...863...24S} shows that the peak intrinsic magnitude was $M_B\approx-19.25$ mag, somewhat fainter than its spectral match, the peculiar SN\,1999aa (which exhibited $M_B \approx M_V \approx -19.8 \pm 0.2$ mag and $\Delta m_{15}(B) \approx 0.746 \pm 0.02$ \citealt{2000ApJ...539..658K}).
\citet{2018ApJ...863...90W} also show that the width and luminosity of 2017cbv's light curve is consistent with the Phillips relation \citep{1993ApJ...413L.105P,2009ApJ...700..331H}.
A further analysis of SN\,2017cbv's light curve by \citet{2020ApJ...895..118B} also matches with the re-analysis of \citet{2018ApJ...863...90W}, indicating that SN\,2017cbv was a normal SN\,Ia.

However, SN\,2017cbv was special in one respect: \citet{2017ApJ...845L..11H} show that the early-time light curve of SN\,2017cbv exhibited a ``blue bump" during the first five days.
``Blue bumps" at early times could be due to the ejecta interacting with a non-degenerate binary companion star or its CSM \citep{2010ApJ...708.1025K}, or due to the mixing of $^{56}$Ni to the outer layers of the ejecta \citep{2012ApJ...759...83P}.
\citet{2017ApJ...845L..11H} find that the ``blue bump" of SN\,2017cbv is consistent with the predicted signature of the SN's ejecta impacting a non-degenerate companion star along the line-of-sight, but alternate explanations such as interaction with nearby circumstellar material (CSM) or the mixing of $^{56}$Ni to the outer layers of the ejecta could not be ruled out. 
They also show that the very early optical spectra of 2017cbv exhibited strong \ion{C}{II} absorption and appeared more similar to the  normal SN\,2013dy than SN\,1999aa.

\subsubsection{Late-time SN\,2017cbv}\label{sssec:disc_17cbv_late}

The nebular-phase spectrum for SN\,2017cbv at $466$ days after peak brightness is shown in the top-left panel of Figure~\ref{fig:nebspec_wSN11fe}.
For comparison, two late-time spectra of the normal SN\,2011fe at phases $378$ and $574$ days past peak are also shown (these are $-88$ and $+108$ days relative to the SN\,2017cbv spectrum).
As expected, the peak flux of SN\,2017cbv's [\ion{Fe}{III}]~$\lambda\sim4700$~\AA\ line appears to be intermediate between those of the two SN\,2011fe spectra, and SN\,2017cbv's [\ion{Co}{III}] line has declined beyond visibility, similar to SN\,2011fe's $574$-day spectrum.

SN\,2017cbv's early-time ``blue bump" suggests that we might see late-time H$\alpha$ emission if this feature is due to a non-degenerate binary companion.
\citet{2018ApJ...863...24S} present nebular spectroscopy at $\sim$300 days after peak brightness of SN\,2017cbv and constrain the mass of hydrogen in the system to $\leq 10^{-4}$ $\rm M_\odot$.
This limit is about three orders of magnitude lower than predicted by models in which a non-degenerate companion star's hydrogen is swept up in the SN\,Ia ejecta \citep[e.g.][]{2018ApJ...852L...6B}.
Additionally, \citet{2020MNRAS.493.1044T} find no evidence of stripped companion emission for SN\,2017cbv.

The lack of H$\alpha$ in the nebular-phase spectrum of SN\,2017cbv might suggest that mixing of $^{56}$Ni into the outer layers of the ejecta is more likely as the root cause of the ``blue bump".
The presence of $^{56}$Ni at higher velocities could lead to broader nebular-phase lines for the nucleosynthetic decay products, which we can check with our nebular-phase spectrum.
In Tables \ref{tab:linepars_direct} and \ref{tab:linepars_gaussian} (in Appendix~\ref{app:tables}) we report that the iron and nickel lines exhibit a FWHM of $8100$ to $9500$~$\rm km\ s^{-1}$, which are average values and not especially broad.
At 466 days the cobalt is too weak to be included in this line-width analysis, so instead we directly measure the width of the [\ion{Co}{III}]~$\lambda$5890~\AA\ feature in the 302 day spectrum from \citet{2018ApJ...863...24S}, and find a FWHM of $\sim$10400~$\rm km\ s^{-1}$, which is about a median value of FWHM for our sample (see the middle panel of Figure \ref{fig:line_meas}).
The caveat here is that average-width nebular lines are not strong evidence {\it against} $^{56}$Ni mixing, because observed spectral parameters are degenerate with other physical qualities, such as the density profile \citep{2017ApJ...845..176B}.
Additional (and more reliable) evidence against $^{56}$Ni mixing is that it could cause increased line blanketing in the weeks around peak brightness, depressing the luminosity especially in the bluer filters, which was not observed. 
In general we find no direct evidence in our nebular-phase spectrum that $^{56}$Ni mixing was the root cause of the ``blue bump".

Alternatively, early-time ``blue bumps" might be a signature of the sub-$\rm M_{Ch}$ double detonation model \citep{2017MNRAS.472.2787N}.
As discussed in Section~\ref{ssec:ana_NiFe}, the Ni/Fe ratio in nebular-phase SN\,Ia spectra can be used to distinguish between the delayed detonation and (low-metallicity) sub-$\rm M_{Ch}$ model.
As we were unable to robustly measure the Ni/Fe ratio for our 466 day spectrum of SN\,2017cbv, we applied the multi-component Gaussian fits used in Section \ref{ssec:ana_NiFe} to the 302 day spectrum of SN\,2017cbv from \citet{2018ApJ...863...24S}.
We found a Ni/Fe ratio of $\sim$0.6 from the best fit, and of $\sim$0.14 from the \textit{minimum-Ni} fit method.
These values lie above and below the Ni/Fe ratio of $\sim$0.4 which approximately distinguishes delayed detonation and (low-metallicity) sub-$\rm M_{ch}$ models (Figure \ref{fig:FeNicomp}).
Thus, no conclusions about the potential connection between SN\,2017cbv's ``blue bump" and its explosion mechanism can be drawn from the nebular-phase spectra.

\subsection{SN\,2017ckq, a Potential HVG SN\,Ia} \label{ssec:disc_17ckq}

\subsubsection{Early-time SN\,2017ckq}

SN\,2017ckq was discovered on 2017-03-27 UT at 18.11 mag (AB in the cyan-ATLAS filter; \citealt{2017TNSTR.361....1T}), and was classified as a Type Ia with an optical spectrum from the ESO New Technology Telescope \citep{2017TNSCR.467....1G}.
It is located  in host galaxy ESO-437-G-056 at $z=0.01\pm0.001$ \citep{2000ApJ...529..786M} with a Tully-Fisher distance modulus of $33.2\pm0.2$ mag and a distance of $42.9$ Mpc \citep{2013AJ....146...86T}.
The Las Cumbres optical photometry that we obtained around the time of peak brightness indicates a $B$-band decline rate of $\Delta m_{\rm 15}(B) \approx 0.96 \pm 0.2$ mag and an intrinsic peak brightness of $M_B\approx-18.9\pm0.3$ mag (Section~\ref{ssec:sne_earlyphot}).
Both the peak brightness and decline rate are on the low side, but given the uncertainties they are not inconsistent with the Phillips relation \citep{1999AJ....118.1766P}, and so SN\,2017ckq appears to be a normal SN\,Ia.
We also found that SN\,2017ckq exhibited a photospheric \ion{Si}{II} absorption feature with high velocity and a high velocity gradient (Table \ref{tab:early}), and classified it as a potential HVG event (Table~\ref{tab:early}).

\subsubsection{Late-time SN\,2017ckq}

As a potentially HVG event, we expect the nebular-phase spectrum of SN\,2017ckq to exhibit a redshifted [\ion{Fe}{II}]+[\ion{Ni}{II}] feature in accordance with the asymmetric explosion model discussed in Section~\ref{ssec:ana_asym}.
Unfortunately, we only obtained a partial (blue-side) optical nebular spectrum of SN\,2017ckq at 369 days after peak brightness (top-right panel of Figure~\ref{fig:nebspec_wSN11fe}), which does not include the [\ion{Fe}{II}]+[\ion{Ni}{II}] nebular feature.
From Figure~\ref{fig:nebspec_wSN11fe} we see that the blue-side features of SN\,2017ckq resemble SN\,2011fe, except for the noticeably flatter shape in the peak of the Fe~$\lambda$5250~\AA\ feature.

A unique aspect of the $\lambda$5250~\AA\ feature for SN\,2017ckq also manifests in the top panel of Figure~\ref{fig:nebevolspec}, in which SN\,2017ckq has the distinction of exhibiting the lowest 5250/4700 flux ratio of all the SNe\,Ia in that sample. 
The $\lambda$5250~\AA\ feature is a blend of [\ion{Fe}{II}] and [\ion{Fe}{III}], whereas the $\lambda$4700~\AA\ feature is primarily [\ion{Fe}{III}].
As described in Section~\ref{ssec:ana_nebevol}, the 5250/4700 flux ratio appears to rise in spectra $>$350 days after peak brightness, which could be due to the nebula cooling and the ionization state transitioning from doubly- to singly-ionized iron.
We hypothesized that SNe\,Ia with larger $^{56}$Ni masses (and lower light-curve decline rates) might exhibit a lower 5250/4700 flux ratio for longer into the nebular phase due to a longer cooling timescale and/or a later transition in the ionization state.
By exhibiting the lowest 5250/4700 flux ratio of all the SNe\,Ia in our sample, exhibiting that ratio at $>$350 days past peak brightness, and also exhibiting $\Delta m_{15}(B)\sim0.96\pm0.2$ mag at early times, SN\,2017ckq fits the hypothesis. 
However, the other HVG SNe\,Ia considered in Figure~\ref{fig:nebevolspec} (SNe 2012hr, 2013cs, and 2017erp) do not exhibit this flattened shape for the $\lambda$5250~\AA\ feature, and neither do any of the other spectra in this sample (nor in the sample of \citealt{2017MNRAS.472.3437G}).
The physical origin of the flattened shape of SN\,2017ckq's $\lambda$5250~\AA\ feature remains unclear.

\subsection{SN\,2017erp, an NUV-Red High-Velocity SN\,Ia} \label{ssec:disc_17erp}

\subsubsection{Early-time SN\,2017erp}

SN\,2017erp was discovered on 2017-06-13 UT at 16.8 mag (Vega; clear filter) by K. Itagaki \citep{2017TNSTR.647....1I}, and classified as a Type Ia in NGC~5861 at $z=0.006174$ \citep{2005A&A...430..373T} with an optical spectrum from the South African Large Telescope \citep{2017ATel10490....1J}.
\citet{2019ApJ...877..152B} present an analysis of 2017erp with UV and optical photometry and spectra spanning the first $\sim120$ days after explosion.
They find that SN\,2017erp exhibited a light curve apparent peak brightness of $m_B=13.27\pm0.01$ mag and decline rate of $\Delta m_{\rm 15}(B) \approx 1.1$ mag.
\citet{2019ApJ...877..152B} provide an extensive analysis of the light curve shape and color, which we  summarize as indicating a host-galaxy extinction of $A_V \approx 0.45\pm0.05$ mag.
This, along with the host's distance modulus $\mu = 32.34 \pm 0.097$ mag \citep{2020ApJ...892..121K}, suggests a peak intrinsic magnitude of $B \approx -19.5 \pm 0.15$ mag.
We note also that \citet{2020ApJ...892..121K} report a synthesized $^{56}$Ni mass of $0.975 \pm 0.083$ $\rm M_{\odot}$, the highest in their sample of 17 SNe\,Ia. 
Altogether, SN\,2017erp was a normal, slightly overluminous Type Ia in terms of its optical light curve shape and spectra.

After their careful evaluation of the host galaxy's contribution to the line-of-sight extinction and reddening, \cite{2019ApJ...877..152B} show that SN\,2017erp exhibited a depressed NUV flux and an intrinsically redder NUV-optical color than other SNe\,Ia, such as the NUV-blue SN\,2011fe.
They suggest that this might indicate that the progenitor of SN\,2017erp had a higher progenitor metallicity than SN\,2011fe, but note that mixing of nucleosynthetic products into the outer layers cannot be ruled out.
\citet{2019ApJ...877..152B} demonstrate how the spectra of SN\,2017erp prior to peak brightness exhibit a high-velocity \ion{Si}{II}~${\lambda}6355$~\AA\ feature and a \ion{C}{II}~${\lambda}6580$~\AA\ feature which disappears by phase $-12$ days.
They also show how the photospheric velocity is higher than 2011fe at early times, but declines to be of a similar value by peak brightness (their Fig. 9).
With spectroscopy from Las Cumbres Observatory, we have classified SN\,2017erp as a potential member of the HVG-class (Table~\ref{tab:early}).

\subsubsection{Late-time SN\,2017erp}

If SN\,2017erp had a higher progenitor metallicity, the additional neutrons available might lead to a higher ratio of stable-to-radioactive iron and nickel in the nucleosynthetic products \citep[e.g.,][]{2003ApJ...590L..83T}.
During the nebular phase, all of the radioactive nickel has decayed and only stable nickel remains, and so a higher progenitor metallicity could manifest in our nebular-phase spectrum as a higher Ni/Fe ratio than in typical SNe\,Ia.
However, in Figure~\ref{fig:nebspec_wSN11fe} (second-row left panel) we can see that the Ni/Fe ratio exhibited by the nebular-phase spectrum of SN\,2017erp is similar to that of the typical SN\,2011fe at a similar phase.
Furthermore, as described in Section~\ref{ssec:ana_NiFe} we find a significant degeneracy in the relative contributions of nickel and iron that can provide good fits to the $\lambda$7200~\AA\ Ni+Fe blended feature (as is the case for most of the SNe\,Ia in our sample).
Thus, the nebular-phase spectra of SN\,2017erp do not help to confirm or reject the hypothesis of a higher progenitor metallicity.

If the depressed NUV flux of SN\,2017erp was instead due to the mixing of nucleosynthetic products to the outer layers, where it could absorb the NUV photons, then we might expect to find particularly broad nebular-phase emission lines.
As shown in Tables \ref{tab:linepars_direct} and \ref{tab:linepars_gaussian} in Appendix~\ref{app:tables}, SN\,2017erp's nebular-phase emission lines of [\ion{Fe}{III}] and [\ion{Co}{III}] were of average width, but as previously mentioned (Section~\ref{sssec:disc_17cbv_late}), average-width lines are not strong evidence against mixing.
In general, we find that SN\,2017erp exhibited a typical, SN\,2011fe-like nebular-phase spectrum, and that we are unable to draw any further conclusions about the physical origin of its depressed NUV flux.

SN\,2017erp is one of the few SNe\,Ia which exhibited a high velocity gradient for its photospheric \ion{Si}{II} absorption feature in the two weeks after peak brightness.
The explosion asymmetry model of \citet{2010Natur.466...82M} suggests that detonations which are offset away from the observer along the line of sight can lead to both a photospheric-phase HVG and a nebular-phase redshift in the [\ion{Fe}{II}]+[\ion{Ni}{II}]~$\lambda$7200~\AA\ blended feature (Section~\ref{ssec:ana_asym}). 
The nebular-phase spectra for SN\,2017erp (Figure~\ref{fig:nebspec_wSN11fe}, second-row left panel) shows that this Fe+Ni feature is redshifted for SN\,2017erp compared to SN\,2011fe.
The location of SN\,2017erp in the plot of photospheric velocity gradient versus nebular-phase line velocity (Figure~\ref{fig:asym}) furthermore demonstrates that SN\,2017erp is consistent with the trend exhibited by SNe\,Ia in general, and that observations of SN\,2017erp support the asymmetric explosion model.

\subsection{SN\,2017fzw, a Subluminous Transitional SN\,Ia}\label{ssec:disc_17fzw}

\subsubsection{Early-time SN\,2017fzw}

SN\,2017fzw was discovered on 2017-08-09 by \citet{2017TNSTR.855....1V} at 17.17 mag (AB; clear filter) as part of the DLT40 survey. 
It was spectroscopically classified as a 91bg-like SNe\,Ia using the FLOYDS-S instrument of the Faulkes Telescope South \citep{2017TNSCR.865....1H}. 
SN\,2017fzw is located in the outskirts of NGC~2217 ($\sim$100\arcsec\ from the host galaxy's center), a face-on barred spiral at $z=0.0054$ \citep{1991rc3..book.....D} with a Tully-Fisher distance modulus of $31.45\pm0.4$ mag and a distance of $19.5$ Mpc \citep{Tully1988}.

At early times SN\,2017fzw is similar to the class of SNe\,Ia that resemble the underluminous, spectroscopically peculiar, fast-evolving SN\,1991bg (``91bg-likes"; \citealt{1992AJ....104.1543F}).
Galbany et al. (in prep.) presents a full analysis of SN\,2017fzw and shows that it peaked on 2017-08-22.9 UT with apparent magnitudes of $B\sim13.25\pm0.16$ and $V\sim13.09\pm0.02$ mag, and a peak color of $B-V=0.17\pm0.16$ mag.
Galbany et al. (in prep.) applies corrections for both MW and host galaxy extinction, the latter with $E(B-V)_{\rm host}=0.182$ and $R_{V,\rm{host}}=3.1$, which a pretty high host galaxy extinction given its location in the outskirts (Figure \ref{fig:stamps}). 
They furthermore show that the light curve of SN\,2017fzw exhibited a decline rate of $\Delta m_{15}(B)=1.60\pm0.02$ mag, and that light-curve fits indicate a distance modulus of $\mu = 32.06\pm0.03$ mag, which implies an absolute $B$-band intrinsic magnitude of $-18.81\pm0.18$ mag -- at the luminous end of the 91bg-like class.
These values from Galbany et al. (in prep.) are quoted in Table~\ref{tab:early}, and used in our analysis.

In early-time optical spectra within $\sim$10 days of peak brightness, 91bg-like SNe\,Ia are distinguished by deep titanium absorption features at $\lambda$4200~\AA\ \citep[e.g.,][]{1997MNRAS.284..151M,2017hsn..book..317T}.
In Figure~\ref{fig:2017fzw_91bg_11fe} we plot a time series of phase-matched spectra of 
normal SN\,2011fe (blue; \citealt{2015MNRAS.450.2631M,2014MNRAS.439.1959M,2013A&A...554A..27P,2012ApJ...752L..26P}),
SN\,2017fzw (green; Galbany et al. {\it in prep.}),
and SN\,1991bg (red; \citealt{1992AJ....104.1543F,1996MNRAS.283....1T}).
The titanium feature at $\lambda$4200~\AA\ is most clearly seen for SNe\,2017fzw and 1991bg in the top set of spectra, at phases of -10 and 0 days respectively. 

\subsubsection{Late-time SN\,2017fzw}

Nebular-phase spectra of 91bg-like events are rarely obtained because they are intrinsically less luminous and their brightness declines rapidly \citep{2017hsn..book..317T}, which makes our 234-day spectrum of SN\,2017fzw both unique and valuable. There are two additional things to note about our nebular spectrum of SN\,2017fzw: (1) It does not exhibit a nebular-phase emission feature at $\lambda{\sim}6300$~\AA\ like those seen for subluminous SN\,2002es-like supernovae, which has been suggested to be due to [\ion{O}{I}] and associated with the violent merger model \citep[e.g.,][]{2013ApJ...775L..43T,2016MNRAS.459.4428K}. (2) It was also used by \citet{2019ApJ...877L...4S} to show that SN\,2017fzw did not exhibit any late-time H$\alpha$ emission which might suggest a non-degenerate companion, as found for the subluminous SN\,Ia ASASSN-18tb \citep{2019MNRAS.486.3041K,2019MNRAS.487.2372V}. Additionally, \citet{2020MNRAS.493.1044T} find no evidence of stripped companion emission for SN 2017fzw.

\begin{figure}
\includegraphics[width=\columnwidth]{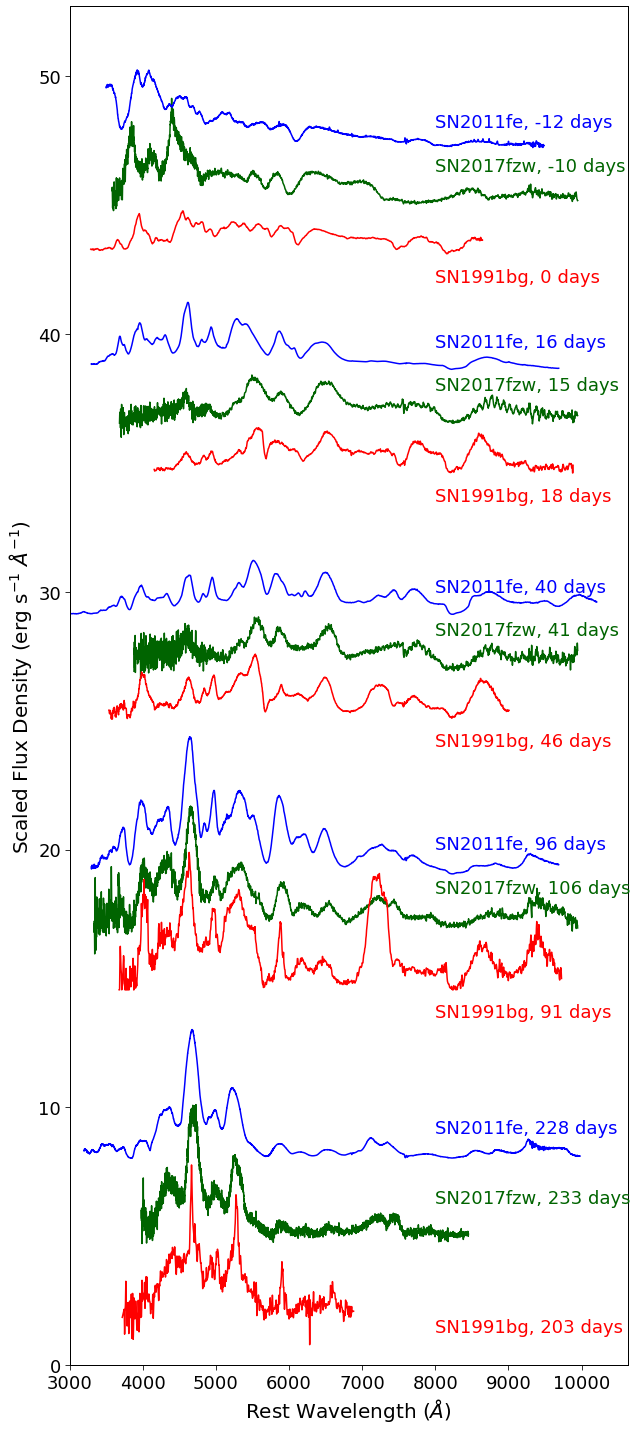}
\caption{A time-series of phase-matched spectra for normal SN\,2011fe (blue), transitional SN\,Ia 2017fzw (green), and the canonical sub-luminous SN\,1991bg (red), from early to late times (top to bottom). Fluxes have been scaled and offset to enable the comparison. In the earliest spectra SN\,2017fzw resembles SN\,1991bg but by the nebular phase epoch ($\geq$200 days) it resembles SN\,2011fe.}
\label{fig:2017fzw_91bg_11fe}
\end{figure}

As shown in Figure \ref{fig:2017fzw_91bg_11fe}, at later phases SN\,2017fzw evolves away from exhibiting 91bg-like features and develops characteristics similar to the normal SN\,2011fe.
Specifically, the nebular-phase emission lines of 91bg-like SNe\,Ia become narrower over time, with typical dispersion velocities around or below $2000$ $\rm km\ s^{-1}$ \citep{2012MNRAS.424.2926M}.
This narrowing is seen in the spectra for SN\,Ia 1991bg at 91 and 203 days in Figure~\ref{fig:2017fzw_91bg_11fe}, but in contrast, SN\,2017fzw exhibits emission lines of comparable width to SN\,2011fe in its nebular phase.
This similarity of SN\,2017fzw to SN\,2011fe at late times is also demonstrated by the middle-right panel of Figure~\ref{fig:nebspec_wSN11fe}.

Based on the optical light curve and spectroscopic evolution of SN\,2017fzw, both this work and Galbany et al. (in prep.) find that it is similar in many respects to ``transitional" SNe\,Ia, which is the term used for SNe\,Ia that exhibit signatures similar to both 91bg-like and normal events, such as the transitional SN\,Ia 1986G \citep{1987PASP...99..592P,2016MNRAS.463.1891A}.
The term `transitional' is used to refer to SNe that are in between two classes, and the SNe\,Ia which are `transitional' between 91bg-like and normal also appear to be transitional in the sense that their spectra transition from being more 91bg-like at early times (exhibiting titanium) to more normal during the nebular phase (exhibiting broader iron features than SN\,1991bg, and weaker calcium; e.g., SN\,2012ht, \citealt{2018MNRAS.477.3567M}; and 2015bp, \citealt{2017MNRAS.466.2436S}).

\subsubsection{SN\,2017fzw in Context: Normal, Transitional, and 91bg-like SNe\,Ia}

\citet{2012MNRAS.424.2926M} and \citet{2011MNRAS.416..881M} model the nebular spectra of SN\,Ia 1991bg and the transitional SN\,Ia 2003hv, respectively, and find that a lower mass and/or density of the inner regions provides the best explanation for their late-time emission features.
However, for transitional SN\,Ia 2003hv, their model is reproducing a large [\ion{Fe}{III}]:[\ion{Fe}{II}] flux ratio, whereas the [\ion{Fe}{III}]:[\ion{Fe}{II}] flux ratio for 2017fzw is not dissimilar to 2011fe or other normal SNe at its epoch (Figure~\ref{fig:2017fzw_91bg_11fe}, and the top panel of Figure~\ref{fig:nebevolspec}). 
The spectral evolution for SN\,2017fzw presented above leads to a hypothesis that the physical qualities that cause a transitional SN\,Ia to appear 91bg-like at early times might originate in the white dwarf's outer layers, and not in its nucleosynthetic products.

\begin{figure}
\includegraphics[width=\columnwidth]{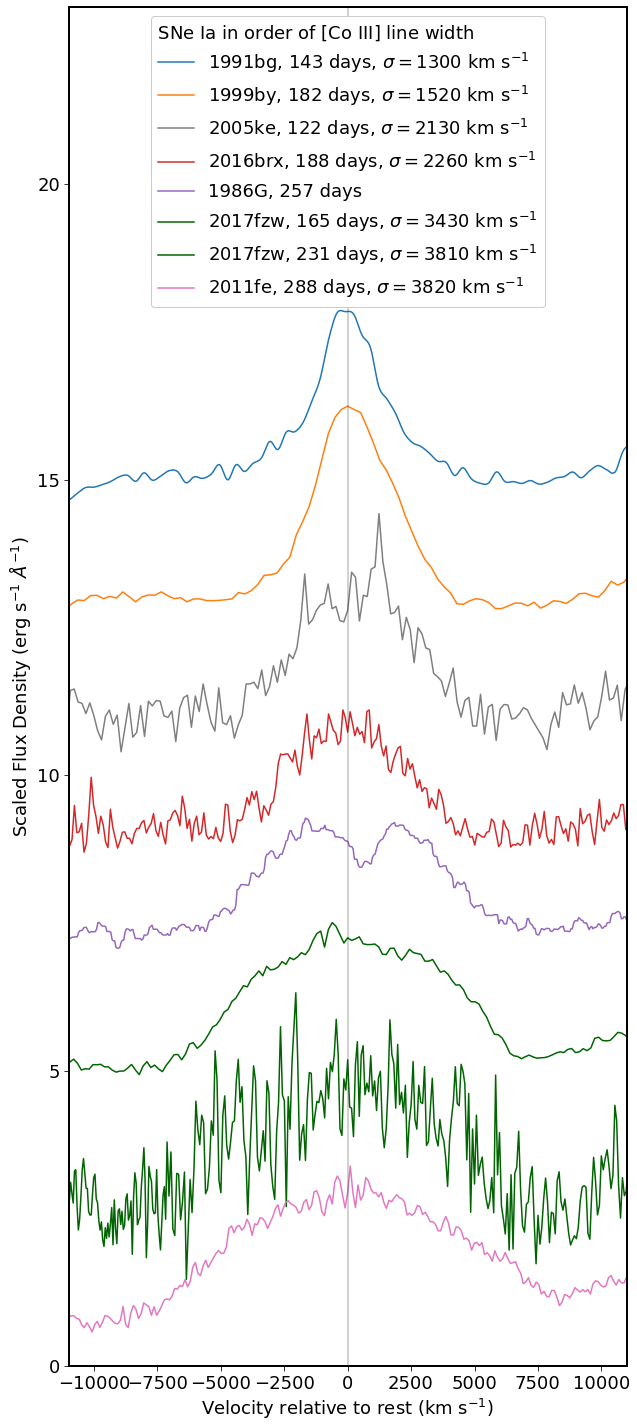}
\caption{The nebular-phase [\ion{Co}{III}]~$\lambda$5250~\AA\ feature for a sample of 91bg-like (91bg, 99by, 05ke, 16brx) and transitional (86G, 17fzw) SNe\,Ia, and the normal SN\,2011fe, ordered by increasing line width from top to bottom.
Fluxes have been scaled and offset for visibility, and the line widths are provided in the legend.
The transitional SN\,2017fzw (green) exhibits a [\ion{Co}{III}] feature which is almost as broad as that of normal SN\,2011fe (pink), and is dissimilar to the 91bg-like events.}
\label{fig:2017fzw_CoIIIdisp}
\end{figure}

To confirm or reject this hypothesis, and to explore the similarity of transitional SN\,Ia 2017fzw's nucleosynthetic products with normal SNe\,Ia, in Figure~\ref{fig:2017fzw_CoIIIdisp} we compare the velocity dispersion of the  [\ion{Co}{III}]~$\lambda$5800~\AA\ emission feature for a sample of normal, transitional, and 91bg-like SNe\,Ia. 
This [\ion{Co}{III}] feature is a good signature of the nucleosynthetic products in part because it is generally free of contamination from other species until very late phases, $>$1000 days, when sodium might contribute \citep{2014MNRAS.439.3114D,2015MNRAS.454.1948G}.
In Figure~\ref{fig:2017fzw_CoIIIdisp} we show, from top to bottom, the [\ion{Co}{III}]~$\lambda$5800~\AA\ emission feature for 91bg-like, transitional, and normal SNe\,Ia, ordered from narrowest to broadest emission\footnote{ 
SN\,1991bg at 143 days \citep{1996MNRAS.283....1T}; 
91bg-like SN\,1999by at 182 days \citep{2012MNRAS.425.1789S};
91bg-like SN\,2005ke at 122 days \citep{2013ApJ...773...53F};
91bg-like SN\,2016brx at 188 days \citep{2018MNRAS.479L..70D};
transitional SN\,1986G at 257 days \citep{1992A&A...259...63C};
transitional SN\,2017fzw at 165 days Galbany et al. {\it in prep.};
transitional SN\,2017fzw at 234 days (this work);
and normal SN\,2011fe at 288 days \citep{2015MNRAS.450.2631M}.}.
Figure~\ref{fig:2017fzw_CoIIIdisp} shows that SN\,2017fzw exhibits the broadest [\ion{Co}{III}] nebular emission line of all the 91bg-like and transitional SNe\,Ia in this sample, and that it resembles normal SN\,Ia 2011fe.
As a side note, \citealt{2020MNRAS.492.3553V} classifies the [\ion{Co}{III}] feature of transitional SN\,1986G as tentatively bimodal\footnote{SN\,2007on is another example of a bimodal [\ion{Co}{III}] emission in a transitional SN\,Ia \citep{2018A&A...611A..58G}, which we have not included in our Figure~\ref{fig:2017fzw_CoIIIdisp}.}, after accounting for the \ion{Na}{I}~D absorption feature seen near ${\sim}1000$~$\rm km\ s^{-1}$.
However, we find no evidence of bimodality in the  [\ion{Co}{III}] emission of SN\,2017fzw (Section \ref{ssec:ana_coiii}).

\begin{figure}
\includegraphics[width=8cm]{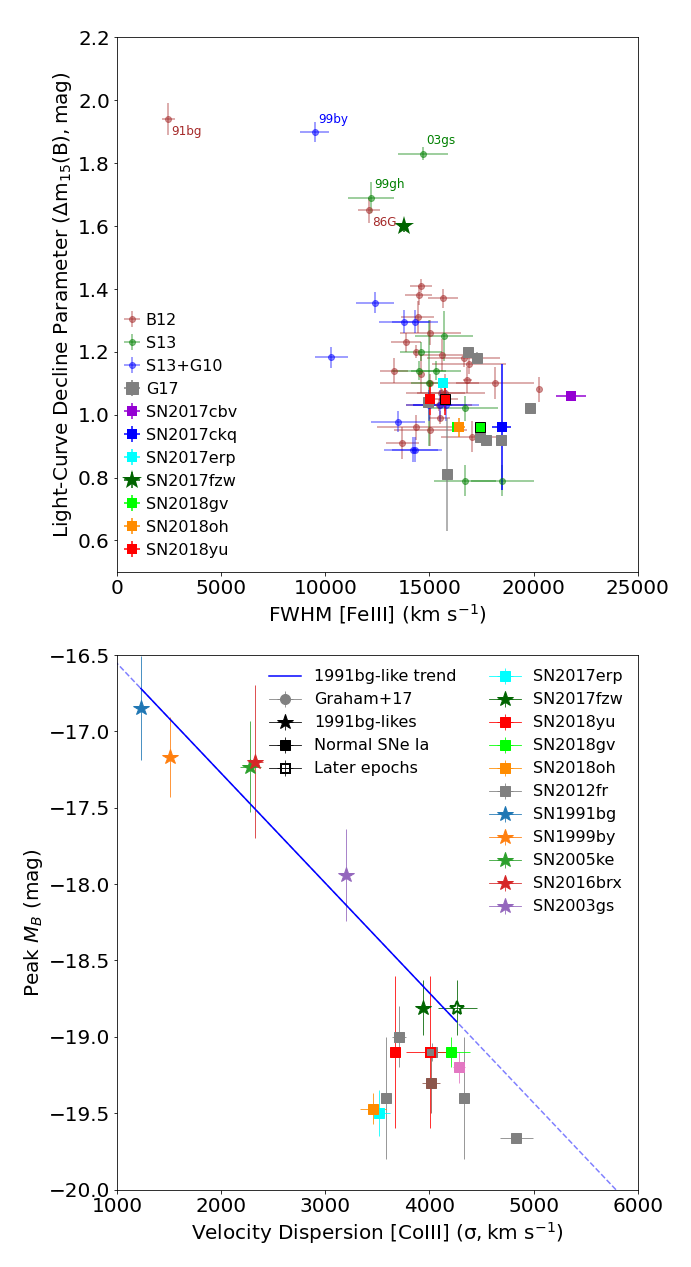}
\caption{\textit{Top:} The optical light-curve decline rate, $\Delta m_{15}(B)$, versus the full-width half-max (FWHM) of the [\ion{Fe}{III}]~$\lambda$4700~\AA\ feature in nebular-phase spectra (from a best-fit Gaussian function).
Data are from \citet[][B13]{2012AJ....143..126B}, \citet[][S13]{2013MNRAS.430.1030S}, \citet[][G10]{2010ApJS..190..418G}, \citet[][G17]{2017MNRAS.472.3437G}, and this work (as labeled in the legend).
The 91bg-like events are individually labeled.
\textit{Bottom:} The light-curve peak $B$-band brightness versus the velocity dispersion of the [\ion{Co}{III}]~$\lambda$5890~\AA\ feature (the standard deviation, $\sigma$).
All velocity widths are from the ``direct measure" method described in Section~\ref{ssec:nebobs_linepars}.
SNe\,Ia that are 91bg-like or transitional are represented with 5-point stars, and normal SNe\,Ia with squares.
Open symbols represent the later epochs for SNe\,Ia with two nebular-phase spectra (SNe\,2017fzw and 2018yu).
The solid blue line is a linear best-fit to the 91bg-like events, and is extended past SN\,2017fzw as a dashed blue line to demonstrate the offset of normal SNe\,Ia from this trend.}
\label{fig:LumxWidth}
\end{figure}

To further explore SN\,2017fzw in the context of 91bg-like, transitional, and normal SNe\,Ia, we consider its location in two phase-space diagrams which compare nebular-phase emission line widths to early-time light curve properties.
In the top panel of Figure~\ref{fig:LumxWidth} we plot the light-curve decline rate parameter $\Delta m_{15}(B)$ as a function of the FWHM\footnote{In order to compare with the FWHM values from \citet{2013MNRAS.430.1030S}, we remeasure the FWHM values for the spectra from \citet{2017MNRAS.472.3437G} and from this work by fitting a single-component Gaussian \textit{without} subtracting the pseudo-continuum, but otherwise it is the same method described in Section~\ref{ssec:nebobs_linepars}.} of the [\ion{Fe}{III}]~$\lambda$4700~\AA\ emission feature \citep[as in][their Figure 22]{2012AJ....143..126B}.
This plot includes data for SNe\,Ia from 
\citet{2012AJ....143..126B}, \citet{2013MNRAS.430.1030S}, \citet{2010ApJS..190..418G}, and \citet{2017MNRAS.472.3437G}, along with the SNe\,Ia presented in this work. 
SNe\,Ia that are classified as 91bg-like or transitional are individually labeled, while SN\,2017fzw is represented as a five-point star.
Compared to the normal SNe\,Ia, these objects appear to occupy a distinct region of the $\Delta m_{15}(B)$ \textit{vs} [\ion{Co}{III}] FWHM parameter space.
Although a potential trend between $\Delta m_{15}(B)$ and [\ion{Co}{III}] FWHM is suggested by this plot, the 91bg-like and transitional events might have a different slope and the normal events have a lot of scatter, as also shown and discussed by \citet{2013MNRAS.430.1030S}.

The advantage of using parameters $\Delta m_{15}(B)$ and [\ion{Fe}{III}] FWHM for the top panel of Figure \ref{fig:LumxWidth} is being able to easily include data from previous works, however, there are two main drawbacks. 
First, \citet{2004ApJ...613.1120G} have shown that the light-curve decline rate $\Delta m_{15}(B)$ is not as well correlated with peak brightness for 91bg-likes as it is for normal SNe\,Ia.
Second, the [\ion{Fe}{III}] feature is not isolated and has a significant flux contribution from neighboring features, and because some stable iron is formed in the explosion [\ion{Fe}{III}] is not as good a representative of the distribution of radioactive $^{56}$Ni as [\ion{Co}{III}]. 
Thus, in the bottom panel of Figure~\ref{fig:LumxWidth} we plot the peak absolute $B$-band brightness as a function of the width of the [\ion{Co}{III}]~$\lambda$5890~\AA\ feature (using the direct measure method with psuedo-continuum subtraction; Section~\ref{sec:nebobs}).
The sample shown includes the same 91bg-like and transitional SNe\,Ia as in Figure~\ref{fig:2017fzw_CoIIIdisp}, plus SN\,2003gs \citep{2012MNRAS.425.1789S}, along with the normal SNe\,Ia from \citet{2017MNRAS.472.3437G}\footnote{For Figure~\ref{fig:LumxWidth}, we updated the absolute peak brightness for SN\,2015F from the measurement given in \citet{2017MNRAS.472.3437G} to use the host galaxy extinction of $E(B-V)=0.085\pm0.019$ mag from \citet{2017MNRAS.464.4476C} and the Cepheid distance modulus of $\mu = 31.51\pm0.05$ mag from \citet{2016ApJ...826...56R}. The updated absolute peak brightness of SN\,2015F is thus $M_B = -19.1 \pm 0.06$ mag.} and this work.

In the lower panel of Figure~\ref{fig:LumxWidth} we find that the 91bg-like and transitional events (all marked as five-point stars) show a clearer trend between peak brightness and [\ion{Co}{III}] line width, whereas the normal SNe\,Ia (squares) show no obvious trend.
The best linear fit to the 91bg-like and transitional events is shown as a solid blue line, extended past SN\,2017fzw (which was included in the fit) as a dashed blue line. 
The transitional SN\,Ia 2017fzw is unique in this sample because it sits within the region of $\Delta m_{15}(B)$ -- $\sigma_{\rm [CoIII]}$ parameter space occupied by normal SN\,Ia events, yet also right on the best-fit line for transitional and 91bg-like events.
This correlation between early- and late-time features for transitional SNe\,Ia like 2017fzw refutes our earlier hypothesis that the physical origins of transitional SNe\,Ia's ``91bg-likeness" might lay \emph{solely} in the outer layers of the white dwarf.

\subsubsection{The lack of a clear $M_B$ -- [\ion{Co}{III}] line width correlation for normal SNe\,Ia}

This trend prompts the question of why 91bg-like and transitional SNe\,Ia -- fainter events that synthesized less $^{56}$Ni -- exhibit a clearer trend between $^{56}$Ni mass (peak brightness) and nebular-phase velocity dispersion than normal SNe\,Ia.
The $^{56}$Ni mass represents the energy of the explosion, and the line width represents the kinetic energy of the ejecta material, and so these two parameters \textit{should} be correlated for a sample of SNe\,Ia in which the ejecta mass is constant or has a shallow dependence on $^{56}$Ni mass.
Here we turn to \citet{2015MNRAS.454.3816C}, which presented a method to estimate the ejecta mass from the flux of the nebular-phase [\ion{Co}{III}]~$\lambda$~5893~\AA\ line for normal SN\,Ia events.
They found that while $^{56}$Ni mass exhibits a shallow correlation with ejecta mass for fainter, faster-declining SNe\,Ia, the $^{56}$Ni mass was uncorrelated with ejecta mass for the normal and brighter, slowly-declining SNe\,Ia \citep[][their Figure 11]{2015MNRAS.454.3816C}.
Although their analysis did not include SN\,1991bg-like events, they are well-established to have low ejecta masses \citep[e.g., ][]{1997MNRAS.284..151M,2010Natur.463...61P,2011ApJ...732..118S}.
Thus, a common and low ejecta mass could explain why the 91bg-like and transitional events exhibit a clear trend in Figure~\ref{fig:LumxWidth}.

Furthermore, for normal SNe\,Ia, there are likely additional factors that influence the mass of $^{56}$Ni synthesized in the explosion which are independent of the ejecta mass, such as the explosion mechanism itself -- and additional factors which cause a scatter in the relation between peak brightness and $^{56}$Ni (e.g., ejecta density or opacity).
These factors could also explain why the normal events do not exhibit as clear a trend in Figure~\ref{fig:LumxWidth}.
As our collective sample of nebular-phase spectra for faint, transitional, and 91bg-like SNe\,Ia grows, we will be better able to establish and characterize such correlations and connect them to the underlying physical explosion mechanism, and provide better context for unique transitional events like SN\,2017fzw.

\subsection{SN\,2018gv, a Spherically Symmetric SN\,Ia} \label{ssec:disc_18gv}

\subsubsection{Early-time SN\,2018gv}

SN\,2018gv was discovered on 2018-01-15 by K. Itagaki in the outskirts of host galaxy NGC~2525 \citep{2018TNSTR..57....1I}, a barred spiral galaxy at $z=0.00527$ \citep{1991rc3..book.....D} with a distance modulus of $\mu=31.1$ mag and a distance of $16.8$ Mpc \citep{2013AJ....146...86T}.
SN\,2018gv was classified as a young, normal Type Ia supernova from an optical spectrum obtained with Keck Observatory on 2018-01-16 by \citet[][who also report the presence of \ion{C}{II} in this early spectrum]{2018TNSCR..75....1S}, and from an optical spectrum obtained with the ESO New Technology Telescope on 2018-01-16 by \citet{2018ATel11177....1B}.

\citet{2020ApJ...902...46Y} present early-time data for SN\,2018gv, showing that it reached a $B$-band peak of $12.6$ ($M_B = -19.1$) mag on 2018-01-31 and declined with $\Delta m_{15}(B) = 0.96$ mag.
Early optical spectropolarimetry obtained and analyzed by \citet{2020ApJ...902...46Y} show low continuum polarization and moderate line polarization (for the photospheric absorption feature \ion{Si}{II}).
They find that these observations indicate a high degree of spherical symmetry that is inconsistent with double-degenerate violent mergers (e.g., \citealt{2011A&A...528A.117P}), and consistent with both the double-detonation \citep[e.g.,][]{1994ApJ...423..371W} and delayed detonation \citep[e.g.,][]{2013MNRAS.429.1156S} models and predictions of their polarization signatures \citep{2016MNRAS.462.1039B}. 
\citet{2020ApJ...902...46Y} also demonstrate that SN\,2018gv resembles SN\,2011fe for the first $\sim100$ days, and exhibits a low \ion{Si}{II} velocity gradient during the days after peak brightness ($33.3 \pm 6.4$~$\rm km\ s^{-1}\ day^{-1}$), which agrees with the measurement from our early-time optical spectra of $29 \pm 4$~$\rm km\ s^{-1}\ day^{-1}$ in Table~\ref{tab:sne}.

A radio non-detection of SN\,2018gv obtained on 2018-01-18.6 with the Australia Telescope Compact Array did not reveal any CSM, constraining the mass-loss rate of a non-degenerate companion star to $<1.3\times10^{-8}$ $\rm M_{\odot}\ yr^{-1}$ \citep{2018ATel11211....1R,2020ApJ...890..159L}. This radio non-detection for SN\,2018gv rules out a symbiotic progenitor in which a white dwarf accretes material from a giant star, but cannot rule out progenitor systems containing, e.g., Roche Lobe overflow from a giant or main sequence star, or recurrent or quiescent novae \citep{2020ApJ...890..159L}.

\subsubsection{Late-time SN\,2018gv}

The fact that SN\,2018gv appears to be a normal, SN\,2011fe-like SN\,Ia with a high degree of spherical symmetry is well-supported by our nebular-phase spectrum.
In Figure~\ref{fig:nebspec_wSN11fe} we find a good match to the normal SN\,Ia 2011fe at similar phases, confirming that SN\,2018gv is a typical SN\,Ia.
A high degree of spherical symmetry would manifest as low velocity nebular features, and indeed we find that the [\ion{Fe}{II}]+[\ion{Ni}{II}]~$\lambda$7200~\AA\ feature of SN\,2018gv exhibits an absolute velocity $<1000$~$\rm km\ s^{-1}$, which is among the lowest in our sample (Figure~\ref{fig:asym}).

Radio non-detections ruled out a symbiotic system as the progenitor for SN\,2018gv \citep{2020ApJ...890..159L}, and so no nebular-phase H$\alpha$ is expected for SN\,2018gv.
Sand et al. (in prep.) perform a re-analysis of SN\,2018gv's nebular-phase spectrum which prioritizes the mitigation of host-galaxy H$\alpha$ emission, and find no evidence of an H$\alpha$ emission feature similar to expectations of swept-up material from a non-degenerate companion star (FWHM~$\sim$1000 $\rm km\ s^{-1}$, velocity $\pm1000$ $\rm km\ s^{-1}$), in agreement with expectations of the radio non-detection.

\subsection{SN\,2018oh, a SN\,Ia with Early-Time \textit{Kepler} Photometry} \label{ssec:disc_18oh}

\subsubsection{Early-time SN\,2018oh}

SN\,2018oh was discovered as ASASSN-18bt on 2018-02-04 with an apparent $V$-band magnitude of $15.15$ mag (Vega; \citealt{2018TNSTR.150....1S}, and spectroscopically classified as a normal SN\,Ia about one week before peak brightness \citep{2018TNSCR.159....1L}.
SN\,2018oh resides in the central region of dwarf intermediate barred spiral galaxy UGC~04780 at $z=0.01098$ \citep{1990ApJS...72..245S}.

\cite{2019ApJ...870...13S} use time-series optical photometry of SN\,2018oh from the {\it Kepler Space Telescope} (30-minute cadence) to show that the four days after first light exhibit a near linear rise, and that the pre-maximum light-curve is well fit by a double power law which suggests two independent luminosity sources.
They demonstrate that the slower initial rise of SN\,2018oh is not predicted by models of SN\,Ia ejecta impacting a non-degenerate companion, but is instead more compatible with the presence of $^{56}$Ni in the outer layers of the ejecta. 
\cite{2019ApJ...870...13S} also show that the optical data, along with X-ray nondetections from the {\it Swift Space Telescope}, together suggest an absence of any CSM as further evidence against the presence of a non-degenerate companion star.

\cite{2019ApJ...870L...1D} combine the {\it Kepler} data with their own ground-based optical photometry obtained within hours of the explosion, and show that the additional emission causing the early-time light curve bump was blue, with a blackbody temperature of $\sim17500$ $\rm K$.
They find that this evidence more strongly supports the hypothesis that this emission originates due to the impact of the SN\,Ia ejecta on a non-degenerate companion, and anticipate that the late-time spectra might thus exhibit H$\alpha$ emission.
\cite{2019ApJ...870L...1D} make the point that aside from this fleeting evidence in the first four days, SN\,2018oh appears to be a normal SN\,Ia -- and thus that a much higher fraction of SN\,Ia might have non-degenerate companions than is generally thought.

\cite{2019ApJ...870...12L} analyze the optical through near-infrared photometry (including the {\it Kepler} data) for the first 140 days after explosion.
They report that SN\,2018oh reached peak B-band luminosity on 2018 Feb 13.7 with an apparent brightness of $B=14.32\pm0.01$ mag ($M_B= -19.47\pm0.10$ mag), exhibited a decline rate of $\Delta m_{15}(B) = 0.96\pm0.03$ mag, and synthesized a nickel mass of $M_{\rm Ni} = 0.55 \pm 0.04$ $\rm M_{\odot}$.
\citet{2020ApJ...892..121K} also analyze SN\,2018oh's light curve and report a slightly higher $^{56}$Ni mass of $0.598 \pm 0.059$ $\rm M_{\odot}$.

\cite{2019ApJ...870...12L} present and analyze early-time optical spectroscopy of SN\,2018oh and classify it as an LVG event.
In addition, they show how \ion{C}{II} absorption is seen until 3 weeks after peak brightness, the latest such detection yet, indicating that unburned carbon exists deeper in the ejecta than for most SNe\,Ia.
However, whether the carbon's depth could be connected to mixing that brings $^{56}$Ni to the surface, or whether abundance stratification or a low-metallicity progenitor allow the carbon feature to be visible for longer, would require additional modeling.

\subsubsection{Late-time SN\,2018oh}

We find that the nebular-phase spectrum of SN\,2018oh is quite similar to that of the normal SN\,Ia 2011fe (Figure~\ref{fig:nebspec_wSN11fe}), as expected.
If the early-time blue component of SN\,2018oh's light curve, which was revealed by the \textit{Kepler} photometry, is due to the ejecta interacting with CSM or a non-degenerate binary companion then we could expect to find H$\alpha$ emission in the nebular-phase spectrum.
However, both \citet{2019ApJ...872L..22T} and \citet{2019ApJ...870L..14D} present and analyze late-time optical spectra for SN\,2018oh, and place upper limits on hydrogen and helium emission which strongly indicate that SN\,2018oh was a double-degenerate system of two white dwarf stars.
Additionally, \citet{2020MNRAS.493.1044T} find no evidence of stripped companion emission for SN\,2018oh.
This lack of late-time hydrogen or helium emission also agrees with the particular physical interpretation for SN\,2018oh offered by \cite{2019ApJ...870L..14D}, who show that the early-time blue bump could have originated from the interaction of SN\,Ia ejecta with a disk of material formed during the merger process of a white dwarf binary system. 
Sand et al. (in prep.) use the nebular spectra presented in this paper to place upper limits on the hydrogen emission as well.
As with SN\,2018gv, they re-process the spectra to account for host galaxy H$\alpha$ emission, and find no evidence of swept-up hydrogen from a non-degenerate companion star.

As previously discussed for SN\,2017cbv in Section~\ref{ssec:disc_17cbv}, mixing of $^{56}$Ni into the outer layers is a potential alternative explanation for excess blue emission at early times, a physical trait which could manifest as broad emission features such as  [\ion{Co}{III}] at nebular phases.
However, SN\,2018oh's [\ion{Co}{III}] emission line does not appear to be broad, and was actually among the narrowest in our sample (Section~\ref{ssec:nebobs_linepars}, Tables~\ref{tab:linepars_direct} and \ref{tab:linepars_gaussian} (Appendix~\ref{app:tables}), and as seen in the bottom panel of Figure~\ref{fig:LumxWidth}). 
The dense time-series of early-time optical spectra for SN\,2018oh presented by \citet{2019ApJ...870...12L} exhibited the latest ever detection of \ion{C}{II}, at three weeks after peak brightness.
Such deep \ion{C}{II} could be related to mixing in the outer layers of the white dwarf, or could indicate a lower-metallicity progenitor in which there are less iron group elements to ``smear out" the carbon feature \citep{2019ApJ...870...12L,2019ApJ...871..250H}.
Figure~\ref{fig:nebspec_wSN11fe} shows that SN\,2018oh has a smaller [\ion{Ni}{II}] to [\ion{Fe}{II}] ratio than SN\,2011fe, which would be consistent with a lower-metallicity progenitor in which less stable nickel is synthesized.
However, as described in Section~\ref{ssec:ana_NiFe} we were unable to more robustly measure the Ni/Fe ratio for SN\,2018oh from multi-component Gaussian fits.

\subsection{SN\,2018yu, a Normal SN\,Ia with a Near-Infrared Nebular Spectrum} \label{ssec:disc_18yu}

\subsubsection{Early-time SN\,2018yu}

SN\,2018yu was discovered by the DLT40 survey as DLT18i on 2018-03-01 with an $r$-band apparent brightness of $17.6$ mag \citep{2018ATel11371....1S}, and within 24 hours was spectroscopically classified as a young SN\,Ia \citep{2018ATel11374....1Z}.
SN\,2018yu is located in the outer disk of inclined spiral galaxy NGC~1888 at $z=0.00811$ \citep{1991rc3..book.....D}, with a distance modulus of $32.85\pm0.4$ mag and a distance of $37.1$ Mpc \citep{2007A&A...465...71T}.

The Las Cumbres optical photometry obtained around the time of peak brightness indicates a $B$-band decline rate of $\Delta m_{\rm 15}(B) \approx 1.05\pm0.05$ mag and an intrinsic peak brightness of $M_B\approx-19.1\pm0.5$ mag (Section~\ref{ssec:sne_earlyphot}).
The light curve of SN\,2018yu is consistent with a normal SN\,Ia on the Phillips relation  \citep{1999AJ....118.1766P}, and the early time spectra are consistent with a normal, LVG-type SN\,Ia.

\subsubsection{Late-time SN\,2018yu}

SN\,2018yu was a normal SN\,Ia at early times, and it appears to be normal at late times as well: its nebular-phase spectrum in the bottom panel of Figure~\ref{fig:nebspec_wSN11fe} demonstrates that it is very similar to the prototypical SN\,2011fe.
Also, in Section~\ref{sec:ana} we show correlations between several measured parameters for the nebular-phase emission lines of the SNe\,Ia in our sample, and SN\,2018yu always falls in the midst of the range of values.
Sand et al. (in prep.) show that there is no signature of H$\alpha$ to indicate a non-degenerate companion for SN\,2018yu, similar to the other SNe\,Ia in this sample.
Additionally, \citet{2020MNRAS.493.1044T} find no evidence of stripped companion emission for SN\,2018yu.

\begin{figure}
\includegraphics[width=\columnwidth,trim={1cm 0.7cm 1.5cm 1cm},clip]{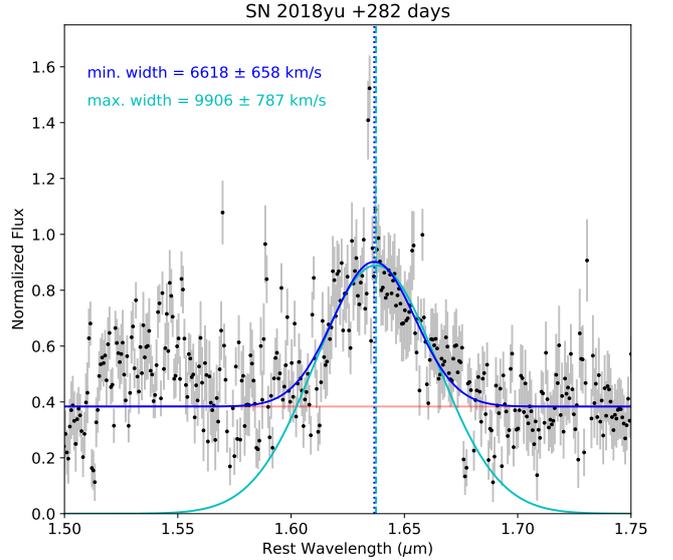}
\caption{The [\ion{Fe}{II}]~$\lambda$1.644~$\mu$m emission line in the +282 days NIR spectrum of SN\,2018yu, obtained using Flamingos-2 on Gemini-South.
As described in the text, the dark blue line is a single-Gaussian fit after subtracting a maximum flat continuum (red line; determined by the red side minimum of the [\ion{Fe}{II}]~$\lambda$1.644~$\mu$m feature), which provides a minimum line width measurement
The lighter teal line is a single-Gaussian fit assuming no underlying continuum, which provides the maximum possible line width.
The minimum and maximum line widths provide constraints on the progenitor's central density \citep{2006NuPhA.777..579H,2015ApJ...806..107D,2018ApJ...861..119D}.
Both fit methods also provide consistent line velocity measurements (dashed vertical lines) that agree with the [\ion{Fe}{II}] velocity measured in the optical.
}
\label{fig:18yuNIR}
\end{figure}

To enrich our analysis of SN\,2018yu, we include an NIR nebular-phase spectrum obtained with Flamingos-2 at Gemini-South Observatory at 282 days past peak brightness (Section~\ref{ssec:nebobs_nir}), which is shown in Figure \ref{fig:18yuNIR}.
One of the benefits of using NIR and optical spectra obtained at a similar nebular-phase epoch is that the [\ion{Fe}{II}]~$\lambda$1.644~$\mu$m emission line is the strongest and most well-isolated iron feature in the NIR \citep{2015ApJ...806..107D,2018ApJ...861..119D}, and originates from the same region of the SN ejecta as the [\ion{Fe}{II}]~$\lambda$7155~\AA\ line.
This NIR emission line can be used to verify optical [\ion{Fe}{II}] line velocities and thus address line blending concerns in the 7200~\AA\ feature.
We measured the [\ion{Fe}{II}]~$\lambda$1.644~$\mu$m line velocity with a least squares fit of a single Gaussian with a flat continuum, finding a velocity of $-1333 \pm 385$~$\rm km\ s^{-1}$.
The uncertainty was determined using a Monte Carlo approach: introducing random noise to the smoothed spectrum and calculating the dispersion of the line velocity from 5000 iterations.
This NIR-derived iron velocity could be used to fix the iron velocity in our multi-Gaussian fits to the iron and nickel feature at $\lambda$7200~\AA.
However, it is not necessary to redo that fit for SN\,2018yu because this NIR velocity is already in very good agreement with the result of that fit, $-1384.03\pm71.11$~$\rm km\ s^{-1}$ (Section~\ref{ssec:ana_NiFe}).

We also use the nebular-phase NIR spectrum of SN\,2018yu to measure the width of the [\ion{Fe}{II}]~$\lambda$1.644~$\mu$m line.
Theoretical models suggest that carbon-oxygen white dwarfs with higher central densities will exhibit broader [\ion{Fe}{II}]~$\lambda$1.644~$\mu$m features: a higher central density leads to a larger volume of stable material in the core, and the radioactive material -- the source of the [\ion{Fe}{II}] emission -- extends to higher velocities \citep[e.g.][]{2004ApJ...617.1258H,2006ApJ...652L.101M,2010ApJ...708.1703M,2015A&A...573A...2S}.
We follow the analysis method of \citet{2018ApJ...861..119D} to measure the range of maximum and minimum possible line widths for the [\ion{Fe}{II}]~$\lambda$1.644~$\mu$m line by assuming no continuum and a maximum flat continuum, respectively, as shown in Figure~\ref{fig:18yuNIR} and described in the figure's caption.
Compared to the models published in \citet{2018ApJ...861..119D}, our range of measured line widths for [\ion{Fe}{II}]~$\lambda$1.644~$\mu$m for SN\,2018yu corresponds to central densities of $\sim0.5$ to $0.9 \times 10^{9}$ $\rm g\ cm^{-3}$, which lies on the lower end of central densities included in these models.
This could imply that the progenitor white dwarf of SN\,2018yu experienced a high accretion rate from its companion, leading to a prompt explosion before it evolved to a higher central density.
In this respect SN\,2018yu is similar to SN\,2014J, which \citet{2018ApJ...861..119D} also suggest to have a low central density (and which, like SN\,2018yu, exploded in the disk of a dusty host galaxy with a young stellar population).

\section{Conclusions}\label{sec:con}

We have presented nine new nebular-phase optical spectra from Gemini Observatory for seven nearby Type Ia supernovae with early-time optical photometric and spectroscopic monitoring from the Las Cumbres Observatory Global Supernova Project.
The combination of early- and late-time data is essential for a comprehensive understanding of the explosion mechanisms and progenitor systems of SNe\,Ia, especially as they continue to be used as cosmological probes.

In this work we have shown that as our collection of nebular-phase observations continue to grow, analysis of ever larger samples continues to reveal how the temperature and ionization state of the nebula evolves with time; whether potential asymmetries in the explosion and nucleosynthetic material exist; consistency with various explosion models and WD progenitor masses; and the probability of WD-WD collisions or the presence of non-degenerate companion stars. 
Although most of the flux in the nebular spectrum is in smooth, broad emission features, our analysis has revealed that high signal-to-noise and high spectral resolution are beneficial for  constraining the shape of the nebular features (i.e., fitting single- versus double- or multi-component features) and the presence of narrow hydrogen emission. 
Our analysis has also revealed the importance of obtaining phase-matched NIR nebular-phase spectra.
Having the relatively isolated iron lines in the NIR would allow the iron line parameters to be held fixed when fitting the iron and nickel blended feature in the optical, and provide more accurate measures of the amount of stable nickel formed in the explosion.

We have also shown that analyses of individual SN\,Ia events remains a useful and enlightening exercise -- especially for SNe\,Ia with unique observations or some kind of peculiarity at early times, such as the ``blue bump" of SN\,2017cbv, the depressed NUV flux of SN\,2017erp, the spectropolarimetry of SN\,2018gv, or the \ion{C}{II} of SN\,2018oh. 
In particular, nebular-phase spectroscopy of subluminous, rapidly-evolving ``91bg-like" and transitional SNe\,Ia such as SN\,2017fzw remain rare and potentially high-impact.
We recommend that the community continue to target subluminous, peculiar, and/or otherwise rare SNe\,Ia for nebular-phase follow-up in order to continue to advance our understanding of their progenitor systems and explosion mechanisms.

\section*{Acknowledgements}

This work was enabled by observations made from the Gemini North telescope, located within the Maunakea Science Reserve and adjacent to the summit of Maunakea.
The authors wish to recognize and acknowledge the very significant cultural role and reverence that the summit of Mauna Kea has always had within the indigenous Hawaiian community.
We are most fortunate to have the opportunity to conduct observations from this mountain.

The international Gemini Observatory, a program of NSF's NOIRLab, is managed by the Association of Universities for Research in Astronomy (AURA) under a cooperative agreement with the National Science Foundation on behalf of the Gemini Observatory partnership: the National Science Foundation (United States), National Research Council (Canada), Agencia Nacional de Investigaci\'{o}n y Desarrollo (Chile), Ministerio de Ciencia, Tecnolog\'{i}a e Innovaci\'{o}n (Argentina), Minist\'{e}rio da Ci\^{e}ncia, Tecnologia, Inova\c{c}\~{o}es e Comunica\c{c}\~{o}es (Brazil), and Korea Astronomy and Space Science Institute (Republic of Korea).
The observations were made under Gemini programs GS-2018A-Q-315, GS-2018B-Q-209, and GN-2018B-Q-213 (optical spectra), and GS-2018B-Q-218 (NIR spectrum).
The data and calibration files were obtained from the Gemini Observatory Archive at NSF's NOIRLab, and processed using the Gemini IRAF package.

This work makes use of observations from the Las Cumbres Observatory global telescope network. The Las Cumbres Observatory team is supported by NSF grants AST-1911225 and AST-1911151.

We acknowledge ESA Gaia, DPAC and the Photometric Science Alerts Team\footnote{ \url{http://gsaweb.ast.cam.ac.uk/alerts}}.

M.L.G. and T.D.K. acknowledge support from the DIRAC Institute in the Department of Astronomy at the University of Washington.
The DIRAC Institute is supported through generous gifts from the Charles and Lisa Simonyi Fund for Arts and Sciences, and the Washington Research Foundation.
M.L.G. and T.D.K. thank Brigitta Sip\H{o}cz for assistance with implementing the Bayesian Information Criterion.

Time domain research by D.J.S. is supported by NSF grants AST-1821987, 1813466, \& 1908972, and by the Heising-Simons Foundation under grant \#2020-1864. 

L.G. acknowledges financial support from the Spanish Ministry of Science, Innovation and Universities (MICIU) under the 2019 Ram\'on y Cajal program RYC2019-027683 and from the Spanish MICIU project PID2020-115253GA-I00.

X.W. is supported by the National Science Foundation of China (NSFC grants 12033003 and 11633002), the Scholar Program of Beijing Academy of Science and Technology (DZ:BS202002), and the Tencent XPLORER Prize.

\section*{Data Availability}

The raw optical spectra and calibration frames from Gemini Observatory are available via the Gemini Observatory Archive at \url{archive.gemini.edu}.
The reduced spectra will be made available in the Weizmann Interactive Supernova Data Repository (WISeREP) at \url{www.wiserep.org}.
The reduced data and derived measurements in this article will be shared on  request to the corresponding author.

\bibliographystyle{mnras}
\bibliography{apj-jour,myrefs}

\appendix
\section{Large Tables}\label{app:tables}

\begin{table*}
\centering
\begin{tabular}{|llclc|}
\hline
SN Name & UT Date$^{(a)}$     & Phase & Instrument \& Configuration$^{(b)}$  & Exposure \\
  &  & [days] &  & Time [s]  \\
\hline
2017cbv & {\bf 2018-07-08} & {\bf 466} & GMOS-S: B600,450,4x4 & 4x1200 \\
          & 2018-07-11 & 469 & GMOS-S: R400,750,4x4 &  4x1200 \\
2017ckq & {\bf 2018-04-12} & {\bf 369} & GMOS-S: B600,450,4x4 & 6x1200 \\
          & 2018-06-17 & 435 & GMOS-S: R400,750,4x4 & 6x1200  \\
2017erp & {\bf 2018-03-16} & {\bf 259} & GMOS-S: B600,520,2x2 & 4x300 \\
          & 2018-03-22 & 265 & GMOS-S: R400,750,2x2 & 2x300 \\
          & 2018-04-12 & 286 & GMOS-S: R400,750,2x2 & 4x300  \\
2017fzw & {\bf 2018-04-13} & {\bf 234} & GMOS-S: B600,520,2x2 & 4x300 \\
          & 2018-05-03 & 254 & GMOS-S: R400,750,2x2 & 4x300 \\
2018gv  & {\bf 2018-11-16} & {\bf 289} & GMOS-S: B600,450,2x2 & 3x400 \\
          & 2018-11-17 & 290 & GMOS-S: R400,750,2x2 & 3x600 \\
          & {\bf 2019-01-10} & {\bf 344} & GMOS-S: B600,450,2x2 & 3x400 \\
          & 2019-01-28 & 362 & GMOS-S: R400,750,2x2 & 3x600 \\
2018oh  & {\bf 2018-11-04} & {\bf 264} & GMOS-N: B600,450,4x4 & 3x1200 \\
          & 2018-12-07 & 297 & GMOS-N: R400,750,4x4 & 3x1200 \\ 
2018yu  & {\bf 2018-10-05} & {\bf 202} & GMOS-S: B600,450,2x2 & 3x400 \\
          & 2018-10-05 & 202 & GMOS-S: R400,750,2x2 & 3x600 \\
          & 2018-10-07 & 204 & SALT: PG0900 & 2386 \\
          & {\bf 2018-12-29} & {\bf 287} & GMOS-S: B600,450,4x4 & 3x1200 \\
          & 2019-01-06 & 295 & GMOS-S: R400,750,4x4 & 4x1200 \\
\hline
\end{tabular}
\caption{Late-time optical spectroscopy for SNe\,Ia presented in this work. (a) Boldface dates and phases represent epochs, for which the R400 was flux-scaled and joined to the B600 spectrum. (b) Configurations for GMOS, the Gemini Multi-Object Spectrograph, list the grating (R400 or B600), the central wavelength in $\rm nm$, and the binning. All GMOS observations were done with the 0.75\arcsec\ longslit. \label{tab:specobs}}
\end{table*}

\begin{table*} 
\centering 
\caption{Directly measured parameters of nebular-phase emission lines.} 
\label{tab:linepars_direct} 
\begin{tabular}{|lllllll|} 
\hline 
\hline 
SN Name & UT Date & Phase & Line & Velocity & FWHM & Integrated Flux \\ 
 & & [days] & & [$\rm km\ s^{-1}$] & [$\rm km\ s^{-1}$] & [$\rm 10^{-15}\ erg\ s^{-1}\ cm^{-2}$] \\ 
\hline 
2017cbv & 2018-07-08 & 466 & [FeIII] & $   919 \pm  138 $ &  $   9496 \pm  324 $ &  $   2.545 \pm  0.608 $ \\ 
\rowcolor[gray]{.92}
2017ckq & 2018-04-12 & 369 & [FeIII] & $  -348 \pm  140 $ &  $  10855 \pm  338 $ &  $   6.212 \pm  1.175 $ \\ 
2017erp & 2018-03-16 & 259 & [FeIII] & $ -1308 \pm   41 $ &  $  11449 \pm   97 $ &  $  35.383 \pm  2.412 $ \\ 
2017erp & 2018-03-16 & 259 & [CoIII] & $  -502 \pm  102 $ &  $   8275 \pm  222 $ &  $   3.116 \pm  0.768 $ \\ 
\rowcolor[gray]{.92}
2017fzw & 2018-04-13 & 234 & [FeIII] & $  -715 \pm   49 $ &  $  10504 \pm  111 $ &  $  19.680 \pm  1.715 $ \\ 
\rowcolor[gray]{.92}
2017fzw & 2018-04-13 & 234 & [CoIII] & $   738 \pm  186 $ &  $  10032 \pm  430 $ &  $   3.052 \pm  1.217 $ \\ 
2018gv & 2018-11-16 & 289 & [FeIII] & $ -1200 \pm   29 $ &  $  12329 \pm   72 $ &  $  55.357 \pm  2.289 $ \\ 
2018gv & 2018-11-16 & 289 & [CoIII] & $  -817 \pm  180 $ &  $   9885 \pm  445 $ &  $   4.416 \pm  1.327 $ \\ 
\rowcolor[gray]{.92}
2018gv & 2019-01-10 & 344 & [FeIII] & $ -1078 \pm   65 $ &  $  13082 \pm  141 $ &  $  22.807 \pm  1.953 $ \\ 
2018oh & 2018-11-04 & 264 & [FeIII] & $ -1867 \pm  119 $ &  $  12106 \pm  283 $ &  $  18.198 \pm  2.362 $ \\ 
2018oh & 2018-11-04 & 264 & [CoIII] & $  -726 \pm  119 $ &  $   8133 \pm  289 $ &  $   2.010 \pm  0.435 $ \\ 
\rowcolor[gray]{.92}
2018yu & 2018-10-05 & 202 & [FeIII] & $ -1573 \pm   39 $ &  $  11927 \pm   91 $ &  $  56.385 \pm  4.225 $ \\ 
\rowcolor[gray]{.92}
2018yu & 2018-10-05 & 202 & [CoIII] & $  -563 \pm   58 $ &  $   8418 \pm  142 $ &  $  10.583 \pm  1.366 $ \\ 
2018yu- & 2018-10-07 & 204 & [CoIII] & $  -512 \pm   41 $ &  $   8635 \pm   99 $ &  $   4.207 \pm  0.459 $ \\ 
\rowcolor[gray]{.92}
2018yu & 2018-12-29 & 287 & [FeIII] & $  -696 \pm   42 $ &  $  11091 \pm   94 $ &  $  22.663 \pm  1.157 $ \\ 
\rowcolor[gray]{.92}
2018yu & 2018-12-29 & 287 & [CoIII] & $  -267 \pm  223 $ &  $   9406 \pm  540 $ &  $   1.547 \pm  0.505 $ \\ 
\hline 
\end{tabular} 
\end{table*} 

\begin{table*} 
\centering 
\caption{Gaussian fit parameters of nebular-phase emission lines.} 
\label{tab:linepars_gaussian} 
\begin{tabular}{|lllllll|} 
\hline 
\hline 
SN Name & UT Date & Phase & Line & Velocity & FWHM & Integrated Flux \\ 
 & & [days] & & [$\rm km\ s^{-1}$] & [$\rm km\ s^{-1}$] & [$\rm 10^{-15}\ erg\ s^{-1}\ cm^{-2}$] \\ 
\hline 
2017cbv & 2018-07-08 & 466 & [FeIII] & $  1541 \pm  170 $ &  $   8188 \pm  386 $ &  $   2.472 \pm  0.077 $ \\ 
2017cbv & 2018-07-08 & 466 & [FeII] & $  1163 \pm  478 $ &  $   9472 \pm  667 $ &  $   6.468 \pm  0.729 $ \\ 
2017cbv & 2018-07-08 & 466 & [NiII] & $  1325 \pm  630 $ &  $   9250 \pm 1024 $ &  $   4.471 \pm  0.740 $ \\ 
\rowcolor[gray]{.92}
2017ckq & 2018-04-12 & 369 & [FeIII] & $  -384 \pm  119 $ &  $  10372 \pm  317 $ &  $   6.180 \pm  0.125 $ \\ 
2017erp & 2018-03-16 & 259 & [FeIII] & $ -1147 \pm   35 $ &  $  10919 \pm   90 $ &  $  35.566 \pm  0.207 $ \\ 
2017erp & 2018-03-16 & 259 & [CoIII] & $  -286 \pm   92 $ &  $   7570 \pm  221 $ &  $   3.190 \pm  0.069 $ \\ 
2017erp & 2018-03-16 & 259 & [FeII] & $  1210 \pm   93 $ &  $   7885 \pm  196 $ &  $   9.931 \pm  0.252 $ \\ 
2017erp & 2018-03-16 & 259 & [NiII] & $  1545 \pm  138 $ &  $   7778 \pm  263 $ &  $   6.961 \pm  0.244 $ \\ 
\rowcolor[gray]{.92}
2017fzw & 2018-04-13 & 234 & [FeIII] & $  -581 \pm   58 $ &  $   9714 \pm  148 $ &  $  19.556 \pm  0.203 $ \\ 
\rowcolor[gray]{.92}
2017fzw & 2018-04-13 & 234 & [CoIII] & $  1072 \pm  172 $ &  $   8945 \pm  443 $ &  $   3.030 \pm  0.109 $ \\ 
\rowcolor[gray]{.92}
2017fzw & 2018-04-13 & 234 & [FeII] & $  2469 \pm  131 $ &  $  10802 \pm  343 $ &  $   6.341 \pm  0.161 $ \\ 
\rowcolor[gray]{.92}
2017fzw & 2018-04-13 & 234 & [NiII] & $  2986 \pm   86 $ &  $   5252 \pm  170 $ &  $   2.929 \pm  0.129 $ \\ 
2018gv & 2018-11-16 & 289 & [FeIII] & $ -1181 \pm   26 $ &  $  11519 \pm   77 $ &  $  55.748 \pm  0.234 $ \\ 
2018gv & 2018-11-16 & 289 & [CoIII] & $  -728 \pm  141 $ &  $   9730 \pm  334 $ &  $   4.459 \pm  0.113 $ \\ 
2018gv & 2018-11-16 & 289 & [FeII] & $  -761 \pm   46 $ &  $   7028 \pm  124 $ &  $   8.704 \pm  0.131 $ \\ 
2018gv & 2018-11-16 & 289 & [NiII] & $  -214 \pm   82 $ &  $   7280 \pm  229 $ &  $   5.087 \pm  0.128 $ \\ 
\rowcolor[gray]{.92}
2018gv & 2019-01-10 & 344 & [FeIII] & $  -940 \pm   46 $ &  $  12108 \pm  124 $ &  $  22.923 \pm  0.157 $ \\ 
\rowcolor[gray]{.92}
2018gv & 2019-01-10 & 344 & [FeII] & $  -874 \pm  154 $ &  $   7075 \pm  326 $ &  $   5.121 \pm  0.293 $ \\ 
\rowcolor[gray]{.92}
2018gv & 2019-01-10 & 344 & [NiII] & $  -416 \pm  259 $ &  $   9013 \pm  565 $ &  $   4.897 \pm  0.281 $ \\ 
2018oh & 2018-11-04 & 264 & [FeIII] & $ -1678 \pm   90 $ &  $  12165 \pm  235 $ &  $  18.188 \pm  0.238 $ \\ 
2018oh & 2018-11-04 & 264 & [CoIII] & $  -999 \pm   97 $ &  $   8039 \pm  247 $ &  $   1.994 \pm  0.047 $ \\ 
2018oh & 2018-11-04 & 264 & [FeII] & $ -1489 \pm   87 $ &  $   6212 \pm  256 $ &  $   3.620 \pm  0.143 $ \\ 
2018oh & 2018-11-04 & 264 & [NiII] & $ -1546 \pm  219 $ &  $   7445 \pm  536 $ &  $   2.234 \pm  0.141 $ \\ 

\rowcolor[gray]{.92}
2018yu & 2018-10-05 & 202 & [FeIII] & $ -1592 \pm   38 $ &  $  10823 \pm  108 $ &  $  57.145 \pm  0.397 $ \\ 
\rowcolor[gray]{.92}
2018yu & 2018-10-05 & 202 & [CoIII] & $  -600 \pm   53 $ &  $   7957 \pm  137 $ &  $  10.601 \pm  0.124 $ \\ 
\rowcolor[gray]{.92}
2018yu & 2018-10-05 & 202 & [FeII] & $  -924 \pm   60 $ &  $   6511 \pm  134 $ &  $   7.872 \pm  0.166 $ \\ 
\rowcolor[gray]{.92}
2018yu & 2018-10-05 & 202 & [NiII] & $ -1154 \pm  121 $ &  $   7857 \pm  269 $ &  $   5.308 \pm  0.174 $ \\ 

2018yu & 2018-10-07 & 204 & [CoIII] & $  -391 \pm   39 $ &  $   8037 \pm  106 $ &  $   4.239 \pm  0.039 $ \\ 
2018yu & 2018-10-07 & 204 & [FeII] & $  -992 \pm   67 $ &  $   6340 \pm  129 $ &  $   4.059 \pm  0.097 $ \\ 
2018yu & 2018-10-07 & 204 & [NiII] & $ -1384 \pm  136 $ &  $   8001 \pm  314 $ &  $   2.857 \pm  0.098 $ \\ 

\rowcolor[gray]{.92}
2018yu & 2018-12-29 & 287 & [FeIII] & $  -653 \pm   38 $ &  $  10842 \pm  106 $ &  $  23.012 \pm  0.140 $ \\ 
\rowcolor[gray]{.92}
2018yu & 2018-12-29 & 287 & [CoIII] & $  -344 \pm  194 $ &  $   8540 \pm  624 $ &  $   1.571 \pm  0.071 $ \\ 
\rowcolor[gray]{.92}
2018yu & 2018-12-29 & 287 & [FeII] & $  -880 \pm   91 $ &  $   6631 \pm  264 $ &  $   4.315 \pm  0.167 $ \\ 
\rowcolor[gray]{.92}
2018yu & 2018-12-29 & 287 & [NiII] & $  -780 \pm  166 $ &  $   7994 \pm  404 $ &  $   3.585 \pm  0.163 $ \\ 
\hline 
\end{tabular} 
\end{table*}

\bsp	
\label{lastpage}
\end{document}